 \documentclass[final,3p,times]{elsarticle}

\usepackage{amssymb}
 \usepackage{amsthm}

\usepackage{appendix}
\usepackage{xcolor}
\journal{Journal of Fluids and Structures}

\begin{document}

\begin{frontmatter}



\title{Effect of structural parameters on the synchronization characteristics in a stall-induced aeroelastic system}


\author[label1]{Dheeraj Tripathi}
\address[label1]{Department of Mechanical Engineering, Shiv Nadar
Institute of Eminence, Greater Noida 203207, India}

\author[label2]{Chandan Bose}
\address[label2]{Aerospace Engineering, School of Metallurgy and Materials, The University of Birmingham, United Kingdom}

\author[label3]{Sirshendu Mondal}
\address[label3]{Department of Mechanical Engineering, National Institute of Technology Durgapur, Durgapur, 713209, India}

\author[label4]{J Venkatramani\corref{mycorrespondingauthor}}
\ead{vramani465@gmail.com}
\address[label4]{Department of Aerospace Engineering, Amrita School of Engineering, Amrita Vishwa Vidyapeetham, Coimbatore 641112, India}

\begin{abstract}
This study focuses on discerning the role of structural parameters on the bifurcation characteristics and the underlying synchronization mechanism in an aeroelastic system undergoing nonlinear stall behavior. To that end, wind tunnel experiments are performed on a NACA 0012 airfoil capable of undergoing bending (plunging) and torsional (pitching) oscillations under scenarios involving nonlinear aerodynamic loads, \textit{i.e.}, dynamic stall conditions. Flow conditions under both deterministic/sterile flows and fluctuating/stochastic flows are fostered. The structure possesses continuous or polynomial-type stiffness nonlinearities and therefore is an aeroelastic experiment involving both structural and aerodynamic nonlinearities. We discern the bifurcation routes for a range of key structural parameters, such as frequency ratio, static imbalance, and the extent of structural nonlinearity. In addition to interesting and atypical routes to stall-induced instabilities, we systematically demonstrate the role of modal interactions - via a synchronization analysis - over the manifestation of these instabilities. To the best of the authors' knowledge, this is perhaps the first study to document the role of multiple structural parameters on a stall-induced aeroelastic system and in turn cast the physical mechanism behind these dynamical transitions from the vantage of synchronization.
\end{abstract}




\end{frontmatter}


\section{Introduction}

Aeroelastic systems, being rife with nonlinearities and multiple modes, are in turn susceptible to instabilities such as flutter. In scenarios involving nonlinear aerodynamic loads, often under dynamic stall conditions, the aeroelastic system undergoes violent pitch-driven limit cycle oscillations (LCOs) called stall flutter. A key feature of stall flutter is the dominance of pitch mode over the other modes \cite{dimitriadis2009bifurcation,vsidlof2016experimental,goyaniuk2020pitch,vishal2021routes}. Typically viewed as a single degree of freedom (DoF), torsionally dominant oscillatory instability, stall flutter is frequently encountered in wind turbine blades, helicopter blades, aircraft wings, micro aerial vehicles (MAVs), and similar slender structures subjected to nonlinear aerodynamic loads \cite{fung2008introduction}. However, recent studies by Venkatramani and co-workers \cite{bethi2020response,vishal2021routes,tripathi2022stall,tripathi2023frequency} have underscored the role of plunge DoF as well in a stall flutter problem. This end of discerning the rather complex nonlinear fluid-structure interaction was done by casting the problem from the vantage of synchronization \cite{raaj2019synchronization,vishal2021routes,tripathi2022experimental,tripathi2023frequency}.

The description from a synchronization point of view - though descriptive - remains far from complete as the underlying bifurcation characteristics and the modal interactions are sensitively dependent on the aeroelastic structural parameters \cite{poirel2018frequency,benaissa2021beating}. Changes in the structural parameters may lead to altered bifurcation scenarios, and in turn, give rise to qualitatively diverse dynamical behaviours which have been shown both via wind tunnel experiments \cite{razak2011flutter,goyaniuk2020pitch,benaissa2021beating} and numerical simulations \cite{vishal2022numerical,dos2022dynamical}. For instance, wind tunnel experiments from Razak \textit{et al.}\cite{razak2011flutter} showed that the bifurcation path changes from a degenerate Hopf bifurcation (classical linear ﬂutter) to a sub- or supercritical Hopf bifurcation (stall ﬂutter) as the static Angle of Attack (AoA) is varied. Poirel \textit{et al.}\cite{poirel2018frequency} experimentally observed lock-in phenomena for plunge-to-pitch frequency ratio ($\bar{\omega}$) close to 1 whereas, for $\bar{\omega}$ away from 1, the pitch mode drives the plunge which is a general stall flutter mechanism. Benaissa \textit{et al.}\cite{benaissa2021beating} also reported a similar behaviour and attributed it to a \textit{resonance type} phenomenon between the plunge and the inflow. Vishal \textit{et al.}\cite{vishal2022numerical} and dos Santos \textit{et al.}\cite{dos2022dynamical} showed the dependence of flutter speed, flutter frequency, and associated bifurcation routes on structural mass. The effect of elastic axis location was analyzed and was shown to alter the flutter mechanism by virtue of static imbalance and associated inertial/aerodynamic coupling \cite{pigolotti2017experimental,goyaniuk2023energy}. Wind tunnel experiments from our previous studies \cite{tripathi2022experimental,tripathi2023frequency} presented the aeroelastic responses for a specific set of structural parameters. In Tripathi \textit{et al.} \cite{tripathi2022experimental}, a subcritical Hopf bifurcation resulting in stall flutter via suppression of plunge dynamics by dominant pitch mode was observed for a structurally linear aeroelastic system for a frequency ratio ($\bar{\omega}$) of 0.57. Tripathi \textit{et al.}\cite{tripathi2022experimental} demonstrated that stall flutter onset is via both perfect and imperfect synchronization between the pitch and plunge modes.
Subsequently, in Tripathi \textit{et al.}\cite{tripathi2023frequency}, an altered bifurcation scenario was shown as the $\bar{\omega}$ changes to 0.44 for a nonlinear system, where the dominance of pitch mode diminished, and the plunge mode was no longer pitch-governed.

A key gap that arises at this juncture is that even a synchronization-based description varies on the type of nonlinearity in the problem \cite{liu2014synchronization,antonio2015nonlinearity,vishal2021routes}. The influence of aerodynamic and structural nonlinearities on aeroelastic systems can vary significantly across different flow speed regimes \cite{vishal2022numerical,dos2022dynamical} resulting in an elusive bifurcation scenario. This concern is exacerbated under situations involving stochastic/random wind gusts as the input flow \cite{ghommem2012aeroelastic,venkatramani2016precursors,tripathi2022stall}. To summarize the lacuna in a nutshell, hand-in-hand with the developments in the hitherto literature, a reader may note the following.

\begin{enumerate}
    \item Though conceived as a torsionally dominant problem, recent literature has underscored the role of plunge DoF in a stall flutter by invoking the concept of synchronization between the pitch and plunge modes.
    \item Stall being highly nonlinear and complex in the mathematical description, has no single unified bifurcation theory (in comparison to its classical counterpart with a structural nonlinearity).
    \item Based on the system/structural parameters, the bifurcation routes in a stall flutter problem are non-unique \cite{razak2011flutter,dos2017limit,goyaniuk2020pitch}, physical mechanisms are distinct \cite{poirel2018frequency,goyaniuk2023energy}, can be sensitive to triggering via subcriticality \cite{rangarajan2023non} and display atypical routes like internal resonance post-stall flutter LCOs \cite{tripathi2023frequency}.
    \item Fatigue damage incurred is the highest in a stall flutter scenario \cite{tripathi2022experimental} than its classical counterpart, and a range of engineering systems ranging from helicopter blades/wind turbine blades to MAVs are highly susceptible to stall \cite{bose2019transition} and each of the engineering system is rid with a different parameter space and in turn substantiating the non-unique bifurcation routes and mechanisms. 
    \item The understanding one gain about a stall flutter problem - irrespective of the sensitivity to structural parameters - is jeopardized under the influence of randomly fluctuating input wind \cite{venkatramani2016precursors,tripathi2022experimental}. Traditional paradigms of bifurcation terminologies are no longer applicable and even the underlying synchronization description are distinct - see for example, the presence of phase flip bifurcations in stall problems under the stochastic wind, presented by Tripathi \textit{et al.} \cite{tripathi2022stall}.
\end{enumerate}

Hence, a comprehensive understanding of the interplay between structural parameters, structural nonlinearities, and dynamic stall, that govern the synchronization mechanism in aeroelastic systems needs further investigation and is the focus of this paper. To that end, the effect of changes in elastic axis position ($x_{ea}$) and frequency ratio ($\bar{\omega}$) on the aeroelastic responses of a structurally nonlinear aeroelastic system is analyzed through a wide range of experiments under both deterministic and stochastic inflows. The aeroelastic behaviour is found to be sensitive to structural parameters and nonlinearities. For a range of structural parameters, we observe atypical signatures such as the presence of subcriticality, manifestation of period doubling behavior, and intermittent switching between period-1 and period-2 behavior. Wind tunnel experiments show the susceptibility of the aeroelastic system to often undergo stall flutter, however, for some combination of structural parameters the experiments show qualitatively different instability routes marked by the appearance of beats-like response via an internal resonance. Likewise, under stochastic flow, a noise-induced intermittency regime is observed to presage the random LCOs. Each of these different dynamics and instability routes is systematically characterized over its physical mechanisms by invoking the concepts of synchronization. The documentation of these dynamical transitions, hand-in-hand with the synchronization insights, is employed to gain a deeper understanding of bifurcations in nonlinear aeroelastic systems and in turn, can augment a safer design of the same.

The rest of the paper is as follows. Section 2 details the wind tunnel experiments, in terms of the set-up mechanism, static parameter identification, and data acquisition methodology. The methodology to post-process the acquired aeroelastic time responses using the concepts of synchronization is elaborated in Section 3. The key findings that emerge from frequency ratio variation and elastic axis variation are presented in detail with relevant discussions in Sections 4 and 5, respectively. Under salient structural parameters identified from the above experiments, the bifurcation routes and synchronization characteristics that emerge under wind tunnel experiments with stochastic input wind are presented in Section 6. The main findings, their importance, and the future avenues that emerge are all summarized in Section 7.

\begin{figure}
\begin{center}
\includegraphics[width=6in,height=4in]{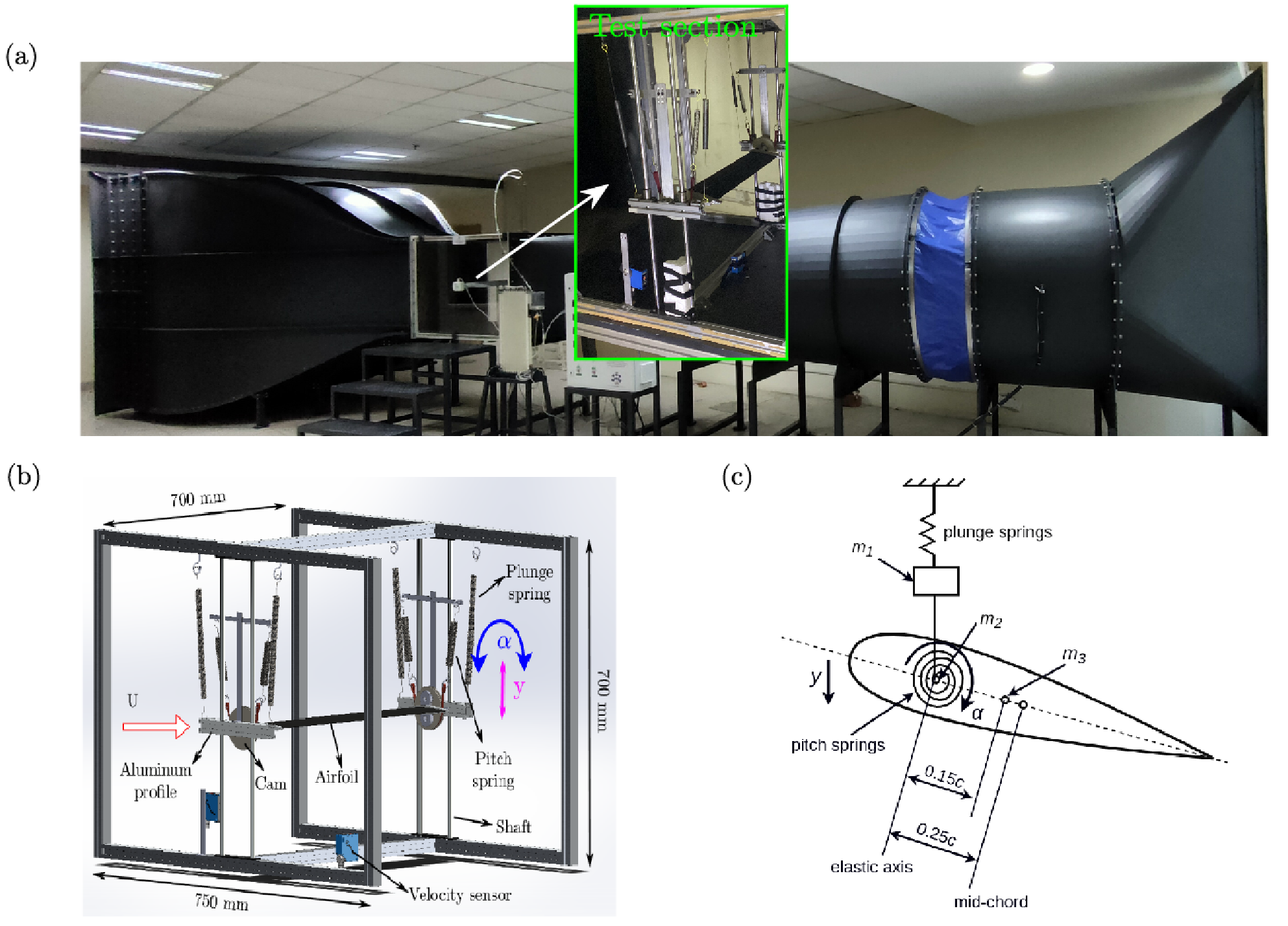}  
\caption{\label{setup} Experimental setup: (a) wind tunnel image, (b) schematic of the pitch-plunge aeroelastic system, and (c) wing cross-section.}
\end{center}
\end{figure}

\section{Experimental setup}\label{exp}

The experimental setup is similar to that used in our previous studies \cite{tripathi2022stall,tripathi2022experimental,tripathi2023frequency}. Experiments are conducted in a low-speed open-type wind tunnel at Shiv Nadar University (see Fig.\ref{setup}(a)), with a test section measuring 0.8 m x 0.8 m x 1.2 m. The setup for the experiment consists of a NACA 0012 profile with a chord length ($c$) of 100 mm and a span of 500 mm mounted horizontally on a support mechanism enabling 2-DoF for the airfoil motion, namely, pitch ($\alpha$) and plunge ($y$), as depicted in Fig.\ref{setup}(b). The cross-section of the aeroelastic system, detailing key structural parameters is illustrated in Fig.~\ref{setup}(c). The masses considered included the mass of the plunging frame ($m_1$), the mass of the pitching mechanism ($m_2$), and the airfoil mass ($m_3$). The total mass in the plunge mode was given as $m_y = m_1 + m_2 + m_3$, and in the pitch mode, it was $m_{\alpha} = m_2 + m_3$.

Restoring force is supplied by tension springs for both pitch and plunge movements. The springs possess nonlinear characteristics which are revealed from the static tests and described in the following section. Plunge motion is facilitated by a translating carriage, while the pitching motion is enabled by a pulley-like cam. Airfoil motion was measured using two NCDT-type laser displacement sensors, with a resolution of 1 micron and a range of 50-350 mm. Delta HD 4V3 TS3 air velocity sensor is employed to measure the flow velocities in the wind tunnel test section. Data from the sensors is acquired using an 8-channel Data Acquisition system with a 24-bit resolution.

The wind tunnel operates in two modes: suction and blowing. The suction mode features streamlined flow with minimal fluctuations due to a honeycomb mesh at the entrance, while the blowing mode introduces continuous flow disturbances, directly fed from the fan to the test section. More details on the wind tunnel flow characteristics can be seen in Tripathi \textit{et al.}\cite{tripathi2022experimental}. The experiments cover an airspeed range from rest up to 16 m/s, corresponding to Reynolds numbers up to $1 \times 10^{5}$. The experimental setup demonstrates good repeatability of the response data as detailed in \ref{A0}.

\section{Methodology}

\begin{figure}
\begin{center} \includegraphics[width=3.6in,height=1.8in]{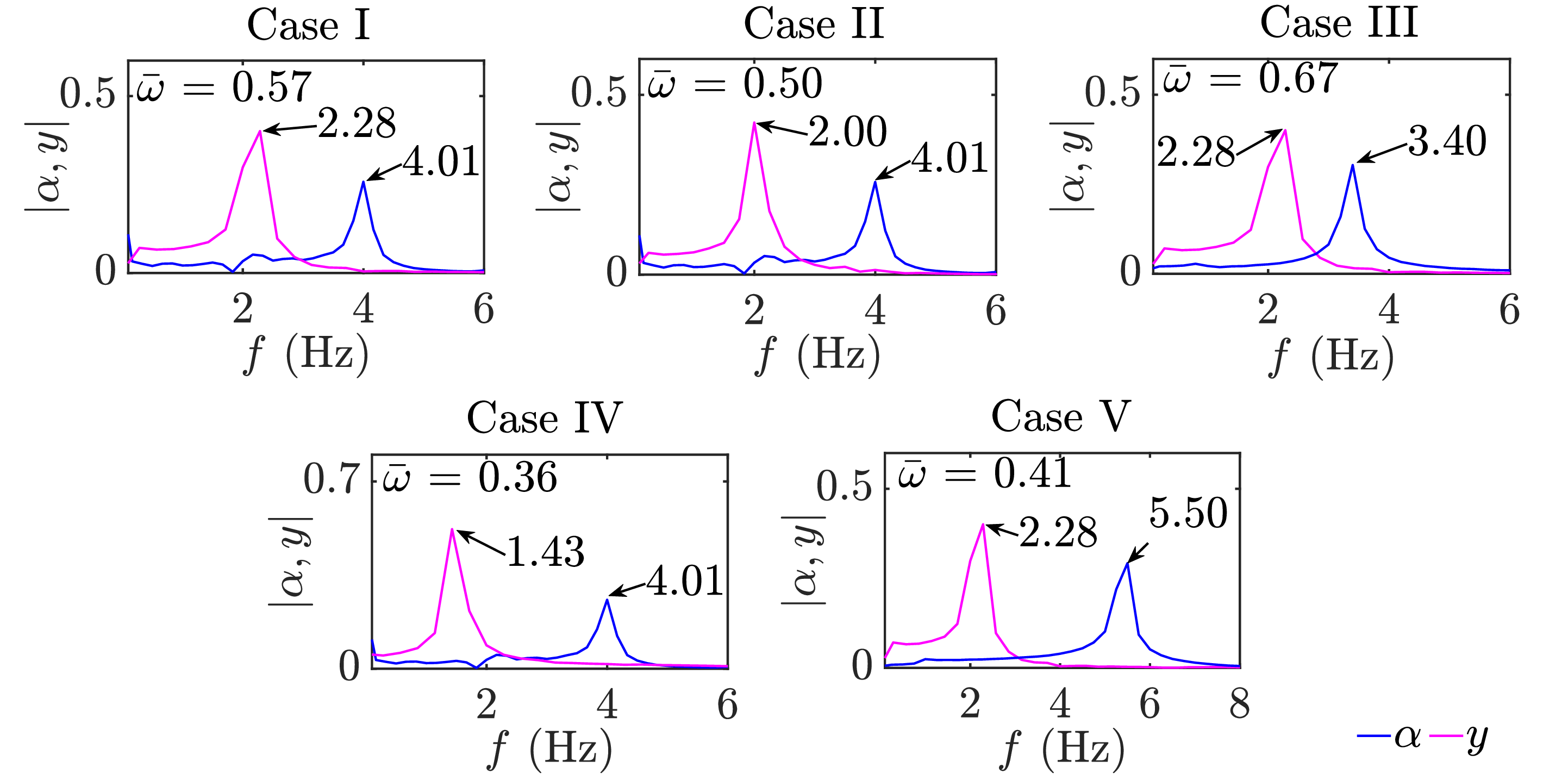}  
 \caption{\label{nat_wc4} Natural frequencies of plunge and pitch modes for different $\bar{\omega}$. Case I is set as the baseline, with common pitch stiffness in Cases I, II, and IV, and common plunge stiffness in Cases I, III, and IV.}
\end{center}
\end{figure}

Experiments are conducted in three parts. The first two parts are conducted in suction mode: (i) for different $\bar{\omega}$ and (ii) for different elastic axis positions. Next, the salient cases chosen from (i) and (ii) are repeated in blowing mode in part (iii). Frequency ratios in the range of 0.36 - 0.67 are obtained by changing pitch or plunge stiffness systematically. Figure~\ref{nat_wc4} shows the frequency ratios for five cases under consideration along with, natural frequencies of individual pitch ($f_{\alpha}$) and plunge ($f_y$) modes. The values of natural frequencies are obtained from the free vibration test, \textit{i.e.}, in wind-off conditions.

\begin{figure}
\begin{center} \includegraphics[width=3.2in,height=2.4in]{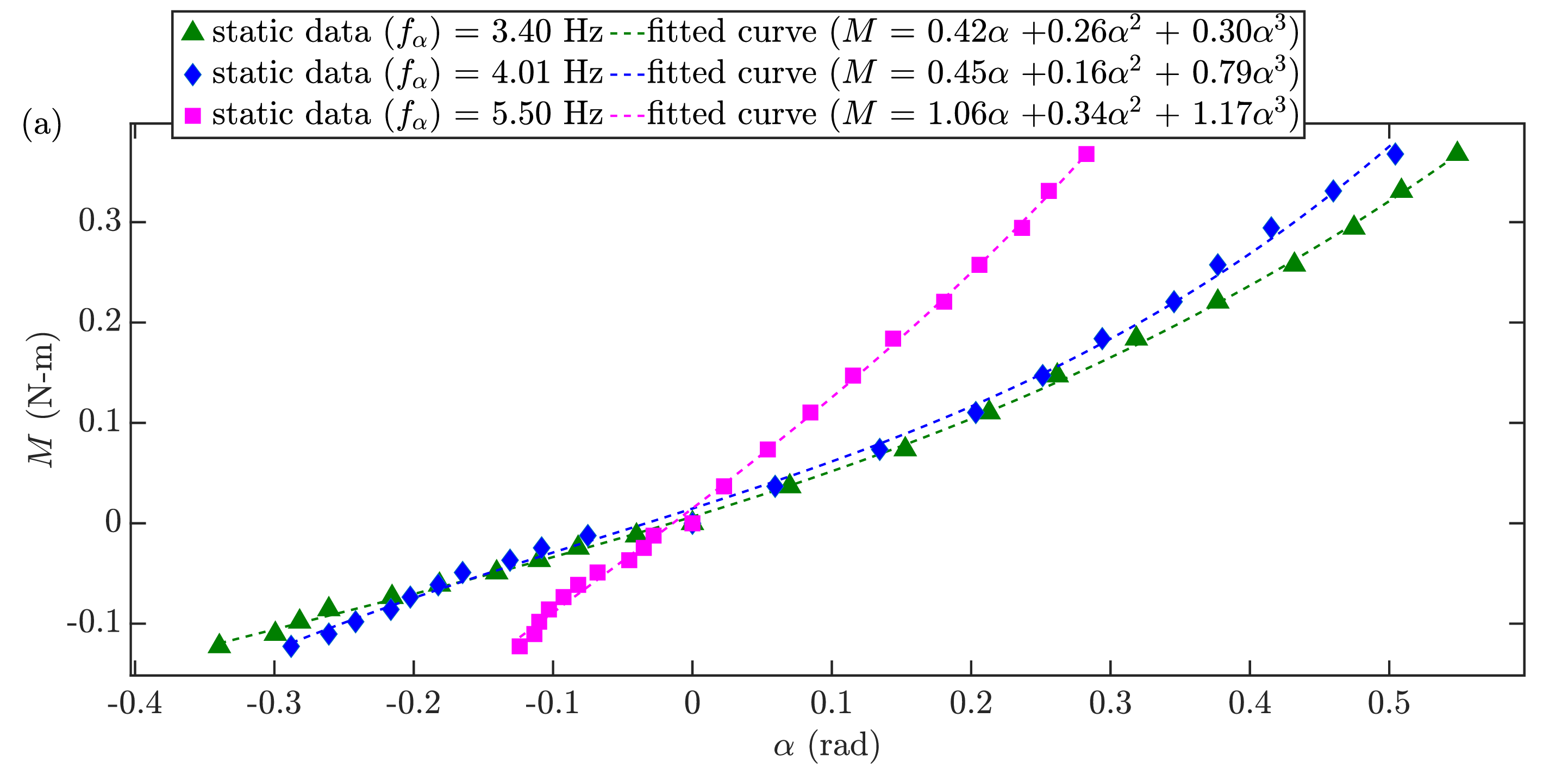}  \includegraphics[width=3.2in,height=2.3in]{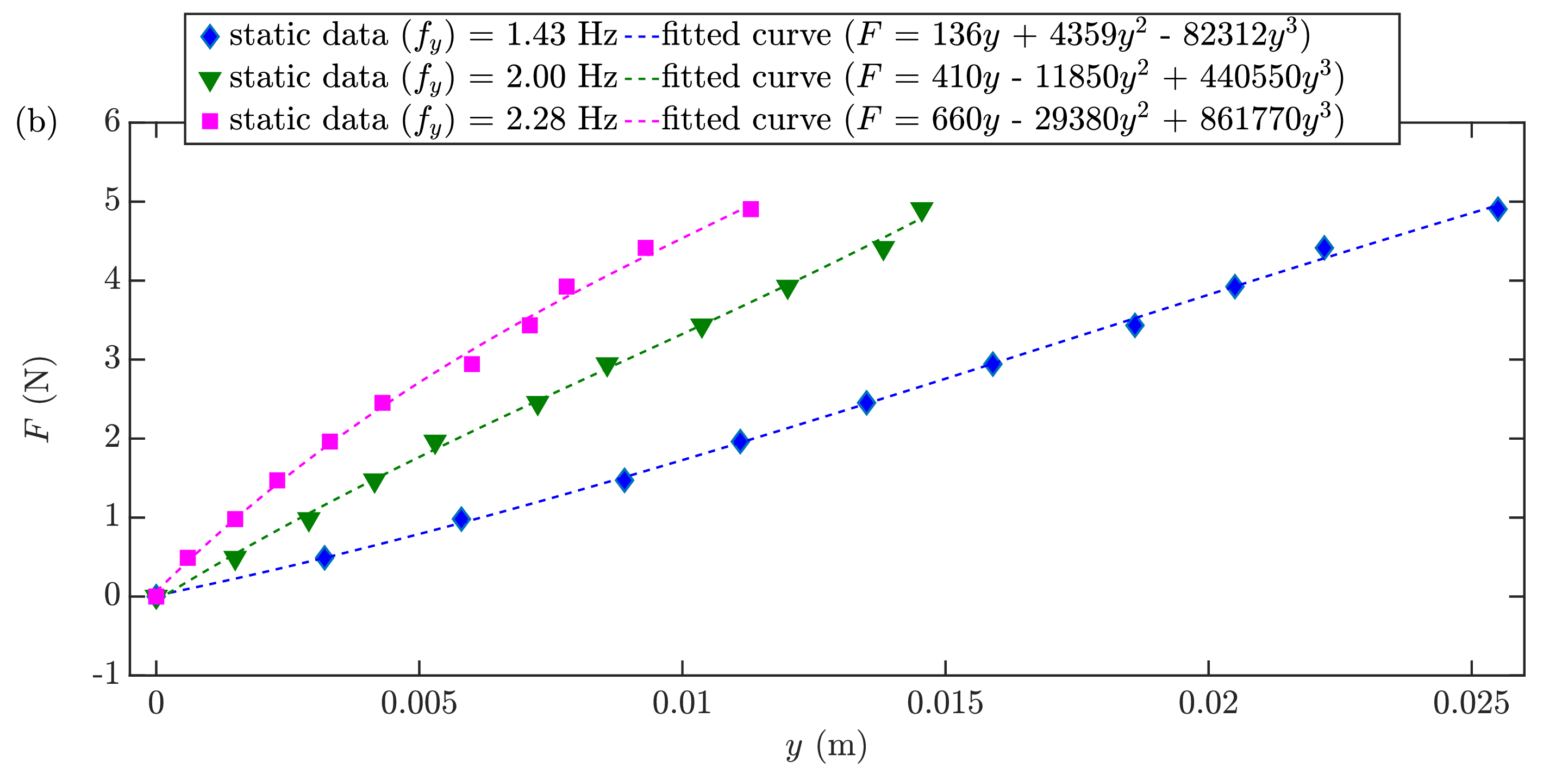}
 \caption{\label{nonlnc4} Characterization of nonlinearities in: (a) pitch stiffness, and (b) plunge stiffness, from static test and cubic polynomial curve fitting.}
\end{center}
\end{figure}

The inherent structural nonlinearities are represented through a third-order polynomial curve-fitting on the load vs. deflection data obtained from the static experiments and are shown in Fig.~\ref{nonlnc4}(a) for pitch stiffness, and in Fig.~\ref{nonlnc4}(b) for the plunge stiffness. For $x_{ea}$ variations, the airfoil rotational axis is systematically fixed at three different locations 0.15$c$, 0.25$c$, and 0.35$c$ (measured from the leading edge).

Wind tunnel experiments are carried out with $U$ as the bifurcation parameter. The bifurcation analysis is two-pronged: (i) $U$ increased from zero to the onset of instability and further to a maximum speed - called the \textit{forward sweep}, (ii) $U$ is reduced from maximum speed to the point where the aeroelastic system comes back to the equilibrium state - called the \textit{backward sweep}. The latter reveals the subcriticality, in terms of hysteretic behaviour, if present in the system.

Typically, the presence of synchronization is analyzed by examining the instantaneous phases and the frequency of the interacting oscillators. The instantaneous phases of the oscillations are generally obtained by adopting an analytic signal approach, wherein the analytic signal, $\zeta(t)$, is a complex quantity with the real part being the original signal, $z(t)$ and the imaginary part being its corresponding Hilbert transform \cite{balanov2009simple} given by

\begin{equation}
 z_{H}(t) = \frac{1}{\pi} P.V. \int_{-\infty}^{\infty}\frac{z(\tau)d\tau}{(t-\tau)}.
\end{equation}
where $P.V.$ is the Cauchy principal value of the integral. Thus, the analytic signal can be written as
\begin{equation}
 \zeta(t) = z(t) + iz_{H}(t) = A(t)e^{i\phi(t)},
 \label{hil}
\end{equation}

\noindent where $\phi(t)$ represents the instantaneous phase, and $A(t)$ is the instantaneous amplitude of the signal. The relative phase value (RPV) or phase difference between two oscillators (for instance plunge and pitch modes of aeroelastic system in this study) at $j^{th}$ instant can be given as $\triangle\phi_j = \phi_{j,plunge} - \phi_{j,pitch}$. To characterize the strength of synchronization, the phase locking value (PLV) of the responses is also estimated as

\begin{equation}
 PLV = N^{-1} |\sum_{j=1}^{N} exp(i\triangle\phi_j)|.  
\end{equation}

\noindent Since a constant phase difference between oscillators is indicative of strong synchronization, a perfectly synchronized state gives a PLV of one, while a completely asynchronous state gives a PLV close to zero. The PLV for an imperfect synchronization gives a value between zero and one \cite{tripathi2022experimental}.

\section{\label{eow} Effect of frequency ratio}
In this part of the investigation, the structural stiffness is changed by changing the pitch or plunge springs. In turn, five different $\bar{\omega}$ cases are obtained which are shown in Fig.~\ref{nat_wc4}. The first three cases depicted in Fig.~\ref{nat_wc4} have $\bar{\omega}$ $\geq$ 0.5 while the cases IV and V have $\bar{\omega}$ $<$ 0.5. The reason for this categorization is due to the fact that similar qualitative behaviours are observed for $\bar{\omega}$ $\geq$ 0.5 and some non-conventional bifurcation routes observed for $\bar{\omega}$ $<$ 0.5. These dynamics will be discussed in detail in this section.

\subsection{\label{cat1c4} Catagory I: frequency ratios equal and above 0.5}
\begin{table*}
\caption{\label{ep5ac4} Structural parameters for the experiment estimated from static tests for $\bar{\omega}$ $\geq$ 0.5 cases.}
\begin{center}
\begin{tabular}{@{}c| c c c c@{}} 
\hline
\hline
Parameter & \multicolumn{3}{c}{Values} \\
\hline
\rule{0pt}{12pt}
Total moving mass in plunge, $m_y$ (kg)& \multicolumn{3}{c}{1.908}\\
Total moving mass in pitch, $m_{\alpha}$ (kg)& \multicolumn{3}{c}{0.937}\\
Airfoil chord length, $c$ (m)& \multicolumn{3}{c}{0.1}\\
Position of rotational (elastic) axis from the leading edge, $x_{ea}$ & \multicolumn{3}{c}{0.25$c$}\\
Position of the mass center from the leading edge, $x_{c}$ & \multicolumn{3}{c}{0.40$c$}\\
Mass moment of inertia in pitch about elastic axis, $I_\alpha$ ($\rm{kg\cdot m^2}$)& \multicolumn{3}{c}{0.0017}\\
\hline
& Case I & Case II & Case III\\
Pitch damping, $\zeta_\alpha$ & 0.01 & 0.01 & 0.01\\
Plunge damping, $\zeta_y$ & 0.05 & 0.05 & 0.05\\
Natural frequency of the plunge mode, {$f_y$ (Hz)}& 2.28 & 2.00 & 2.28 \\
Natural frequency of the pitch mode, {$f_\alpha$ (Hz)}& 4.01 & 4.01 & 3.40\\
Plunge to pitch natural frequency ratio, {$\bar{\omega}$} & 0.57 & 0.50 & 0.67\\
\hline
\hline
\end{tabular}
\end{center}
\end{table*}

\begin{figure}
\centering
\includegraphics[width=5in, height=2in]{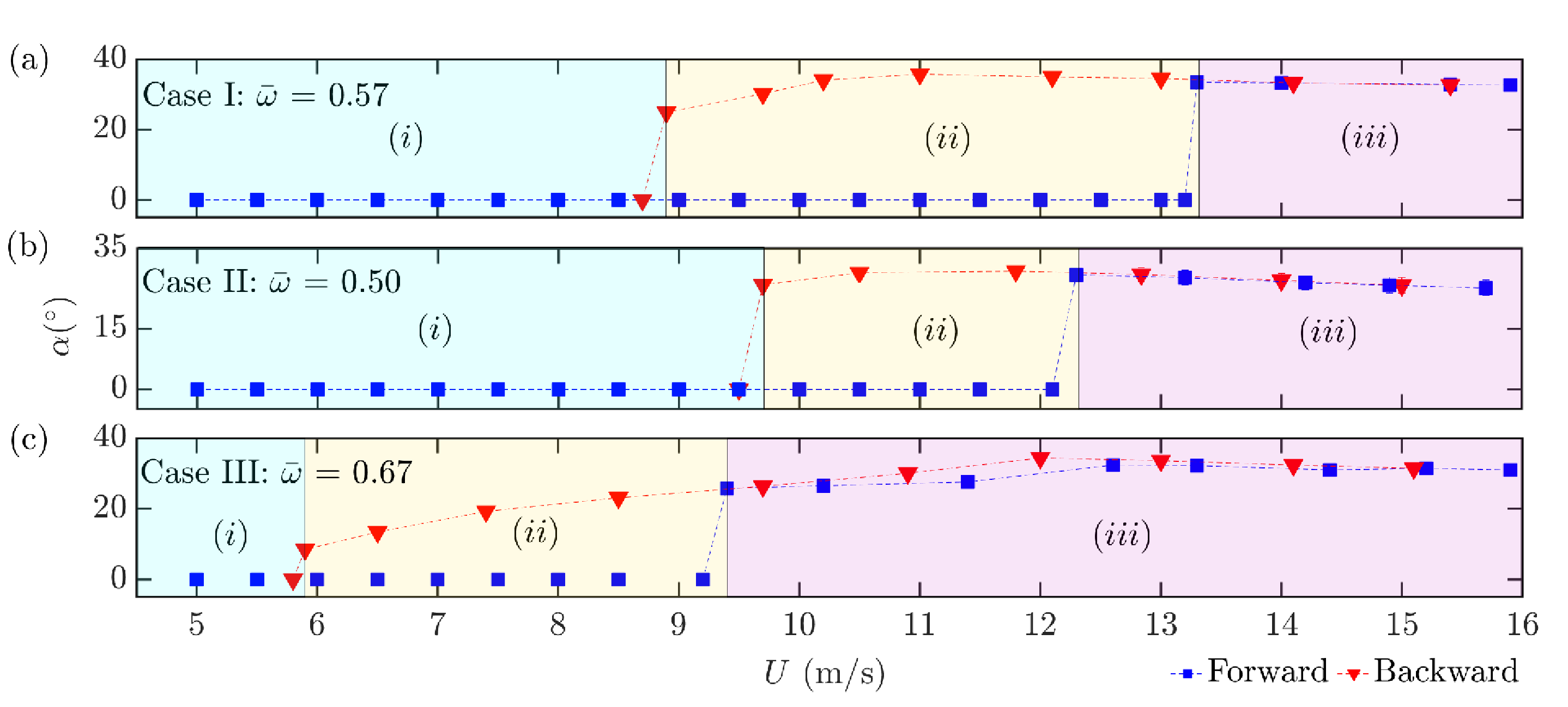}
\caption{Bifurcation plots for $\bar{\omega}$ $\geq$ 0.5. Regime I - stable fixed point (FP) attractor, Regime II - coexisting stable FP and stable LCO attractors, and Regime III - stable LCO attractor.}
\label{bif_wc4a}
\end{figure}

\begin{figure}
\centering
\includegraphics[width=5in, height=2.2in]{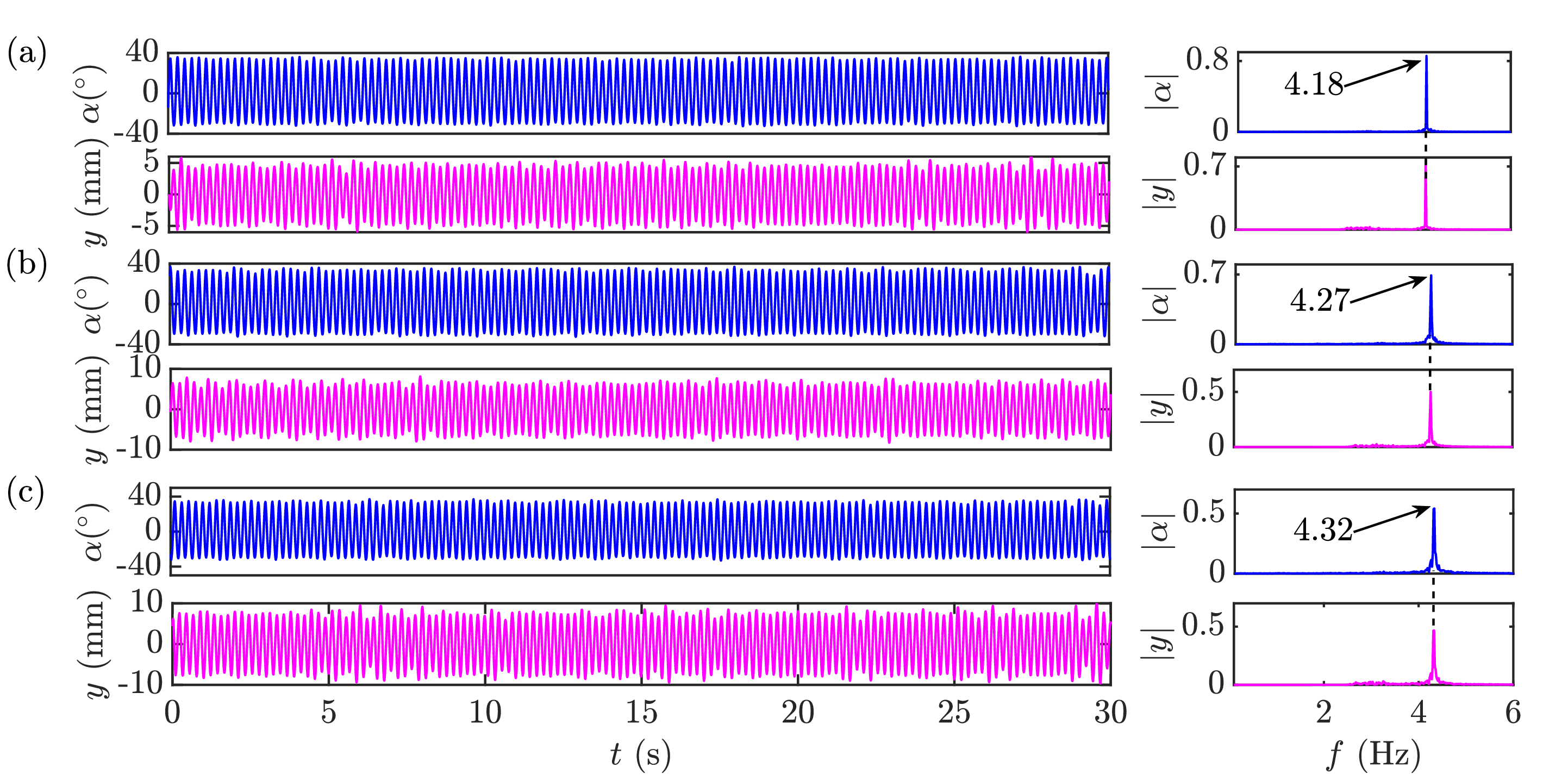}
\caption{Response dynamics post Hopf point (Case I: $\bar{\omega}$ = 0.57, $f_y$ = 2.28 Hz, $f_{\alpha}$ = 4.01 Hz) (a) $U$ = 13.3 m/s, (b) $U$ = 14.9 m/s, (c) $U$ = 15.9 m/s.}
\label{thistyyc4}
\end{figure}

Three cases are examined in this category and the corresponding structural parameters are listed in Table~\ref{ep5ac4}. The bifurcation diagrams for these three cases, \textit{i.e.}, $\bar{\omega}$ = 0.57, 0.50, and 0.67 are shown in Fig.~\ref{bif_wc4a}. The flutter onset is marked by a subcritical Hopf bifurcation (obtained from the forward sweep) and a fold bifurcation (obtained from the backward sweep). The bifurcation route is qualitatively similar to the one observed in our previous study\cite{tripathi2022experimental}. Case I carries the same $\bar{\omega}$ value as presented in Tripathi \textit{et al.}\cite{tripathi2022experimental}, with the only distinction being that, in our prior study, the stiffness in both pitch and plunge was linear. The structural nonlinearities are developed over time and repeated utilization, and hence the experiments are redone here for this $\bar{\omega}$ value. For this case, the flutter boundaries are marginally shifted as compared to those reported in Tripathi \textit{et al.}\cite{tripathi2022experimental}. The Hopf point ($U_{cr}$) is observed at 13.3 m/s and the fold point ($U_{fld}$) is observed at 8.9 m/s (see Fig.~\ref{bif_wc4a}(a)), whereas in earlier study\cite{tripathi2022experimental} $U_{cr}$ = 13.7 m/s and $U_{fld}$ = 9.1 m/s. The response dynamics post the Hopf point for this case is shown in Fig.~\ref{thistyyc4}. The flutter onset is observed at 13.3 m/s with pitch-dominant LCOs with pitch amplitudes ranging up to 35$^\circ$ - indicating stall flutter. The aeroelastic response dynamics do not show any qualitative change further in the regime $U$ = 13.3 - 15.9 m/s, with flutter frequency having a slight increase with flow speed. 

\begin{figure}
\centering
\includegraphics[width=5in, height=2.2in]{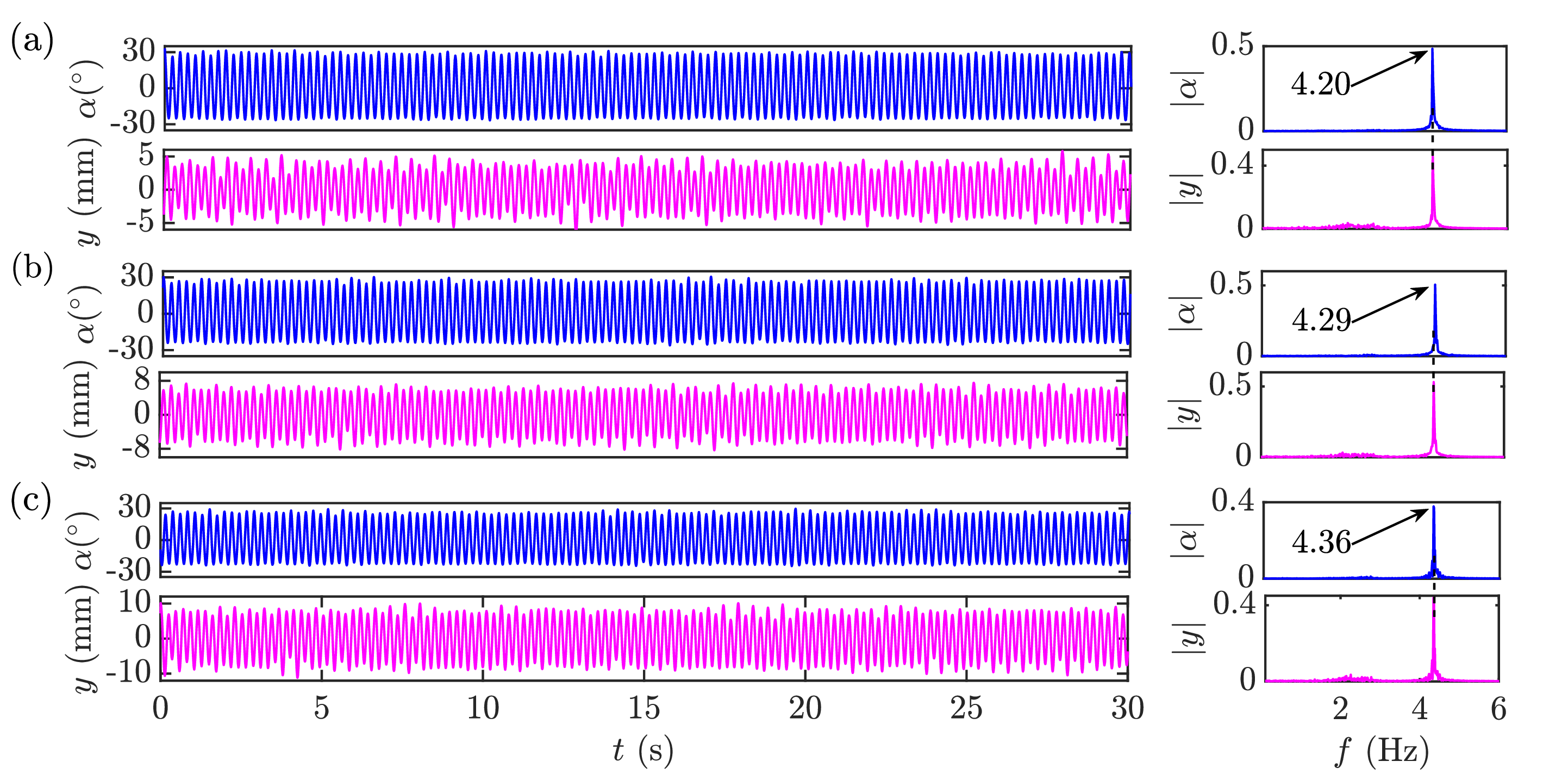}
\caption{Response dynamics post Hopf point (Case II: $\bar{\omega}$ = 0.50, $f_y$ = 2.00 Hz, $f_{\alpha}$ = 4.01 Hz) (a) $U$ = 12.3 m/s, (b) $U$ = 14.2 m/s, (c) $U$ = 15.7 m/s.}
\label{thistywc4}
\end{figure}

In the second case ($\bar{\omega}$ = 0.50), the plunge stiffness is reduced to 2.00 Hz, keeping the pitch stiffness the same (4.01 Hz) as case I (see Table~\ref{ep5ac4}). The resulting aeroelastic response dynamics are shown in Fig.~\ref{thistywc4}. The flutter onset is observed at 12.3 m/s with a pitch amplitude close to 30$^\circ$ (see Fig.~\ref{thistywc4}(a)). The plunge amplitude is close to 6 mm and the flutter frequency is 4.20 Hz. Similar to Case I, the flutter can be characterized as stall flutter as the flutter frequencies of pitch and plunge modes are close to $f_{\alpha}$ and pitch amplitudes are higher than the static stall values ($\approx$ 12$^\circ$) \cite{dos2021improvements}. The flow speed is further increased up to 15.7 m/s with no significant qualitative change in the dynamics (see Fig.~\ref{thistywc4}(b)-(c)).

\begin{figure}
\centering
\includegraphics[width=5in, height=2.2in]{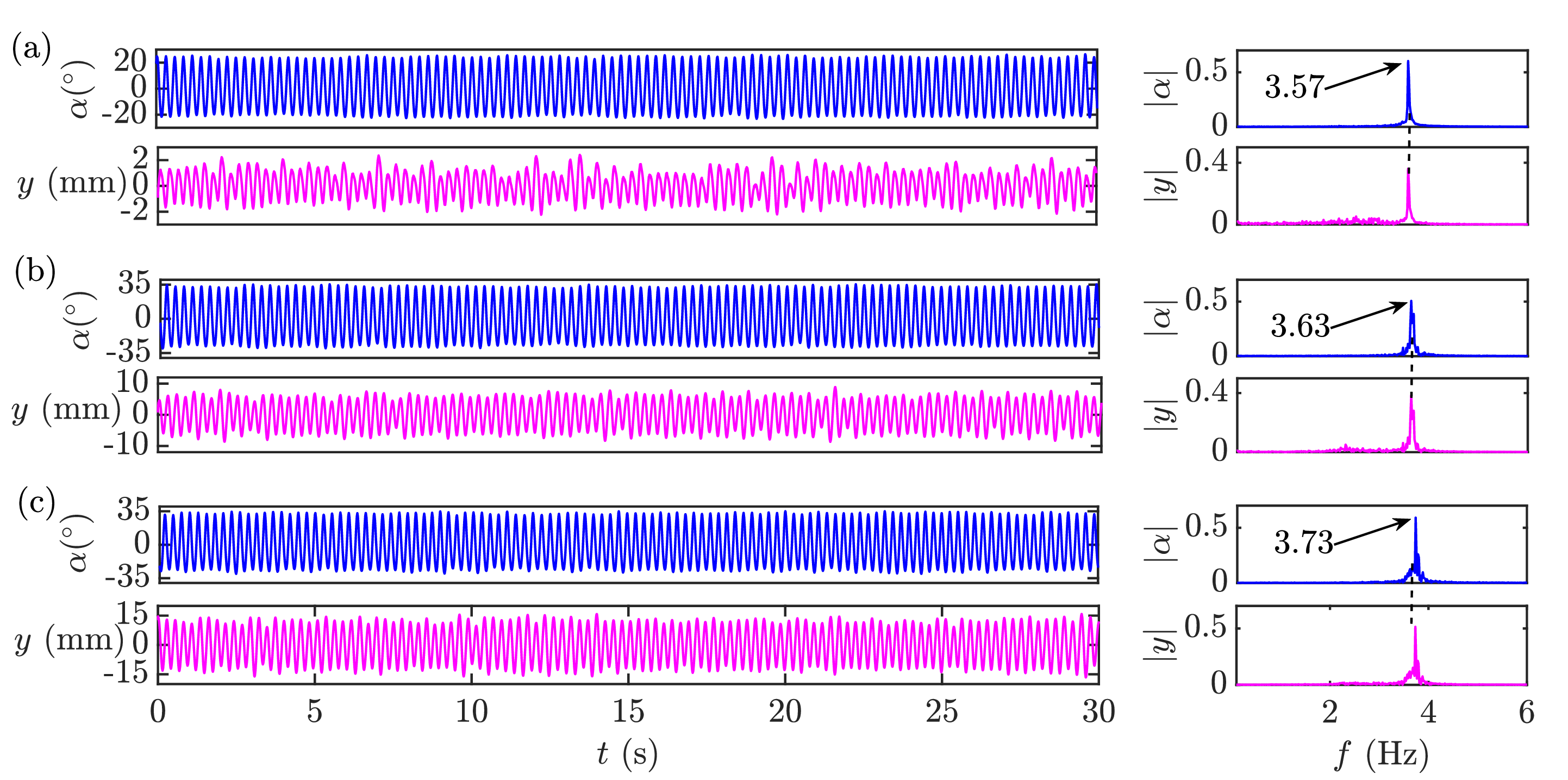}
\caption{Response dynamics post Hopf point (Case III: $\bar{\omega}$ = 0.67, $f_y$ = 2.28 Hz, $f_{\alpha}$ = 3.4 Hz) (a) 9.4 m/s, (b) 12.6 m/s, (c) 15.9 m/s.}
\label{thistr3c4}
\end{figure}

In case III ($\bar{\omega}$ = 0.67), the $f_y$ is 2.28 Hz, kept the same as Case I, and $f_{\alpha}$ is reduced to 3.40 Hz (see Table~\ref{ep5ac4}). The aeroelastic responses are shown in Fig.~\ref{thistr3c4}. The $U_{cr}$ is noted at 9.4 m/s, with the flutter frequency 3.57 Hz - close to $f_{\alpha}$, for both pitch and plunge modes. This advancement in flutter onset (as compared to Case I with $U_{cr}$ = 13.3 m/s) can be corroborated by the study from Goyaniuk \textit{et al.}\cite{goyaniuk2020pitch}, who demonstrated LCO onset delays to higher Reynolds numbers as the pitch stiffness is increased, for a 1-DoF stall flutter problem. The pitch and plunge amplitudes are noted as approximately 25$^\circ$ and 2 mm. The aeroelastic response dynamics remain the same qualitatively up to $U$ = 15.9 m/s with a slight increase in flutter frequencies, akin to the observations from case I and case II.

The three reported cases ($\bar{\omega}$ $\geq$ 0.5) thus far illustrate conventional routes to stall flutter, consistent with the recent stall flutter experiments\cite{dimitriadis2009bifurcation,razak2011flutter,vsidlof2016experimental,tripathi2022experimental}. Here, the pitch mode is the driving mode as both the pitch and plunge modes oscillate near the natural frequency of the pitch mode. This means that the plunge mode essentially has negligible participation as far as the energy extraction from the flow is concerned\cite{dimitriadis2009bifurcation,benaissa2021beating}. Wind tunnel experiments by Dimitriadis and Li \cite{dimitriadis2009bifurcation} showed similar findings where the plunge mode oscillates close to the pitch natural frequency. The authors attributed this to much higher stiffness in the plunge as compared to that in pitch ($\bar{\omega}$ = 6.94), where the plunge is only driven due to the aerodynamic coupling between the two modes. However, a subsequent study by Razak \textit{et al.}\cite{razak2011flutter} shows that even for closely spaced pitch and plunge natural frequencies ($\bar{\omega}$ = 0.77), stall flutter can be observed at sufficiently high AoAs. Wind tunnel experiments from Poirel and co-workers\cite{poirel2018frequency,goyaniuk2020pitch,benaissa2021beating,goyaniuk2023energy} show that the pitch heave flutter mechanism is essentially akin to a 1-DoF stall flutter for all the $\bar{\omega}$ values above and below 1, except the cases where it is very close to 1. For very closely spaced natural frequencies ($\bar{\omega}$ $\approx$ 1), they observed the participation from plunge mode to be very prominent. It is worth noting that, this participation from the plunge mode was not due to a coalescence flutter mechanism but was rather some form of \textit{resonance} phenomenon where pitch LCO frequency acts as a forcing frequency\cite{goyaniuk2020pitch}. The resulting aeroelastic response appears to be a beating response\cite{benaissa2021beating}. In our work, later in Section~\ref{cat2c4} we present the findings where the beating response is observed, which is attributed to an \textit{internal resonance}.

Thus, all the experiments for the range of $\bar{\omega}$ considered in Category I, \textit{i.e.}, 0.5 - 0.67, indicate a stall flutter for the nonlinear structure at the specified wind speeds. Since the plunge remains a passive (driven) mode, altering plunge stiffness has a minimal impact on flutter frequency compared to pitch stiffness. For $\bar{\omega}$ = 0.57 and 0.5 (Case I and II, respectively), maintaining $f_{\alpha}$ while adjusting $f_y$ results in flutter frequencies at $U_{cr}$ of 4.18 Hz and 4.20 Hz. In contrast, comparing Case I ($\bar{\omega}$ = 0.57) with Case III ($\bar{\omega}$ = 0.67), with constant $f_y$ but varying $f_{\alpha}$, yields flutter frequencies changing from 4.01 Hz to 3.57 Hz, respectively.

\begin{figure}
\centering
\includegraphics[width=3.4in, height=1.6in]{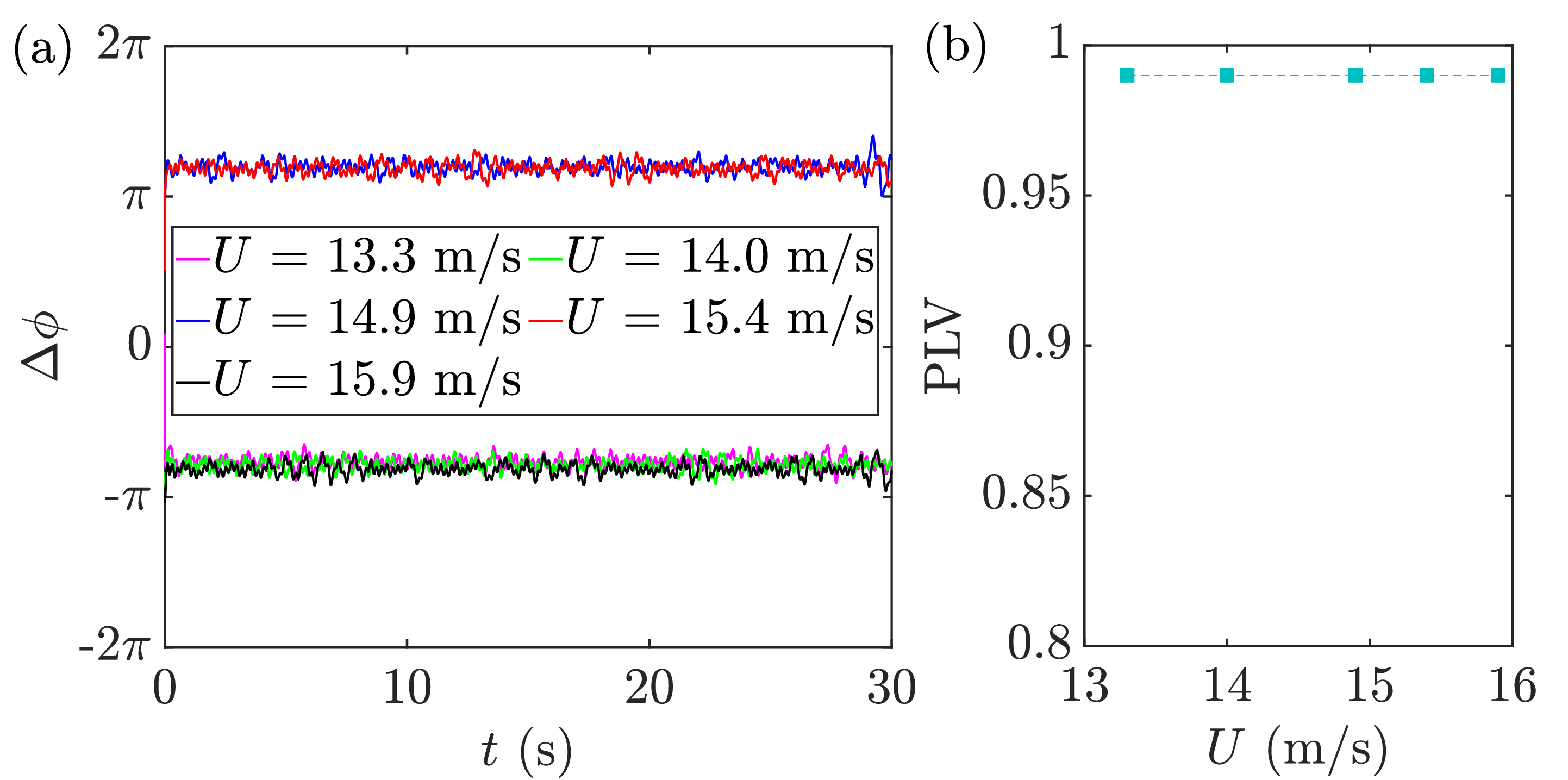}
\caption{Synchronization analysis post Hopf point (Case I: $\bar{\omega}$ = 0.57, $f_y$ = 2.28 Hz, $f_{\alpha}$ = 4.01 Hz). (a) RPV time histories, and (b) PLV variation.}
\label{phist_yyc4}
\end{figure}



\begin{figure}[htbp]
  \begin{minipage}[t]{0.48\textwidth}
    \centering
\includegraphics[width=3.2in, height=1.8in]{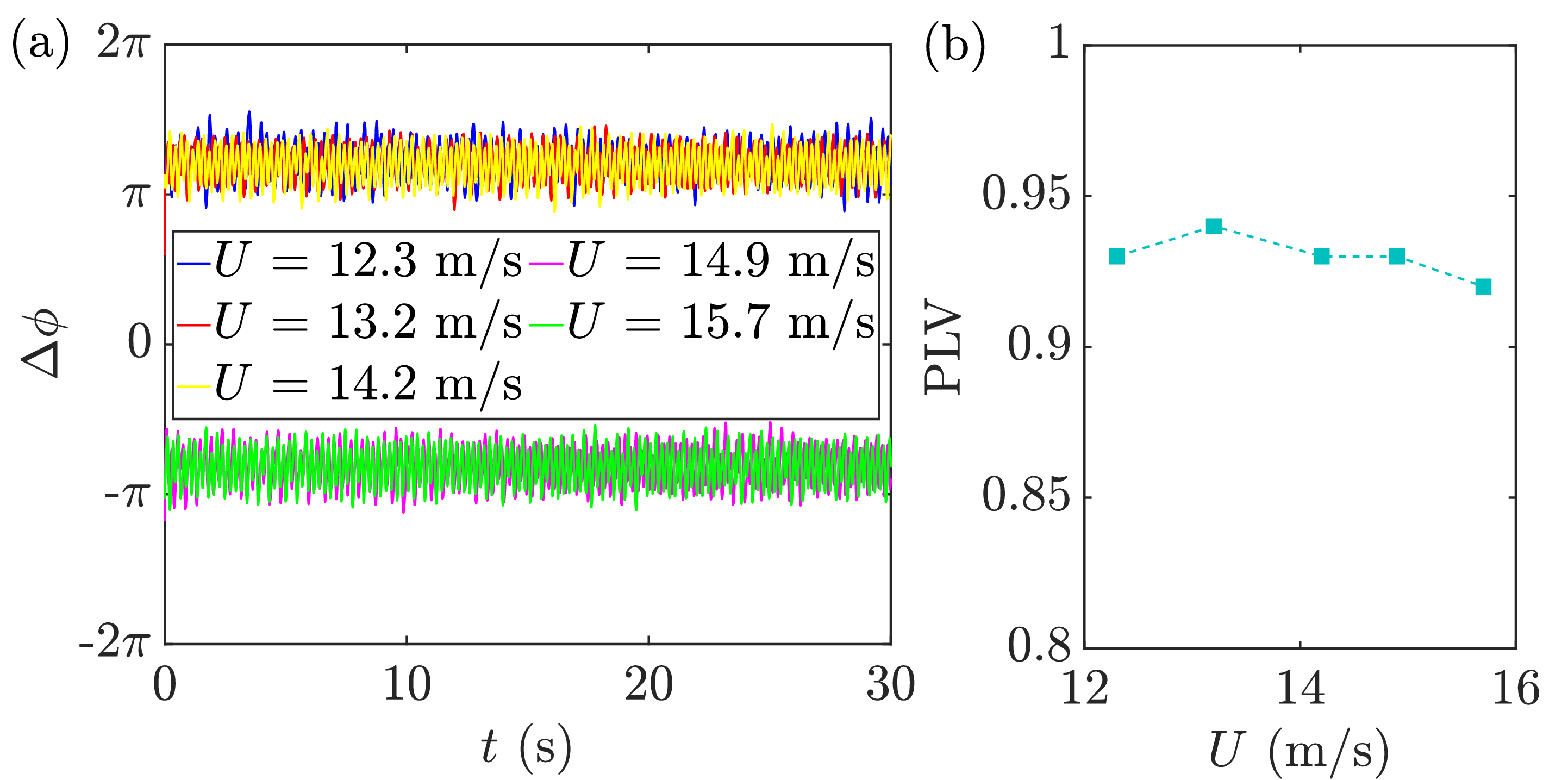}
\caption{Synchronization analysis post Hopf point (Case II: $\bar{\omega}$ = 0.50, $f_y$ = 2.00 Hz, $f_{\alpha}$ = 4.01 Hz). (a) RPV time histories, and (b) PLV variation.}
\label{plvywc4}
  \end{minipage}\hfill
  \begin{minipage}[t]{0.48\textwidth}
    \centering
 \includegraphics[width=3.2in, height=1.8in]{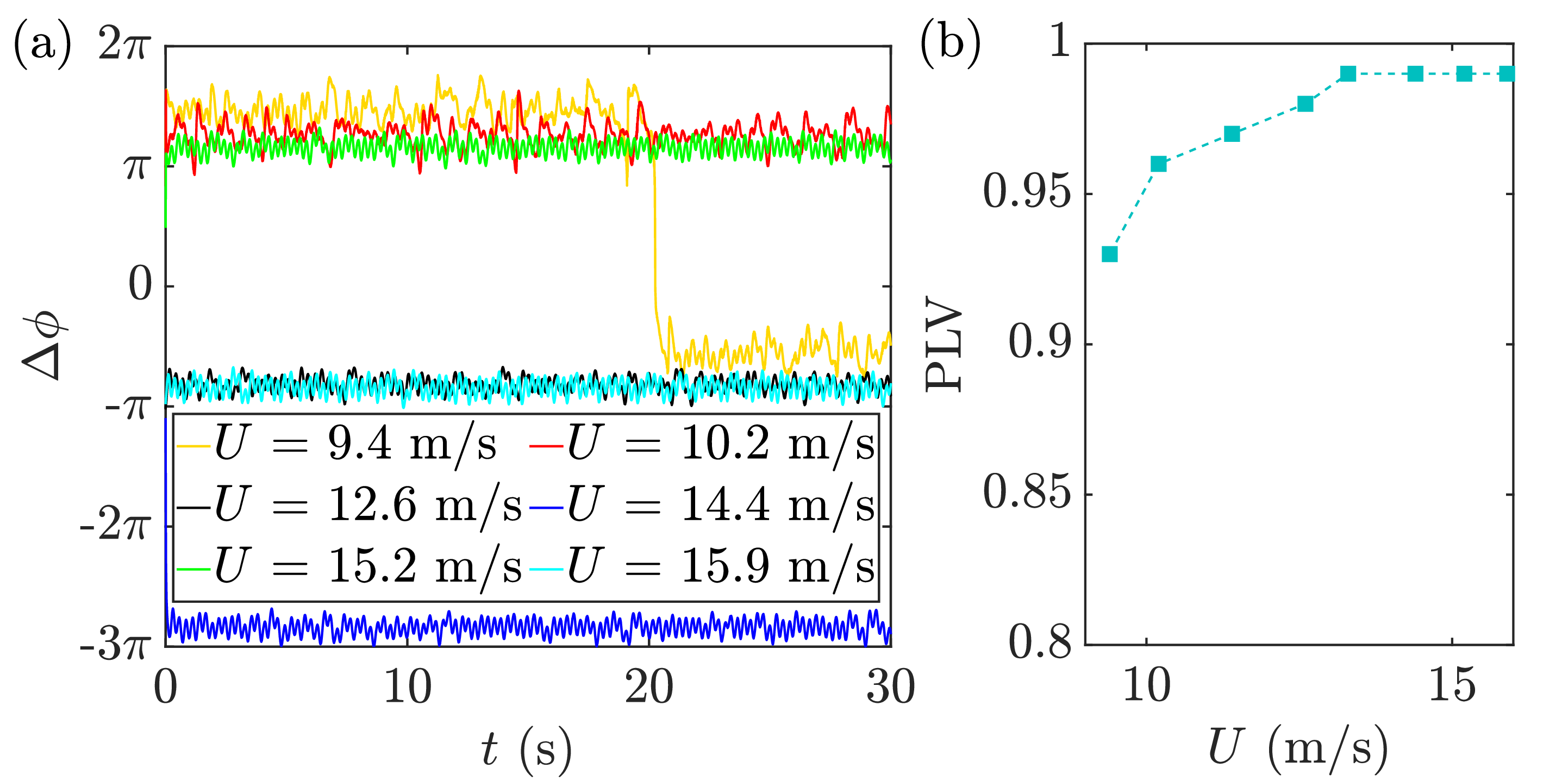}
\caption{Synchronization analysis post Hopf point (Case III: $\bar{\omega}$ = 0.67, $f_y$ = 2.28 Hz, $f_{\alpha}$ = 3.4 Hz).}
\label{phistr3c4}
  \end{minipage}
\end{figure}

The corresponding synchronization analysis post the Hopf point is shown in Figs.~\ref{phist_yyc4} -\ref{phistr3c4} for Cases I-III, respectively. The phases of pitch and plunge modes depict a strong synchronization at $U_{cr}$ for all three cases. For Case I ($\bar{\omega}$ = 0.57), stall flutter onset at Hopf point is marked by a bounded RPV over the time indicative of a strong synchronization via phase trapping (Fig.~\ref{phist_yyc4}(a)). This synchronization dynamics persists up to $U$ = 15.9 m/s, the maximum speed under consideration. The PLVs for $U$ = 13.3 - 15.9 m/s are very close to 1 (Fig.~\ref{phist_yyc4}(b)), indicative of perfectly synchronized pitch and plunge modes. These synchronization dynamics are quite similar to that reported in our previous findings\cite{tripathi2022experimental} for the same structural parameters, but with a linear structure. This suggests that in this instance, structural nonlinearity does not affect synchronization, meaning that the overall flutter mechanism is still the same as it would be in the case of a linear structure. This conjecture, however, does not apply to every scenario, as will be seen in the next subsection.

For Case II ($\bar{\omega}$ = 0.50), the synchronization study reveals a phase trapping at the $U_{cr}$ and post that as well (see Fig.~\ref{plvywc4}(a)). This synchronization dynamics closely resembles that of Case I. The PLV in the range $U$ = 12.3 - 15.7 m/s remains between 0.92 - 0.94 indicating strong synchronization (see Fig.~\ref{plvywc4}(b)). Here, the PLV is seen to have a small dip at the higher flow speeds, possibly due to the small broadband present close to 2 Hz (see Fig.~\ref{thistywc4}(c)). In Case III ($\bar{\omega}$ = 0.67), the RPV at $U_{cr}$ is observed with a single instance of phase slip amidst an overall phase trapping regime (see Fig.~\ref{phistr3c4}(a)). The corresponding PLV is 0.93 (see Fig.~\ref{phistr3c4}(b)). Increasing $U$ to 10.2 m/s, the synchronization becomes stronger with PLV 0.96, as the phase trapping is not interrupted by any phase slip. The synchronization strengthens further with flow speed, and a PLV of 0.99 is observed at $U$ = 15.9 m/s. In all three cases the phases of pitch and plunge modes are observed to be in anti-phase (see Figs.~\ref{phist_yyc4}(a),~\ref{plvywc4}(a), and ~\ref{phistr3c4}(a)). Note that neither $U_{cr}$ nor PLV at $U_{cr}$ vary linearly with $\bar{\omega}$, which means that both flutter onset and underlying synchronization are dependent on individual pitch and plunge stiffness and associated nonlinearities. 

\begin{figure}
\centering
\includegraphics[width=5in, height=2.2in]{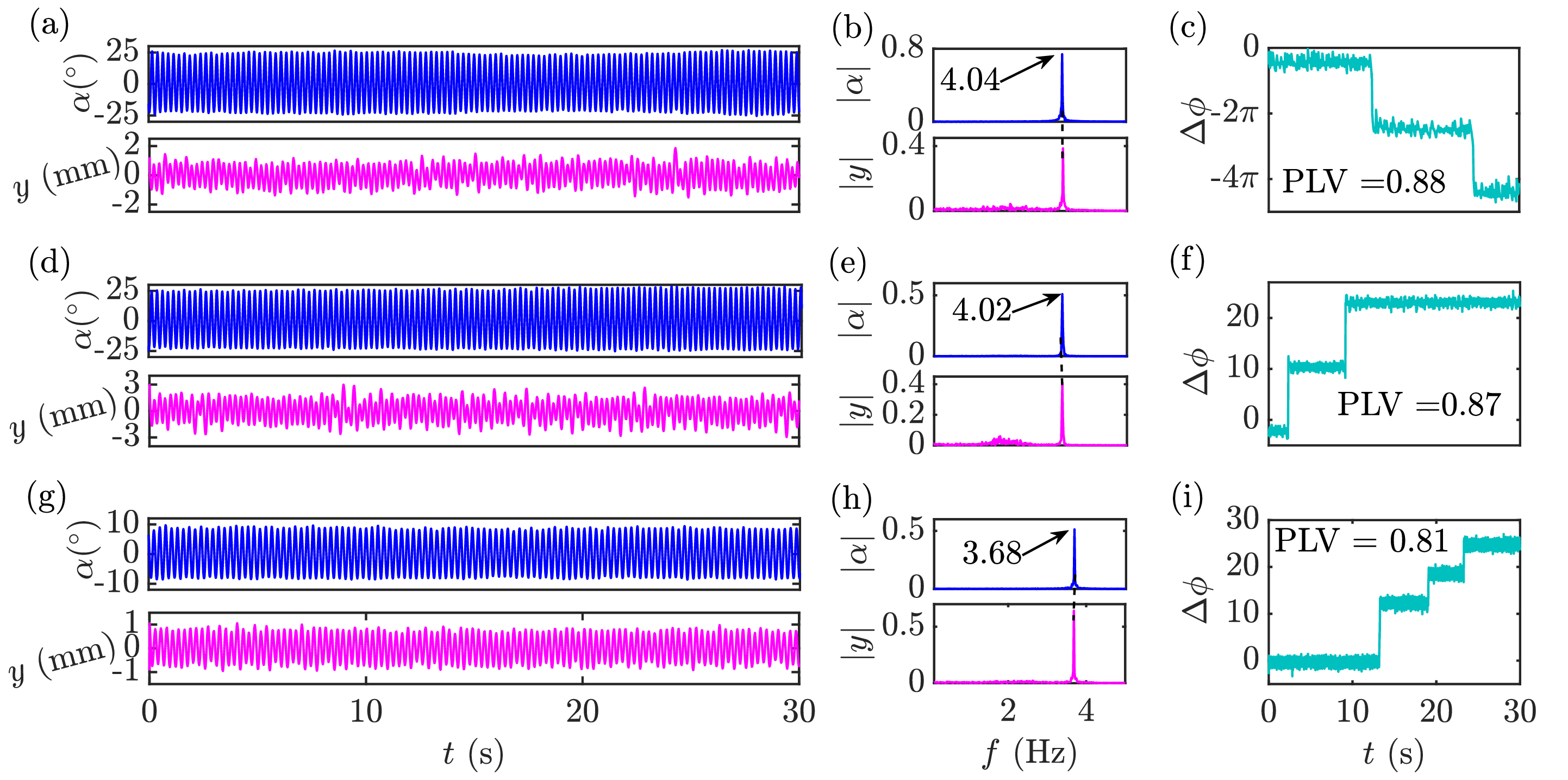}
\caption{Aeroelastic response dynamics, corresponding frequency responses, and synchronization analysis, at $U_{fld}$, for Case I ($\bar{\omega}$ = 0.57, $U_{fld}$ = 8.9 m/s) shown in (a)-(c), respectively; for Case II ($\bar{\omega}$ = 0.50, $U_{fld}$ = 9.7 m/s) shown in (d)-(f), respectively; and for Case III ($\bar{\omega}$ = 0.67, $U_{fld}$ = 5.9 m/s) shown in (g)-(i), respectively.}
\label{fld5a}
\end{figure}

Next, we investigate the dynamics near the fold point, ubiquitous in all three cases, owing to the inherent subcritical nature of the aeroelastic system. For Case I ($\bar{\omega}$ = 0.57), $U_{fld}$ is observed at 8.9 m/s (Fig.~\ref{fld5a}(a)). The pitch amplitude is close to 25$^\circ$ while the plunge amplitudes are close to 2 mm. The flutter frequency at $U_{fld}$ is 4.04 Hz (Fig.~\ref{fld5a}(b)). The PLV at $U_{fld}$ is observed to be 0.88 which indicates a weaker synchronization as compared to that at $U_{cr}$. This is due to the intermittent phase synchronization, where the mutual synchronization between the two modes is disrupted by phase slips (Fig.~\ref{fld5a}(c)). This synchronization behaviour at $U_{fld}$ is consistent with that reported in \cite{tripathi2022experimental}. For Case II, $U_{fld}$ is observed to be at 9.7 m/s with pitch amplitude close to 25$^\circ$, plunge amplitude close to 3 mm, and a flutter frequency of 4.02 Hz (Figs.~\ref{fld5a}(d)-(e)). The relative phase of pitch and plunge modes are seen to synchronize via an intermittent phase synchronization with a PLV of 0.87 (Figs.~\ref{fld5a}(f). For Case III, $U_{fld}$ is identified at 5.9 m/s, with flutter frequency of 3.68 Hz (Figs.~\ref{fld5a}(g)-(h)). The reduction in both $U_{fld}$ and flutter frequency is due to the reduced pitch stiffness (as compared to Case I), consistent with the dynamics observed at the Hopf point. The pitch amplitudes are close to 10$^\circ$ while the plunge amplitudes are $\approx$ 1 mm. Since the flutter sustains around $f_{\alpha}$, such a small amplitude in pitch LCO might indicate a light stall event. For this case the synchronization at $U_{fld}$ shows an intermittent phase synchronization with PLV = 0.81  (see Figs.~\ref{fld5a}(i)), which is slightly lower than that for the other two cases.

To summarize the dynamics at $U_{fld}$, the flutter frequencies are observed to be pitch-dominant in all three cases. The plunge amplitudes are minimal at $U_{fld}$. This is because the plunge motion is essentially pitch-driven and $U_{fld}$ is the last dynamic point during the backward sweep as the motion comes to an end below it. This leads to a weaker synchronization at the fold point via an intermittent phase synchronization, as compared to that at the Hopf point. 

\subsection{\label{cat2c4} Catagory II: frequency ratios below 0.5}

\begin{figure}
\centering
\includegraphics[width=5in, height=2in]{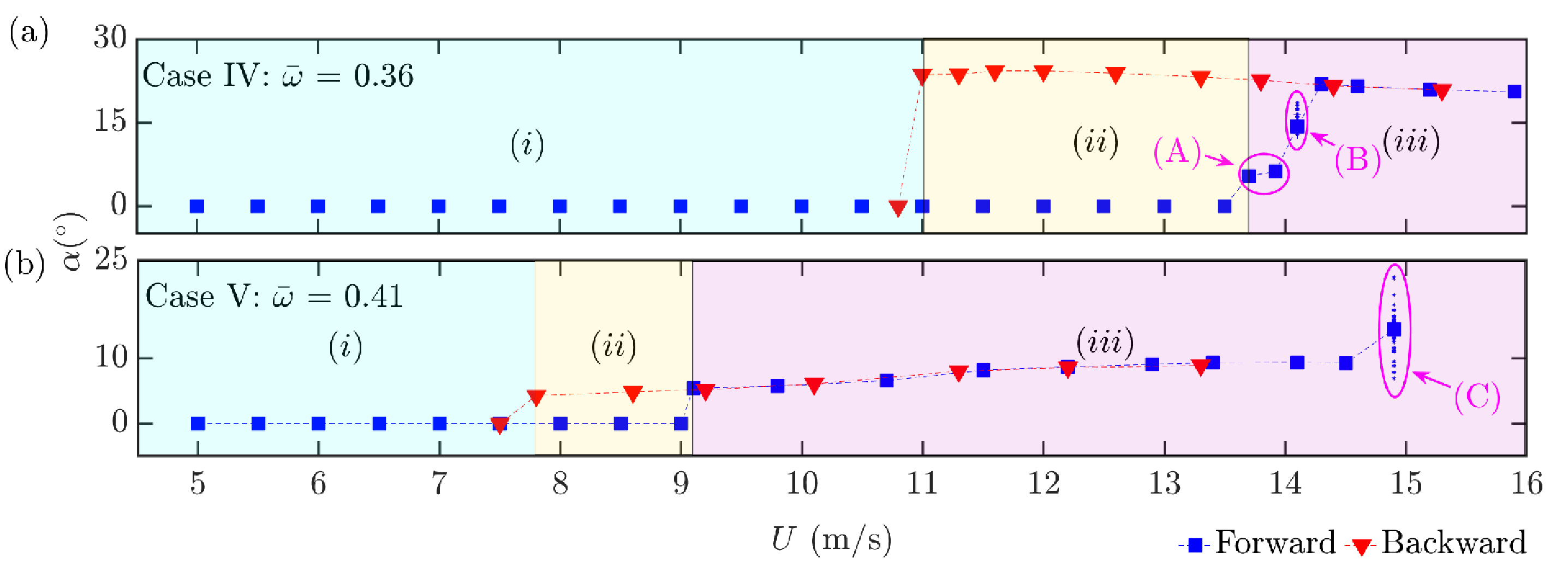}
\caption{Bifurcation plots for $\bar{\omega}$ $<$ 0.5. Regime I - stable FP attractor, Regime II - coexisting stable FP and stable LCO attractors, and Regime III - stable LCO attractor.}
\label{bif_wc4b}
\end{figure}

In this category, two cases (Case IV and V) are tested with $\bar{\omega}$ $<$ 0.5. The structural parameters are listed in Table~\ref{ep5bc4}. In Cases I-III reported in Section~\ref{cat1c4} for $\bar{\omega}$ $\geq$ 0.5, we see a common bifurcation route giving way to stall flutter. Here, the contribution of plunge mode is by and large negligible in the flutter mechanism as the plunging motion takes place due to the energy supplied from the pitch mode by the virtue of coupling between the two. However, for $\bar{\omega}$ $<$ 0.5, some peculiar aeroelastic responses have been observed, altering this bifurcation scenario. The bifurcation diagram for Case IV and Case V are shown in Fig.~\ref{bif_wc4b}.

\begin{table}
\caption{\label{ep5bc4} Structural parameters for the experiment estimated from static tests for $\bar{\omega}$ $<$ 0.5 cases.}
\begin{center}
\begin{tabular}{@{}c| c c c@{}} 
\hline
\hline
Parameter & \multicolumn{2}{c}{Values}\\
\hline
\rule{0pt}{12pt}
$m_y$ (kg)& \multicolumn{2}{c}{1.908} \\
$m_{\alpha}$ (kg)& \multicolumn{2}{c}{0.937} \\
$c$ (m)& \multicolumn{2}{c}{0.1} \\
$x_{ea}$ & \multicolumn{2}{c}{0.25$c$} \\
$x_{c}$ & \multicolumn{2}{c}{0.40$c$} \\
$I_\alpha$ ($\rm{kg\cdot m^2}$)& \multicolumn{2}{c}{0.0017} \\
\hline
&Case IV & Case V\\
$\zeta_\alpha$ & 0.01 & 0.01 \\
$\zeta_y$ & 0.06 & 0.05 \\
{$f_y$ (Hz)}& 1.43 & 2.28 \\
{$f_\alpha$ (Hz)}& 4.01 & 5.50 \\
{$\bar{\omega}$} & 0.36 & 0.41 \\
\hline
\hline
\end{tabular}
\end{center}
\end{table}

For both cases, the bi-stable regime has been observed akin to cases where $\bar{\omega}$ $\geq$ 0.5. In case IV ($\bar{\omega}$ = 0.36), $U_{cr}$ and $U_{fld}$ are observed at 13.7 m/s and 11.0 m/s, respectively (see Fig.~\ref{bif_wc4b}(a)). In Case V ($\bar{\omega}$ = 0.41), $U_{cr}$ and $U_{fld}$ are observed at 9.1 m/s and 7.8 m/s, respectively (see Fig.~\ref{bif_wc4b}(b)), with smallest bi-stable regime among all five cases under consideration. However, in both cases some dynamical scenarios are observed, where the plunge mode plays a significant role in dictating the flutter mechanism, contrary to the conventional definition of stall flutter observed in Case I-III. These responses are marked as (A) and (B) in Fig.~\ref{bif_wc4b}(a) and as (C) in Fig.~\ref{bif_wc4b}(b). In Case IV (see Fig.~\ref{bif_wc4b}(a)) the flutter onset is marked by a classical/coupled-mode flutter (marked as (A)), where the pitch and plunge modes coalesce at a common frequency between $f_{\alpha}$ and $f_y$. Upon increasing the speed, the period-1 LCOs (corresponding to the classical flutter) undergo period doubling, which is marked as (B) in Fig.~\ref{bif_wc4b}(a). In Case V (Fig.~\ref{bif_wc4b}(b)) the flutter onset is characterized as stall flutter akin to Cases I-III, but at a higher flow speed ($U$ = 14.9 m/s), a secondary frequency component is observed close to the plunge natural frequency and subsequently, the aeroelastic responses culminate into a beating response (marked as (C)). A detailed aeroelastic response analysis (in both time and frequency domains) corresponding to these bifurcation scenarios is given in this section. Simultaneously, the underlying synchronization dynamics are also presented for each of these cases in the vicinity of the fold and Hopf points.

\begin{figure}
\centering
\includegraphics[width=5.2in, height=2.6in]{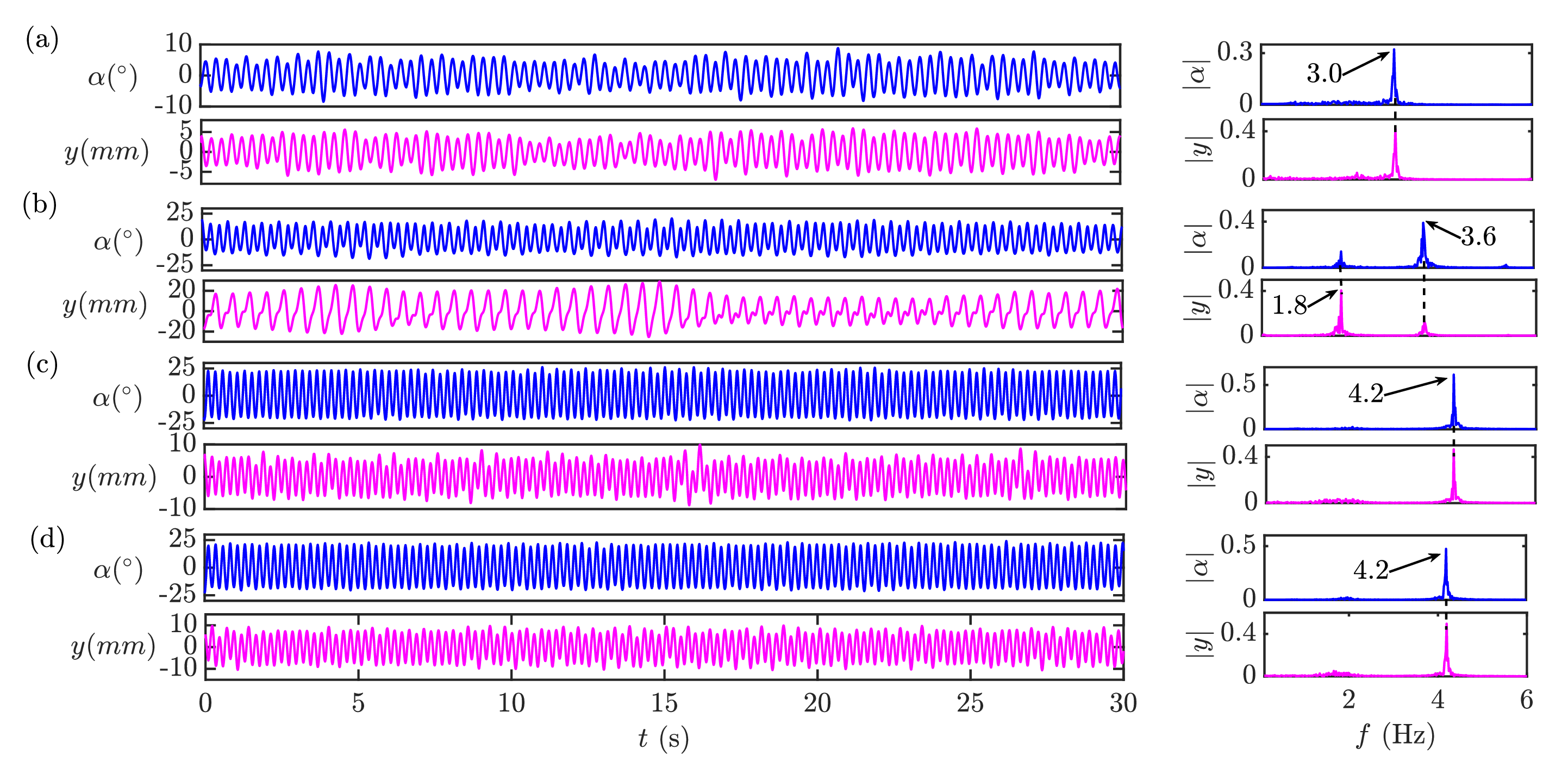}
\caption{Response dynamics post Hopf point (Case IV: $\bar{\omega}$ = 0.36, $f_y$ = 1.43 Hz, $f_{\alpha}$ = 4.01 Hz) (a) $U$ = 13.7 m/s, (b) $U$ = 14.1 m/s, (c) $U$ = 14.3 m/s, (d) $U$ = 15.9 m/s.}
\label{thisty4c4}
\end{figure}

The aeroelastic response dynamics for Case IV ($\bar{\omega}$ = 0.36) is shown in Fig.~\ref{thisty4c4}. In this case, $\bar{\omega}$ is the least among all the $\bar{\omega}$ values considered. Here, the flutter onset is marked by a classical flutter in contrast to other $\bar{\omega}$ cases. The onset of flutter is marked at 13.7 m/s with a pitch amplitude $\approx$ 10$^\circ$ and plunge amplitude $\approx$ 5 mm (Fig.~\ref{thisty4c4}(a)). The flutter frequency of both modes is observed to be 3.00 Hz. Contrary to Cases I-III presented thus far where both pitch and plunge modes oscillate near the pitch natural frequency, the flutter frequency is between $f_{\alpha}$ (4.01 Hz) and $f_y$ (2.28 Hz) and fairly distant from them. This indicates a coupled flutter mechanism where the two modes coalesce at common frequency peaks \cite{vishal2021routes}. Note that in Case I-III as well we observe the two modes to have common frequency peaks, but the main demarcating feature here is that in stall flutter cases (Case I-III) these common frequency peaks are observed close to $f_{\alpha}$, indicating that the plunge natural dynamics is suppressed. On the contrary, for $\bar{\omega}$ = 0.36, it appears that both the modes actively contribute as they come close to coalescing at a common frequency (between $f_{\alpha}$ and $f_y$). This mechanism is characterized as a \textit{frequency locking mechanism} \cite{vishal2021routes}. One possible reason to encounter a coupled flutter at $\bar{\omega}$ can be due to the low stiffness in plunge ($f_y$ = 1.43 Hz). Going by the observations in Dimitriadis and Li\cite{dimitriadis2009bifurcation}, where plunge mode has no contribution due to its high stiffness, conversely, plunge becomes more flexible and hence actively participates in the flutter mechanism. The pitch stiffness in Case IV is the same as in Cases I and II, while the plunge stiffness is significantly lesser as compared to those two (see Tables~\ref{ep5ac4} and ~\ref{ep5bc4}). Consequently, while in Cases I and II the plunge remains a passive/suppressed mode, it actively participates in Case IV, and hence a classical flutter is observed instead of a stall flutter. Another factor that confirms the flutter to be a coupled type is the pitch amplitude which is below the typically reported static stall angles \cite{dos2021improvements} for NACA 0012.

Upon increasing the flow speed, no qualitative change in the dynamics is observed up to $U$ = 14.1 m/s. At $U$ = 14.1 m/s, a secondary bifurcation is observed where a secondary peak emerges in both the modes indicating a period-doubling scenario (Fig.~\ref{thisty4c4}(b)). Interestingly, the two modal frequencies are in a 2:1 ratio (3.60 Hz and 1.80 Hz). This sudden transition is accompanied by a sharp jump in pitch and plunge amplitudes. The pitch amplitudes reach approximately 20$^\circ$ while the plunge amplitude approaches 25 mm. The sudden increase in amplitudes is indicative of resonance as the energy transferred to the system seems to increase significantly. This behaviour closely matches with experimental works by Poirel and co-workers \cite{poirel2018frequency,goyaniuk2020pitch,benaissa2021beating,goyaniuk2023energy} where the authors report resonance between pitch and plunge modes at $\bar{\omega}$ $\approx$ 1. This behaviour was not observed for $\bar{\omega}$ away from 1. While the authors acknowledged the nonlinearities associated with stall flutter, they did not delve into the role of any specific nonlinearity contributing to such nonlinear behavior.

On the other hand, works from Gilliatt and co-workers \cite{gilliatt1997nonlinear,gilliatt1998presence,gilliatt2003investigation} describe the nonlinear mechanism giving way to such energy exchange characterized as an internal resonance. The aeroelastic systems are susceptible to an internal resonance if the modal frequencies of the system commensurate \cite{gilliatt2003investigation,lee2005triggering1}, (\textit{i.e.}, ${n\omega_1}$+ $m{\omega_2}$+ ... = 0, where $n$, $m$, ... are integers and $\omega_1$, $\omega_2$, ... are modal frequencies). However, this condition alone does not suffice for the occurrence of internal resonance as a specific nonlinear coupling between the modes is necessary. More precisely, a 2:1 internal resonance is possible if the modal frequencies are tuned in two to one ratio given quadratic nonlinearities are present \cite{lee2005triggering,tripathi2023frequency}, and a 3:1 internal resonance is possible if the modal frequencies are tuned in three to one ratio, given the cubic nonlinearities are present \cite{gilliatt2003investigation,shaw2016periodic}. As far as response at $U$ = 14.1 m/s for Case IV is concerned, it is due to a 2:1 internal resonance as the primary and secondary frequency peaks are in a 2:1 ratio (Fig.~\ref{thisty4c4}(b)) complimented by the presence of quadratic nonlinearity in the system (see Fig.~\ref{nonlnc4}). It is important to mention here that though the natural frequency ratio is 0.36 indicating the natural frequencies of the two modes are incommensurate, the aeroelastic flutter frequencies are controlled by the aerodynamic loading and are tuned in 2:1 at a specific airspeed.

The secondary frequency peak (at 1.80 Hz) is the dominant peak in plunge frequency response and also has significant strength in pitch frequency response (see Fig.~\ref{thisty4c4}(b)). Since this secondary peak is close to $f_y$, the emergence of it underscores the pivotal role of plunge natural dynamics in ongoing motion. This is in contrast with the suppression of plunge natural dynamics observed in Cases I-III. However, this dynamics is short-lived as the secondary peak suddenly disappears and the motion switches back to a period-1 LCO response (Fig.~\ref{thisty4c4}(c)) with a slight increase in $U$ from 14.1 m/s to 14.3 m/s. The frequency response now shows a single dominant frequency peak at 4.20 Hz in both pitch and plunge. This frequency peak is close to $f_{\alpha}$ = 4.01 Hz, which indicates that the aeroelastic responses are now driven by pitch natural dynamics, relegating plunge to be a passive mode, akin to the behaviour observed in Cases I-III. The plunge amplitudes do not exceed 10 mm and are far lesser than those at $U$ = 14.1 m/s during an internal resonance ($\approx$ 25 mm). Here, it is evident that the energy that is supplied to plunge during the internal resonance is no longer available. The pitch amplitudes on the other hand remain close to 20$^\circ$. The pitch amplitudes and pitch dominant flutter frequencies indicate the flutter mechanism now transits to stall flutter.

A question that arises at this juncture is how the dynamics changes from a period-2 response to a stall flutter (period-1 LCOs). One possible explanation of this can be drawn from \v{s}idlof \textit{et al.}\cite{vsidlof2016experimental}, who explained that the classical flutter can give way to stall flutter by triggering the dynamic stall. In the current scenario, the pitch and plunge amplitude suddenly grow as the internal resonance takes place. As the pitch amplitudes increase, the flow is possibly separated at high AoA and allows the dynamic stall to manifest. This explanation can be rationalized by considering the proximity of the flow speed associated with stall flutter ($U$ = 14.3 m/s) to that linked with internal resonance ($U$ = 14.1 m/s). This means that with a slight increase in flow speed, the aerodynamic loading changes and modal frequencies move slightly away from a 2:1 relationship, and given the dynamic stall conditions provided due to high pitch oscillations, stall flutter is manifested. 

No further qualitative change in the dynamics is noted up to $U$ = 15.9 m/s (Fig.~\ref{thisty4c4}(d)). The dynamic transitions observed in Case IV can be summarized as: i) flutter onset is marked by a classical flutter at $U$ = 13.7 m/s, ii) at $U$ = 14.1 m/s, we observe a period-doubling and a 2:1 internal resonance - pitch and plunge amplitude grow significantly, and iii) flow separation at such high AoA ($\approx$ 20$^\circ$) possibly triggers the dynamic stall and the motion transforms to a period-1 stall flutter LCO. 

\begin{figure}
\centering
\includegraphics[width=5in, height=2.2in]{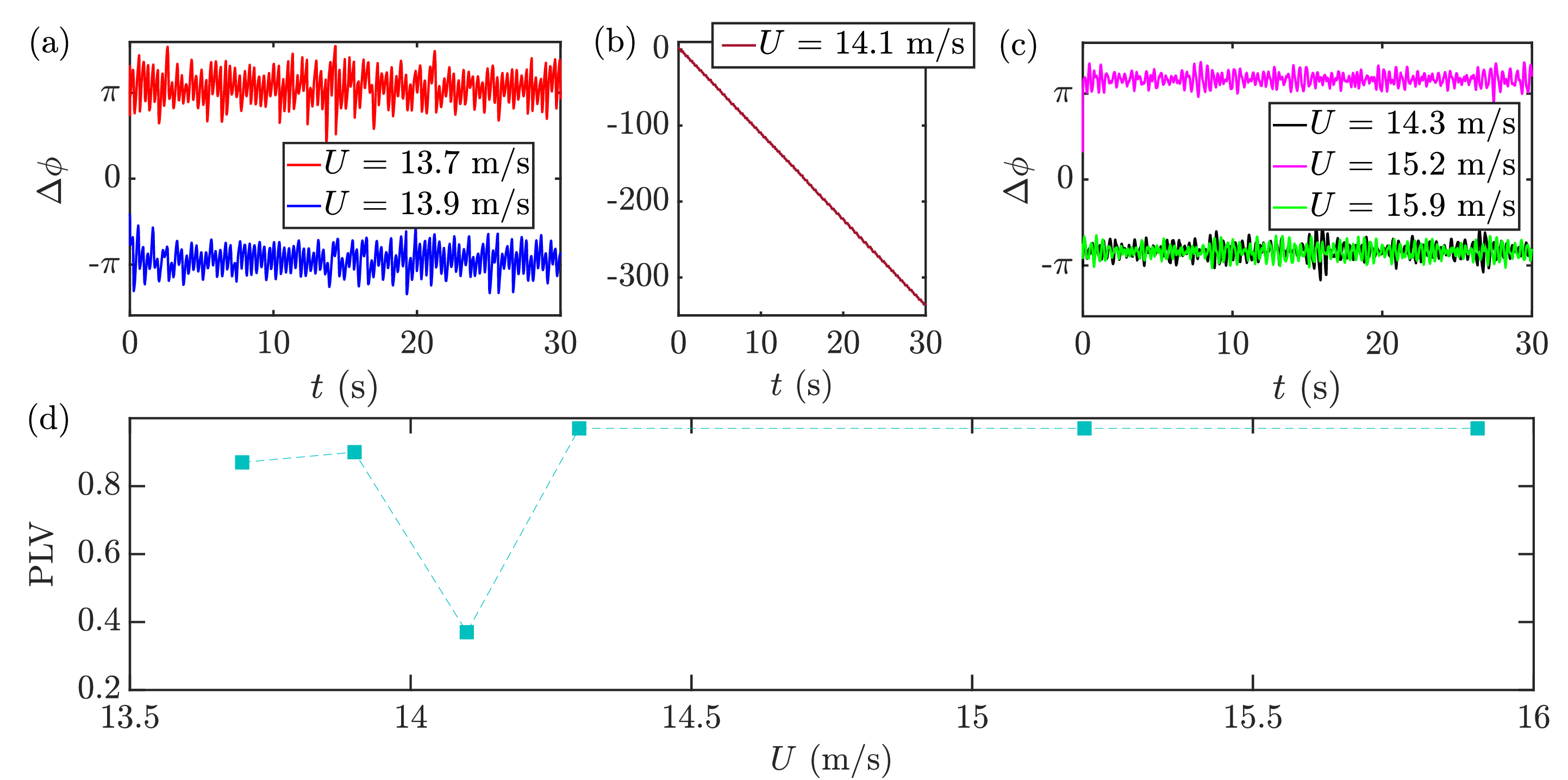}
\caption{Synchronization analysis post Hopf point (Case IV: $\bar{\omega}$ = 0.36, $f_y$ = 1.43 Hz, $f_{\alpha}$ = 4.01 Hz). (a) RPV during classical flutter, (b) RPV during period-2 oscillations, (c) RPV during stall flutter, and (d) Variation of PLV with the flow speed.}
\label{phisty4c4}
\end{figure}

Underlying synchronization analysis for Case IV reveals an interesting scenario (see Fig.~\ref{phisty4c4}). During the classical flutter $U$ = 13.7 - 13.9 m/s the relative phase is bounded over the time indicating a phase synchronization via phase trapping (Fig.~\ref{phisty4c4}(a)). However, a sudden loss in synchronization is observed at $U$ = 14.1 m/s - corresponding to the period-2 dynamics, as the relative phase of pitch and plunge modes exhibit a steep drift over time (Fig.~\ref{phisty4c4}(b)). The two modes become asynchronous. Interestingly the modes instantly switch back to being phase synchronized as the stall flutter takes place at $U$ = 14.3 m/s (Fig.~\ref{phisty4c4}(c)). This sudden and sharp loss and subsequent gain in synchronization are depicted as a V-shaped recovery in PLV variation (Fig.~\ref{phisty4c4}(d)).

\begin{figure}
\centering
\includegraphics[width=5in, height=2.2in]{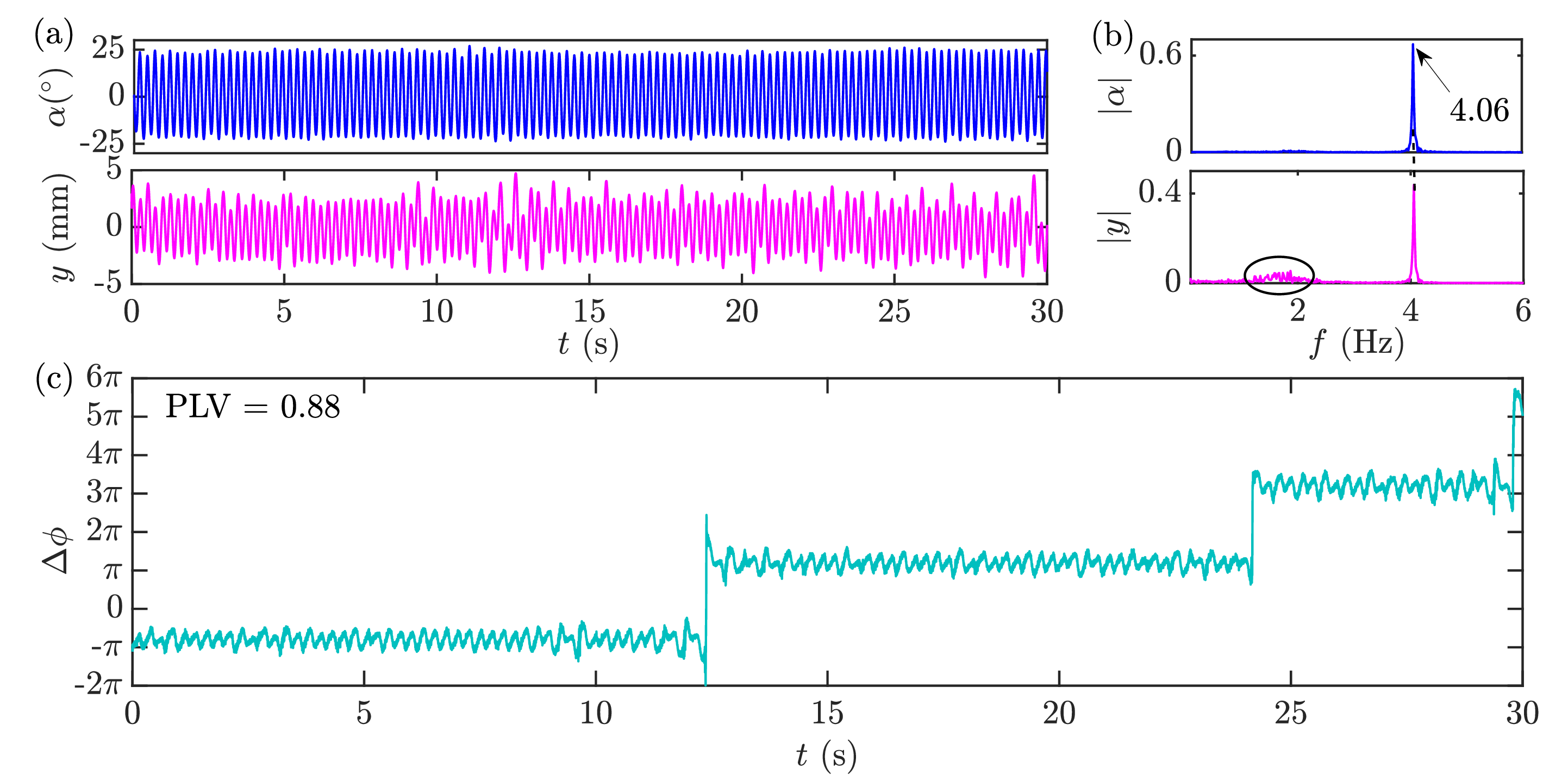}
\caption{(a) Aeroelastic response dynamics, (b) corresponding frequency responses, and (c) synchronization dynamics, at $U_{fld}$ = 11.0 m/s (Case IV: $\bar{\omega}$ = 0.36, $f_y$ = 1.43 Hz, $f_{\alpha}$ = 4.01 Hz).}
\label{u11b_y4c4}
\end{figure}

In the backward sweep experiments, the $U_{fld}$ is observed at 11.0 m/s with pitch amplitude $\approx$ 25$^\circ$ and plunge amplitude $\approx$ 5 mm (Fig.~\ref{u11b_y4c4}(a)). The frequency response shows a single dominant peak at 4.06 Hz, close to $f_{\alpha}$ (Fig.~\ref{u11b_y4c4}(b)). The phase dynamics shows a synchronization via an intermittent phase synchronization with a PLV of 0.88 (Fig.~\ref{u11b_y4c4}(c)). This dynamics is very much similar to that observed in Cases I-III.

\begin{figure}
\centering
\includegraphics[width=5.4in, height=2.6in]{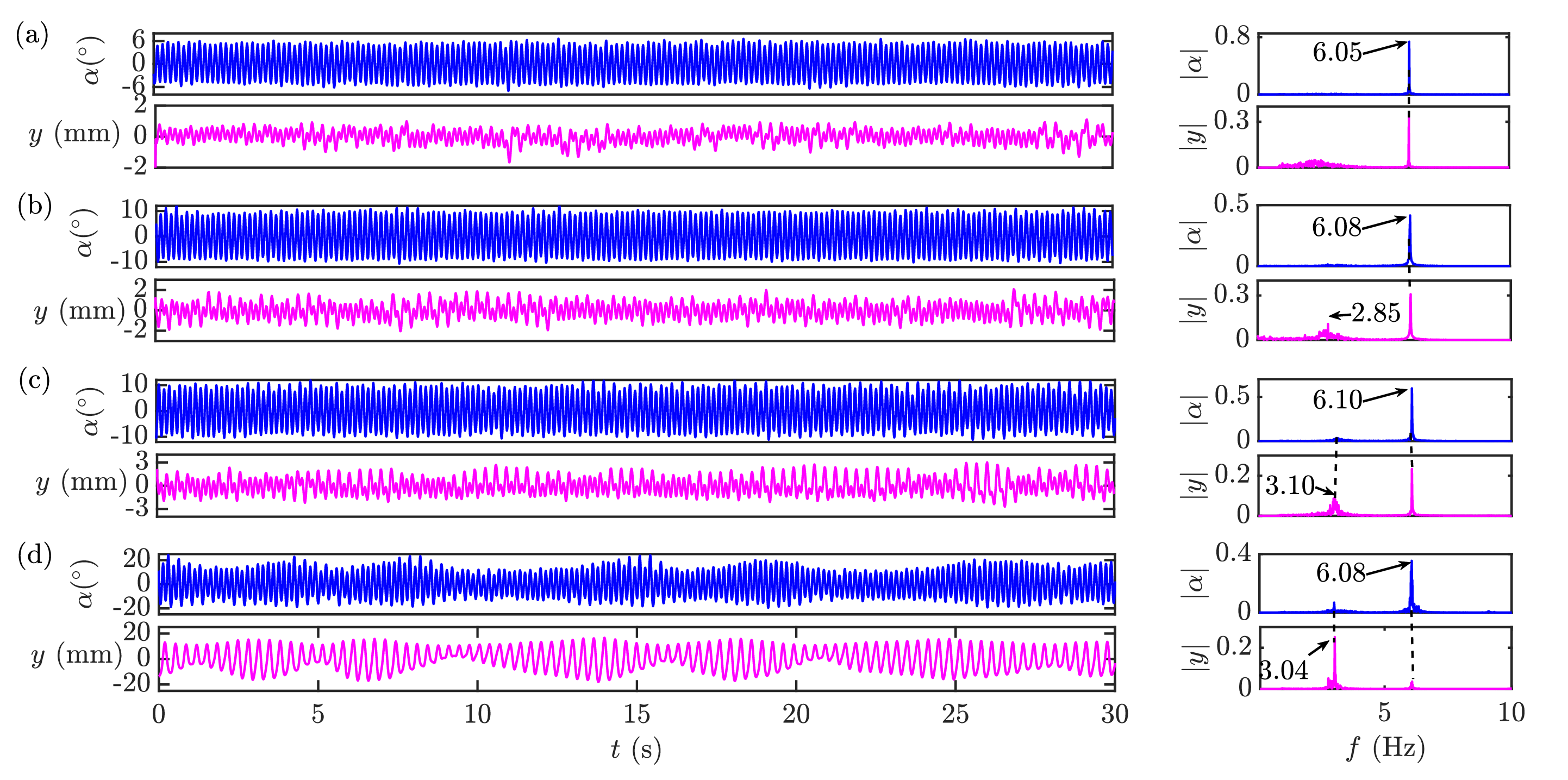}
\caption{Response dynamics post Hopf point (Case V: $\bar{\omega}$ = 0.41, $f_y$ = 2.28 Hz, $f_{\alpha}$ = 5.50 Hz) (a) 9.1 m/s, (b) 13.4 m/s, (c) 14.5 m/s, (d) 14.9 m/s.}
\label{thist2yc4}
\end{figure}

Another interesting bifurcation route is observed in Case V ($\bar{\omega}$ = 0.41), the final case of $\bar{\omega}$ variation. The experimental parameters for this case have been listed in Table~\ref{ep5bc4}. The structural nonlinearities in pitch and plunge contain both cubic and quadratic coefficients (see Fig.~\ref{nonlnc4}). Since $\bar{\omega}$ is close to and below 0.5, and quadratic nonlinearity is present, the prospect of encountering an internal resonance remains viable. It is important to emphasize that the pitch stiffness is the highest in this case among all five cases ($f_{\alpha}$ = 5.50 Hz), while the plunge stiffness is the same as that in Case I ($f_y$ = 2.28 Hz). The aeroelastic response dynamics for this case are shown in Fig.~\ref{thist2yc4}. The flutter onset is market at $U$ = 9.1 m/s (Fig.~\ref{thist2yc4}(a)). Small amplitude LCOs - $\approx$6$^\circ$ in pitch and $\approx$ 1 mm in plunge are observed with a flutter frequency of 6.05 Hz for both modes. Though the flutter frequency is close to $f_{\alpha}$, it is not as close as is observed in Cases I-III. Additionally, close to $f_y$, a broadband of minimal strength is observed in the frequency response of plunge mode. The amplitudes of LCOs are restricted by the high-pitch stiffness. The amplitudes do not increase much even at higher flow speeds.

At 13.4 m/s, the amplitudes are approximately 10$^\circ$ in pitch and 2 mm in the plunge (Fig.~\ref{thist2yc4}(b)). However, the broadband in the plunge now becomes more concentrated and evolves as a secondary peak in the plunge frequency response at 2.85 Hz. This secondary peak is at slightly less than half of the primary peak and hence the two modal frequencies are not commensurate as of now. The plunge displacements show intermittent period-1 - period-2 time responses. At $U$ = 14.5 m/s, the secondary peak - which is now shifted to 3.10 Hz, becomes even more prominent in plunge response (Fig.~\ref{thist2yc4}(c)). A very small secondary frequency peak is shown in the pitch frequency response as well. Note that the primary frequency peak is at 6.10 Hz and hence the dynamics represent a near commensurate situation.

A sudden change in dynamics is observed at $U$ = 14.9 m/s as the aeroelastic responses transition to a beat-like motion (Fig.~\ref{thist2yc4}(c)). The primary and secondary frequency peaks (3.04 Hz and 6.08 Hz, respectively) are now commensurate in 2:1. The frequency response closely resembles a scenario from Case IV where internal resonance is seen. The pitch amplitude is now reached up to 20$^\circ$ despite the high stiffness in pitch. Plunge amplitudes are increased even more drastically and are close to 20 mm (from $\approx$ 3 mm at $U$ = 14.5 m/s). This is a 2:1 internal resonance condition and involves a high amount of energy exchange between the two modes.

\begin{figure}
\centering
\includegraphics[width=5in, height=2.2in]{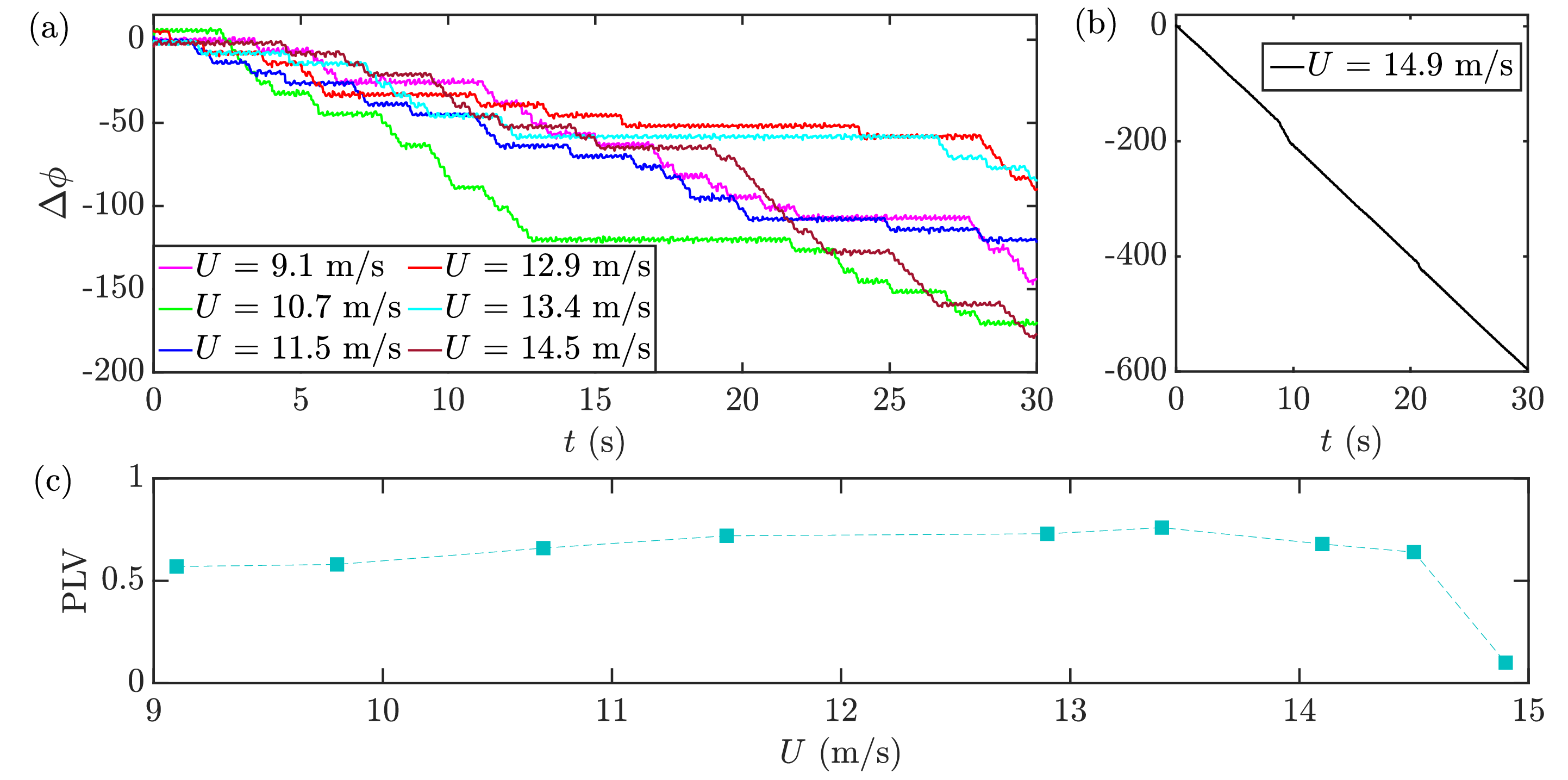}
\caption{Synchronization analysis post Hopf point (Case V: $\bar{\omega}$ = 0.41, $f_y$ = 2.28 Hz, $f_{\alpha}$ = 5.50 Hz). (a) RPV before the onset of 2:1 internal resonance, (b) RPV at the onset of 2:1 internal resonance, (c) Variation of PLV.}
\label{phist_2yc4}
\end{figure}

Corresponding synchronization analysis for the dynamics observed in Case V is carried out next (Fig.~\ref{phist_2yc4}). Figure~\ref{phist_2yc4}(a) shows the RPV dynamics at the flutter onset ($U$ = 9.1 m/s) which depicts an intermittent phase synchronization. The synchronization dynamics remain qualitatively the same up to $U$ = 14.5 m/s. At $U$ = 14.9 m/s, which is the onset of beats, the pitch and plunge become asynchronous (see Fig.~\ref{phist_2yc4}(b)). This synchronization behaviour corresponds to the internal resonance encountered at $U$ = 14.9 m/s and interestingly is observed in Case IV also during the internal resonance. Variation of PLV (Fig.~\ref{phist_2yc4}(c)) is gradually upwards from flutter onset (PLV = 0.57) up to $U$ = 13.4 m/s, where the maximum PLV of 0.76 is attained. While further increasing the flow speed up to 14.5 m/s the PLV drops minimally and reaches 0.64. This dip in PLV is possibly due to the secondary frequency peak becoming stronger. Interestingly, a sharp drop is observed at $U$ = 14.9 m/s (corresponding to beating) and PLV is dropped to 0.10, indicating asynchrony as seen in Fig.~\ref{phist_2yc4}(b).

The sharp drop in the PLV and sudden shift to asynchrony are the telltales of 2:1 internal resonance. Tripathi \textit{et al.}\cite{tripathi2023frequency} reports a similar behaviour with $\bar{\omega}$ = 0.44, with an entirely different set of pitch and plunge stiffness and nonlinearity. However, in both cases $\bar{\omega}$ is close to and below 0.5 and structural nonlinearity has a quadratic term. The internal resonance between two modal frequencies gives rise to a beat-like response. Internal resonance is typically associated with high energy exchanges between the modes and is quite evident from the sudden increase in the amplitudes in both pitch and plunge modes. Such violent oscillations are sustained via a \textit{frequency-specific synchronization} and are extensively elucidated in Tripathi \textit{et al.}\cite{tripathi2023frequency}.
The term frequency-specific synchronization denotes a scenario where the inherent frequency-specific signals of pitch and plunge modes are synchronized however the overlying original signals of the two modes show asynchronous behaviour. For the present case, \textit{i.e.}, $\bar{\omega}$ = 0.41, the frequency-specific synchronization analysis at $U$ = 14.9 m/s is also presented in \ref{A1}.

\begin{figure}
\centering
\includegraphics[width=5in, height=2.2in]{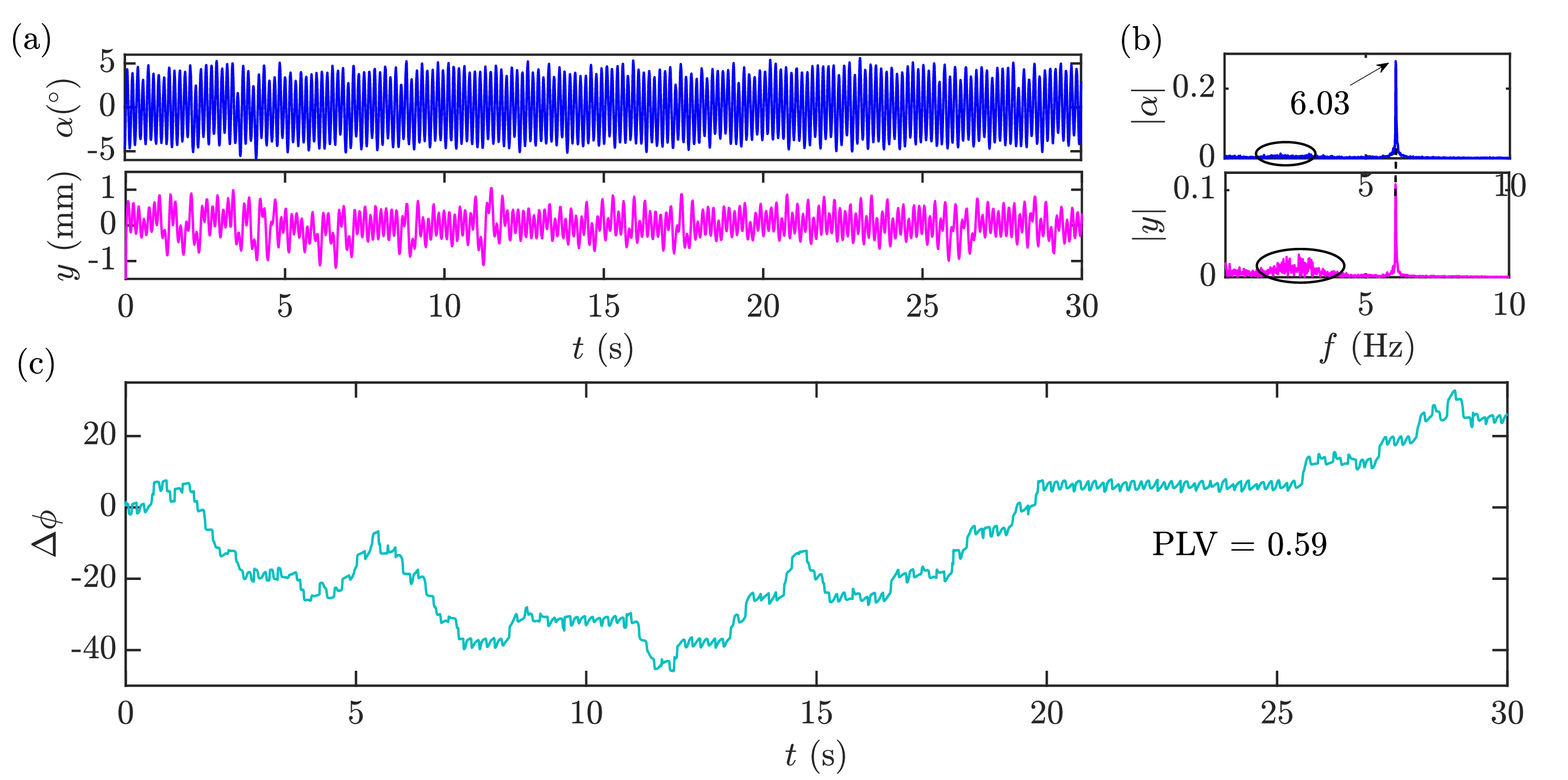}
\caption{(a) Response dynamics, (b) frequency response, and (c) synchronization dynamics, at fold point 7.8 m/s (Case V: $\bar{\omega}$ = 0.41, $f_y$ = 2.28 Hz, $f_{\alpha}$ = 5.50 Hz).}
\label{fld2yc4}
\end{figure}

In the backward sweep exercise, the LCOs are sustained till 7.8 m/s (see Fig.~\ref{fld2yc4}(a)). The pitch oscillations are very small ($\approx$ 5$^\circ$). The plunge motion is almost negligible with the amplitude of below 1 mm. The frequency response shows a single dominant peak at 6.03 Hz accompanied by broadband near $f_y$ (see Fig.~\ref{fld2yc4}(b)). The corresponding RPV dynamics shows an intermittent phase synchronization with arbitrary phase slips (see Fig.~\ref{fld2yc4}(c)). The PLV for this case is noted as 0.59, indicating a weak synchronization at the fold point.

\begin{figure}
\centering
\includegraphics[width=5in, height=2.2in]{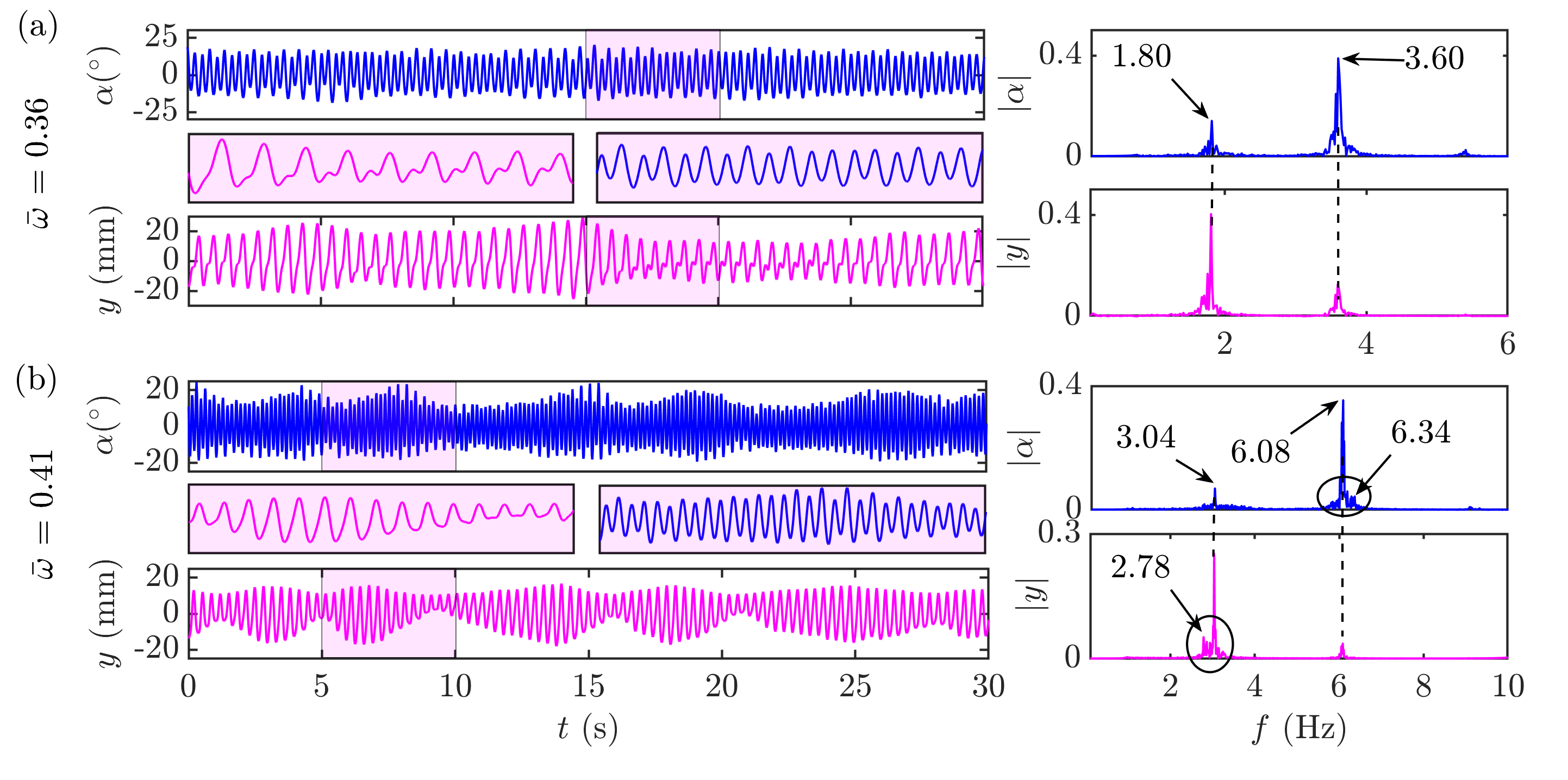}
\caption{Comparison of responses during internal resonance for (a) Case IV: $\bar{\omega}$ = 0.36, and (b) Case V: $\bar{\omega}$ = 0.41.}
\label{cat2_comp}
\end{figure}

\begin{figure}
\centering
\includegraphics[width=5in, height=2.2in]{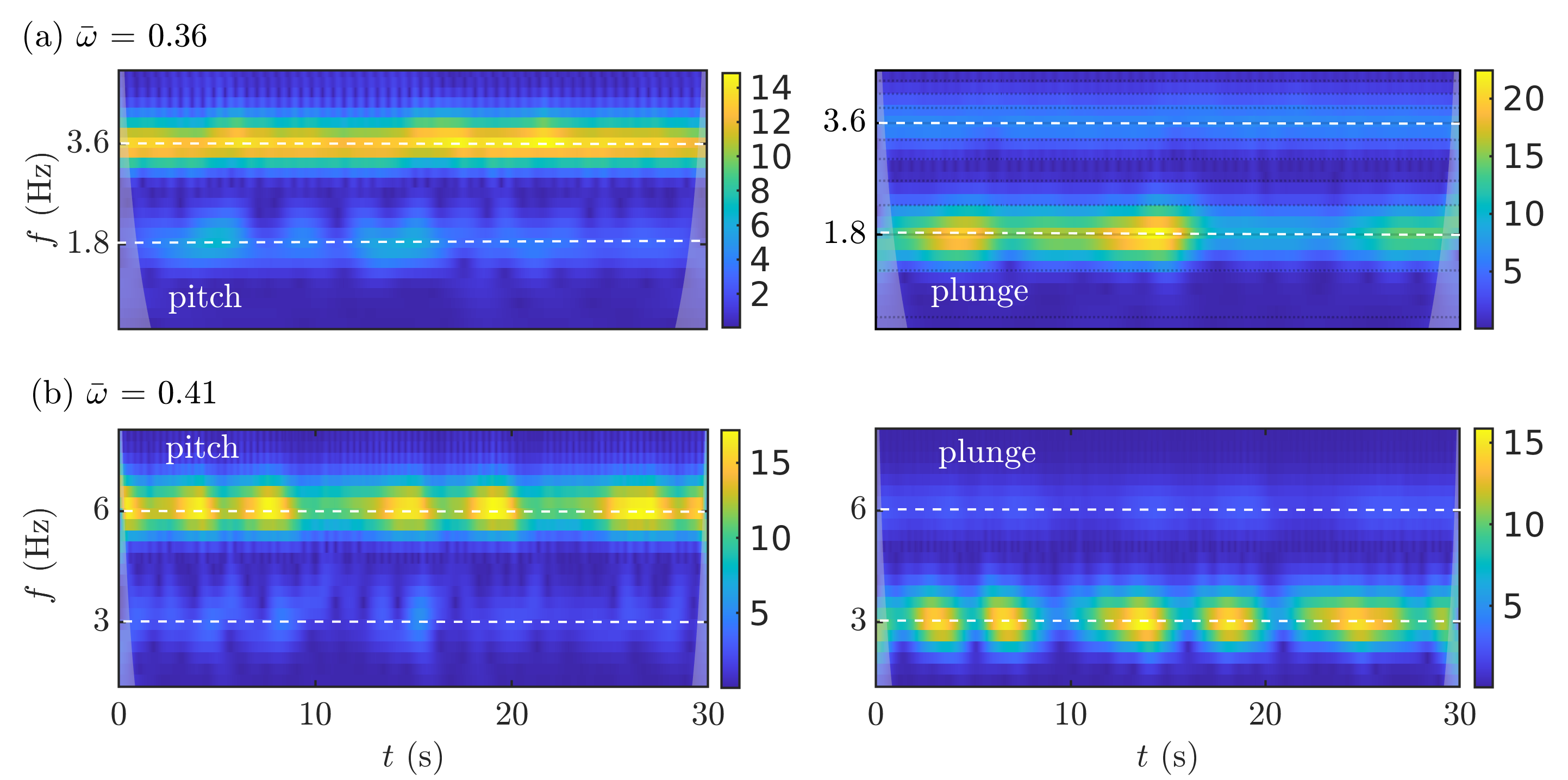}
\caption{CWT scalograms during internal resonance for (a) Case IV: $\bar{\omega}$ = 0.36, and (b) Case V: $\bar{\omega}$ = 0.41.}
\label{cwt_CAT2}
\end{figure}

To compare the responses of internal resonance in Case IV and V, their plots are presented together in Fig.~\ref{cat2_comp}. Additionally, the frequency-time response is also shown in Fig.~\ref{cwt_CAT2} to enhance the visualization of primary and secondary frequency signals. The pitch responses for $\bar{\omega}$ = 0.36 exhibit a period-2 pattern, evident from the time history and its fast Fourier transform (FFT) plots, displaying a larger primary and a smaller secondary peak (Fig.~\ref{cat2_comp}(a)). On the other hand, the plunge response has a modulated amplitude with the secondary peak being dominant. Corresponding continuous wavelet transform (CWT) plot (Fig.~\ref{cwt_CAT2}(a)) for pitch response indicates a consistent strength of the primary frequency signal (at 3.60 Hz) throughout time, while the secondary frequency signal (at 1.80 Hz) shows low and varying strength. In the CWT scalogram of plunge response, the secondary frequency signal is of higher strength albeit with modulation in strength (Fig.~\ref{cwt_CAT2}(a)), while the strength of the primary frequency signal is comparatively far lesser.

The response time histories in Case V on the contrary, are clearly in a beat-like structure (Fig.~\ref{cat2_comp}(b)). Corresponding to the beating response, a small third frequency peak is observed in both pitch and plunge responses at 6.34 Hz and 2.78 Hz, respectively (see Fig.~\ref{cat2_comp}(b)), indicating a beating frequency of approximately 0.26 Hz. The corresponding CWT scalogram reveals a modulated strength in the primary frequency signal of the pitch response and the secondary frequency signal of the plunge response (Fig.~\ref{cat2_comp}(b)).

To summarize, $\bar{\omega}$ variation leads to a phenomenological set of bifurcation scenarios. These experiments are conducted by systematically changing the pitch and plunge stiffness resulting in $\bar{\omega}$ variation in the range 0.35 - 0.67. Note that, changing the stiffness in either mode inherently leads to a change in structural nonlinearities as well. The bifurcation routes remain qualitatively the same where $\bar{\omega}$ is equal to or above 0.5 - that is a subcritical Hopf bifurcation giving way to pitch dominant LCOs, characterized as stall flutter. However, for the cases where $\bar{\omega}$ is below 0.5, though the subcriticality is present, some unique dynamical transitions are observed, altering the bifurcation routes. These results are summarized in Table~\ref{wsum}. 

\begin{table}
\caption{\label{wsum}Summary of the aeroelastic behaviour observed for different $\bar{\omega}$ cases.}
\centering
\begin{tabular}{c| c c c | c c}
\hline
\hline
 & \multicolumn{3}{c|}{$\bar{\omega}$ $\geq$ 0.5} & \multicolumn{2}{c}{$\bar{\omega}$ $<$ 0.5} \\
{Case} & I & II & III & IV & V \\
\hline
$\bar{\omega}$ & 0.57 & 0.50 & 0.67 & 0.36 & 0.41 \\
$U_{cr}$ (m/s) & 13.3 & 12.3 & 9.4& 13.7 & 9.1 \\
$U_{fld}$ (m/s) & 8.9 & 9.7 & 5.9 & 11.0 & 7.8 \\
\hline
Aeroelastic response &  \multicolumn{3}{c|}{Stall flutter} & Classical flutter $\rightarrow$ period & Small amplitude pitch\\
dynamics & \multicolumn{3}{c|}{LCOs} & doubling $\rightarrow$ stall flutter & dominant LCOs $\rightarrow$ beats \\
\hline
\hline
\end{tabular}
\end{table}


\section{\label{eoea} Effect of elastic axis variation}

The next part of the analysis deals with the investigations into the aeroelastic behaviour as the elastic axis position is changed. In this section three elastic axis positions $x_{ea}$ = 0.15$c$, 0.25$c$, and 0.35$c$ are chosen. Accordingly, static imbalance for the three positions is calculated as $S$ = $m_y(x_c - x_{ea})$ and is listed in Table~\ref{eimbc4}. The static imbalance indicates the degree of inertial coupling between the structural modes which reduces as the elastic axis moves closer to the mass center \cite{goyaniuk2023energy} - as in $x_{ea}$ = 0.35$c$, and increases as the elastic axis moves closer to the leading edge - as in $x_{ea}$ = 0.15$c$. As the elastic axis coincides with the mass center, the structural system becomes essentially uncoupled, in this condition any coupling between the two modes will originate solely from the aerodynamics \cite{goyaniuk2020pitch}.

\begin{table}
\caption{\label{eimbc4} Static imbalance for different $x_{ea}$ locations.}
\centering
\begin{tabular}{c| c c c}
\hline
\hline
$x_{ea}$ & 0.15$c$ & 0.25$c$ & 0.35$c$\\
\hline
Static imbalance (kg-m) & 0.048 & 0.029 &  0.010 \\
\hline
\hline
\end{tabular}
\end{table}

\subsection{Change in elastic axis position at frequency ratio 0.57}

The structural parameters detailed in Table~\ref{ep5ac4} are considered here for analysis except that two additional $x_{ea}$ values are considered. While the bifurcation route for $x_{ea}$ = 0.25$c$ is already discussed in Section~\ref{cat1c4} (Case I) which exhibits a subcritical Hopf bifurcation with $U_{cr}$ = 12.3 m/s and $U_{fld}$ = 8.9 m/s (see Fig.~\ref{bif_wc4a}(a)), the dynamics for $x_{ea}$ = 0.15$c$ and $x_{ea}$ = 0.35$c$ is discussed in this part. Changing the elastic axis location results in small changes in $f_{\alpha}$ value which is 4.01 Hz for $x_{ea}$ = 0.25$c$. It shifts to 3.78 Hz for $x_{ea}$ = 0.15$c$, and to 4.22 Hz for $x_{ea}$ = 0.35$c$.

\begin{figure}
\centering
\includegraphics[width=5in, height=2.4in]{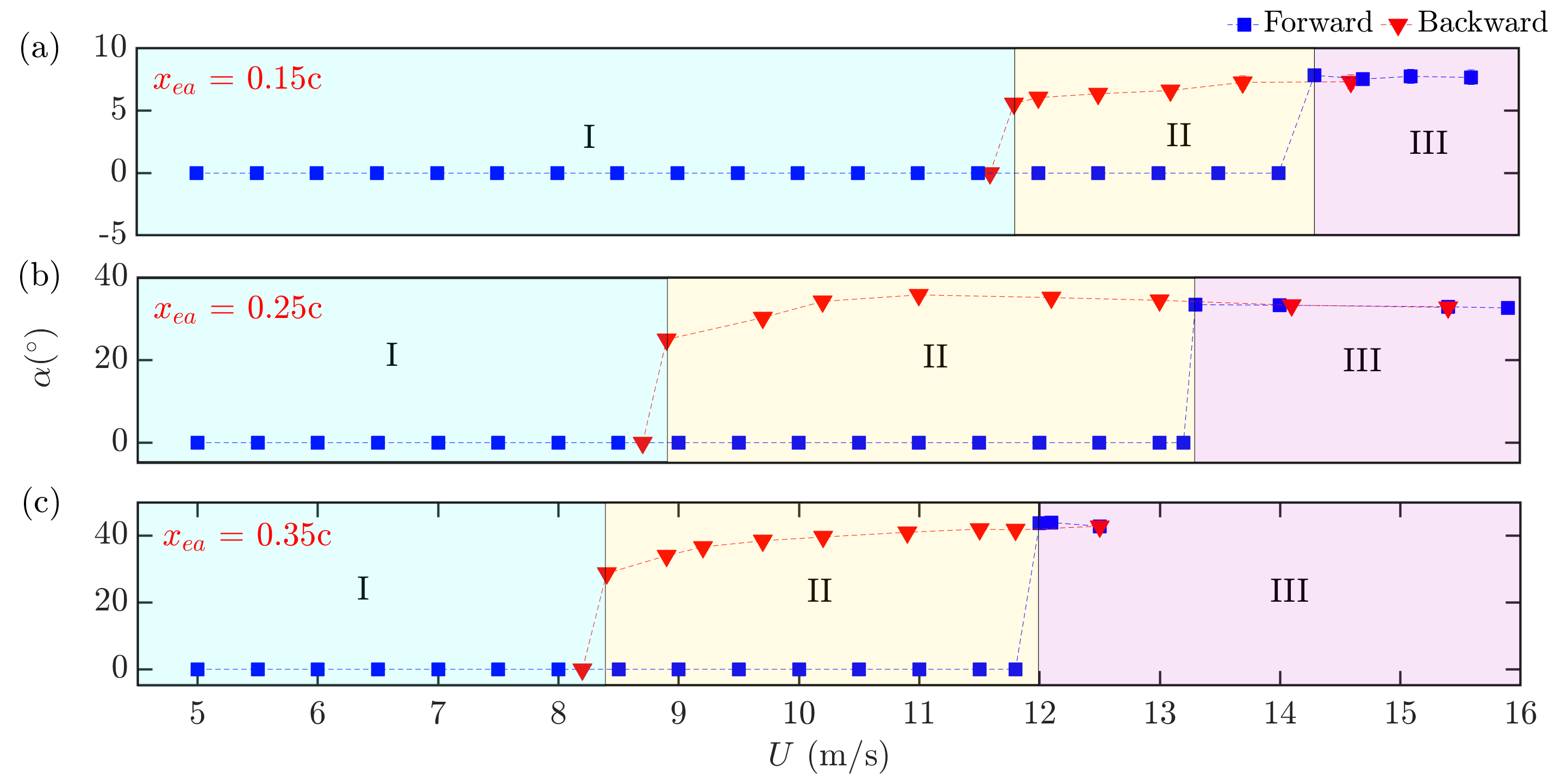}
\caption{Bifurcatiom plots for different $x_{ea}$ values for $\bar{\omega}$ = 0.57. Regime I - stable FP attractor, Regime II - coexisting stable FP and stable LCO attractors, and Regime III - stable LCO attractor.}
\label{bif_ec4}
\end{figure}

\begin{figure}
\centering
\includegraphics[width=5in, height=2in]{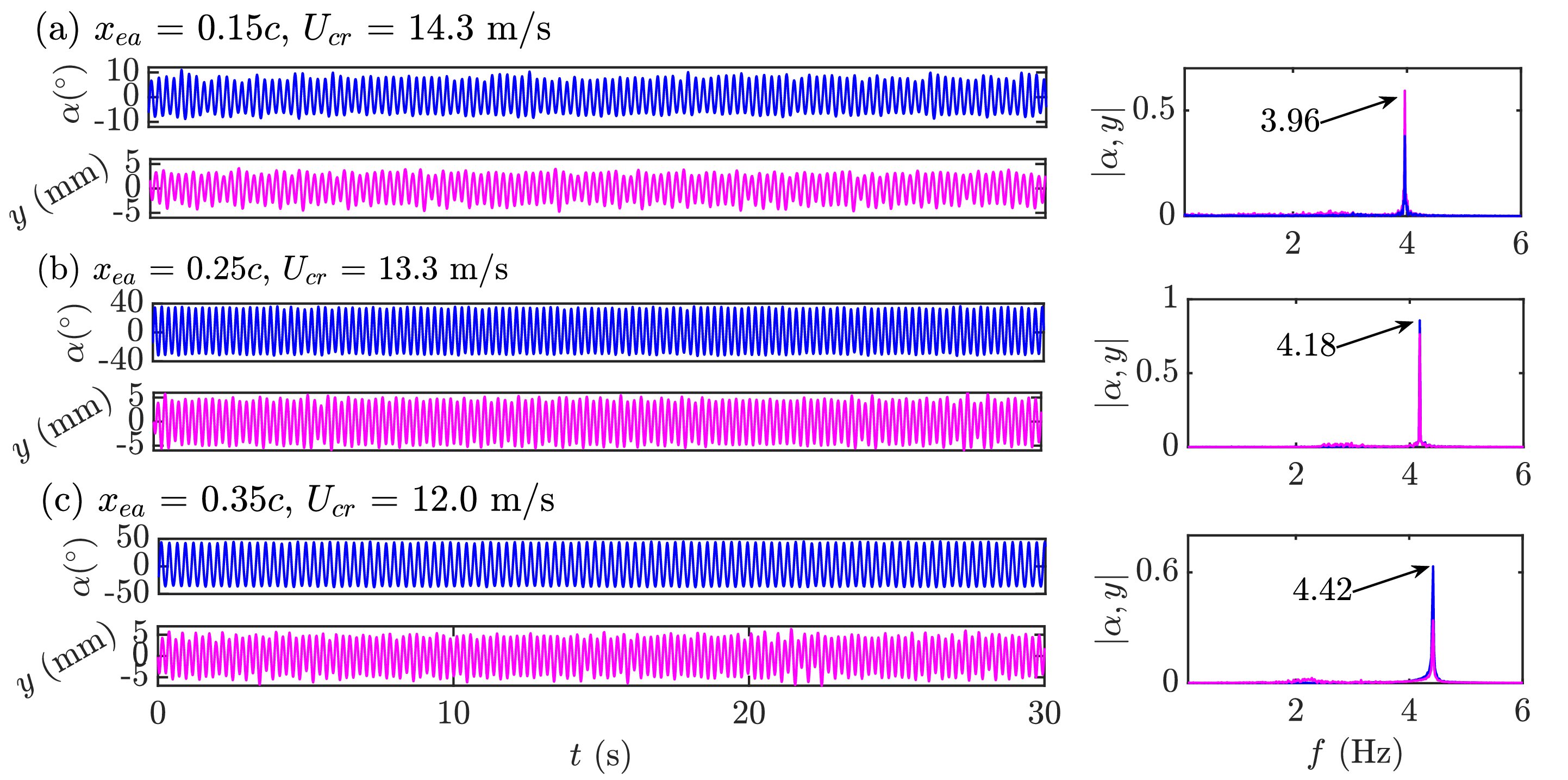}
\caption{Aeroelastic response dynamics at $U_{cr}$ for different elastic axis positions for $\bar{\omega}$ = 0.57.}
\label{xea_thpfc4}
\end{figure}

\begin{figure}
\centering
\includegraphics[width=5in, height=2.2in]{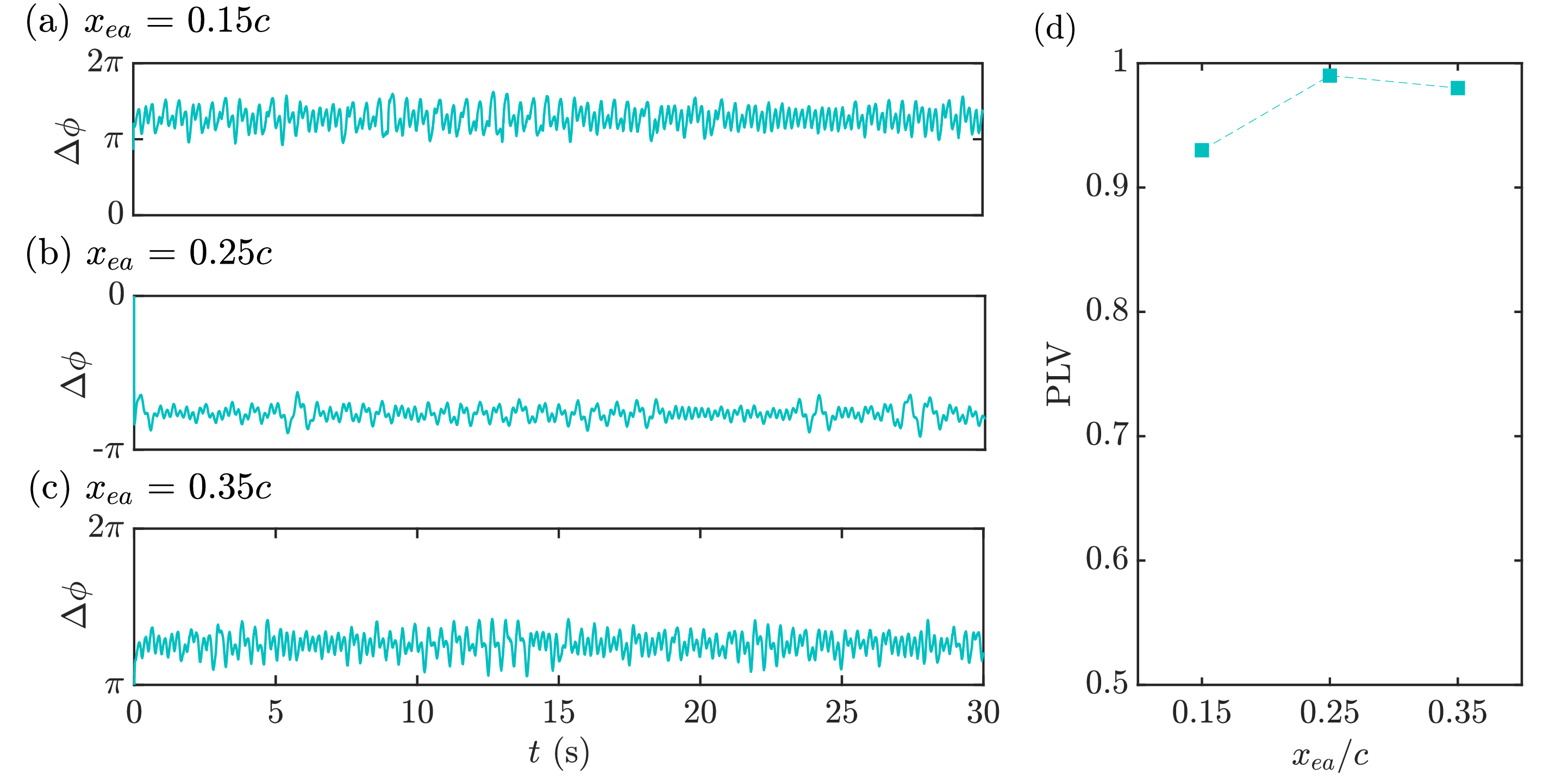}
\caption{The synchronization dynamics at $U_{cr}$ for various elastic axis positions for $\bar{\omega}$ = 0.57.}
\label{xea_phpfc4}
\end{figure}


The bifurcation diagram for three $x_{ea}$ values are collectively shown in Fig.~\ref{bif_ec4}(a)-(c). Akin to $x_{ea}$ = 0.25$c$ case (reproduced in Fig.~\ref{bif_ec4}(b)), the dynamical analysis for $x_{ea}$ = 0.15$c$ (Fig.~\ref{bif_ec4}(a)) and $x_{ea}$ = 0.35$c$ (Fig.~\ref{bif_ec4}(c)) cases also reveals the onset of instability via a subcritical Hopf bifurcation, albeit the flutter boundaries are significantly altered. Figure~\ref{xea_thpfc4} shows a comparison of pitch-plunge responses in time and frequency domains at $U_{cr}$. The flutter onset is found to be delayed as the elastic axis moves closer to the leading edge. For $x_{ea}$ = 0.15$c$, $U_{cr}$ is noted at 14.3 m/s and $U_{fld}$ at 11.8 m/s (Fig.~\ref{bif_ec4}(a)). The pitch LCO amplitudes do not exceed 10$^\circ$ in this case (Fig.~\ref{xea_thpfc4}(a)). It can be deduced from these observations that the aeroelastic system stabilizes as the distance between the elastic axis and mass center is increased and is consistent with the findings from relevant literature \cite{lee1999nonlinear,goyaniuk2023energy}. Note that the pitch and plunge frequency response shows a common peak at 3.96 Hz (Fig.~\ref{xea_thpfc4}(a)) which is close to $f_{\alpha}$ (3.78 Hz) - indicating that the dynamics is governed by the pitch mode. In this case, possibly a light stall event takes place as the pitch amplitudes are small and do not exceed static stall counterparts \cite{dos2021improvements}. A light stall event refers to the conditions where the flow is separated only at the trailing edge and force and moment coefficients do not exceed the static-stall range \cite{mccroskey1981phenomenon,razak2011flutter}.

On the contrary, at $x_{ea}$ = 0.35$c$, an early onset of instability is observed (Fig.~\ref{bif_ec4}(c)). The $U_{cr}$ and $U_{fld}$ in this case are found at 12.0 m/s and 8.4 m/s, respectively. The pitch LCOs in this case are very violently increased post the flutter onset and hence the experiments could not be conducted for flow speeds much higher than $U_{cr}$ in this case. The pitch LCO amplitudes approach $\approx$ 50$^\circ$ at the $U_{cr}$ (Fig.~\ref{xea_thpfc4}(c)). The flutter frequency, in this case, is observed to be 4.42 Hz for both pitch and plunge - again indicative of pitch-dominant dynamics (as $f_{\alpha}$ = 4.22 Hz). 
The synchronization dynamics at $U_{cr}$ (see Fig.~\ref{xea_phpfc4}) depicts strong synchronization for all three cases albeit for $x_{ea}$ = 0.15$c$, the PLV is slightly lower (0.93) as compared to other two cases. 

Note that in all three cases of $x_{ea}$ variation with $\bar{\omega}$ = 0.57, the LCO frequencies at the flutter onset are pitch dominant (Fig.~\ref{xea_thpfc4}). This means that the plunge mode is driven and passive - making it essentially a 1-DoF pitch-only problem. In this scenario, an increase in aerodynamic coupling as in case of $x_{ea}$ = 0.35$c$ results in a reduction in flutter speed and an increase in pitch amplitudes (Fig.~\ref{xea_thpfc4}(c)). While an increase in inertial coupling between the structural modes $x_{ea}$ = 0.15$c$ possibly results in a delayed flutter onset and decrease in pitch oscillations (Fig.~\ref{xea_thpfc4}(a)). This conjecture is also 
supported by the fact that the amplitudes of plunge (driven mode) in all three cases (Fig.~\ref{xea_thpfc4}(a)-(c)) are almost invariant.

\begin{figure}
\centering
\includegraphics[width=5.2in, height=2.4in]{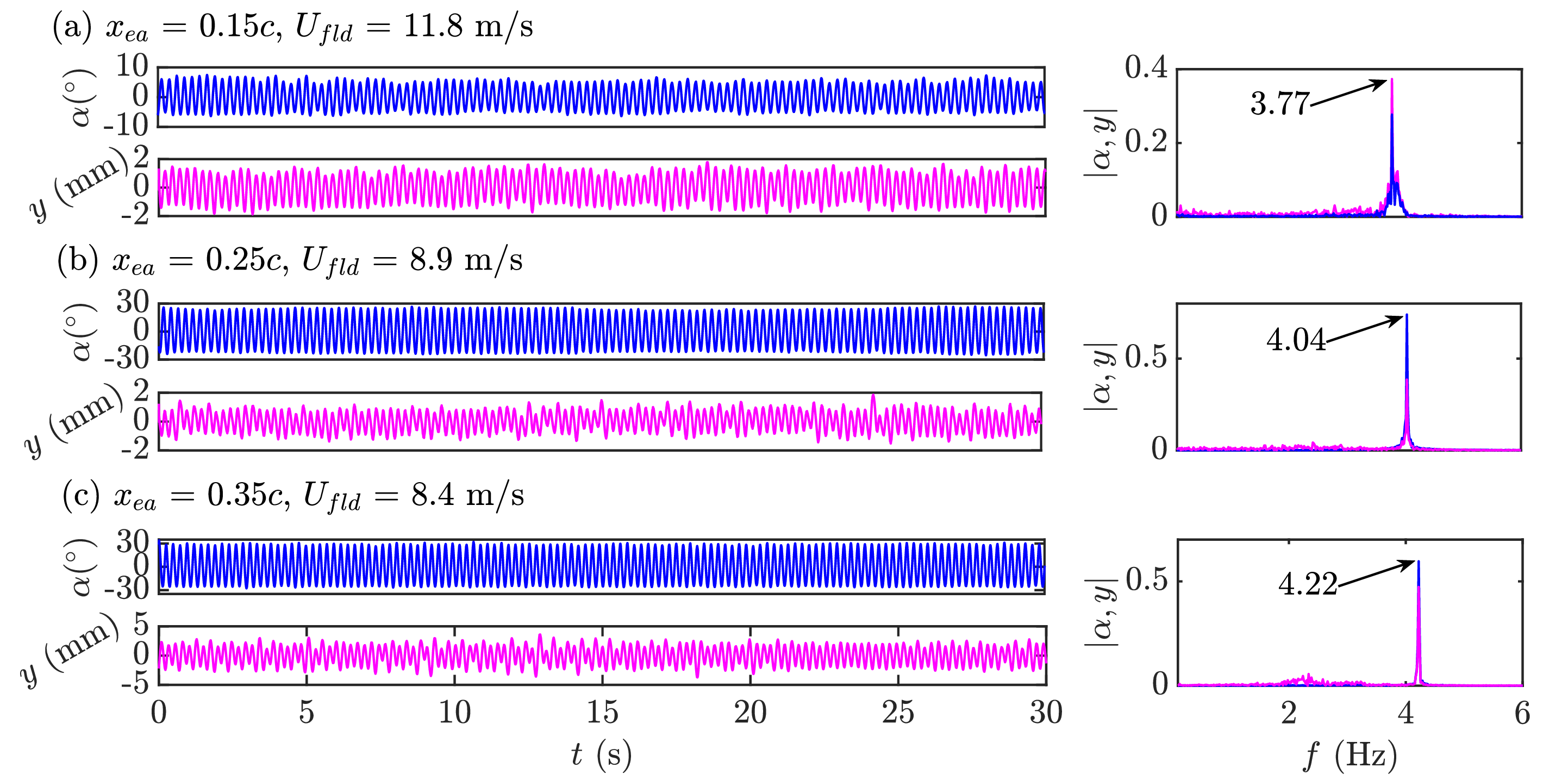}
\caption{Aeroelastic response dynamics at $U_{fld}$ at different elastic axis positions for $\bar{\omega}$ = 0.57.}
\label{xea_tfldc4}
\end{figure}
\begin{figure}
\centering
\includegraphics[width=5in, height=2.2in]{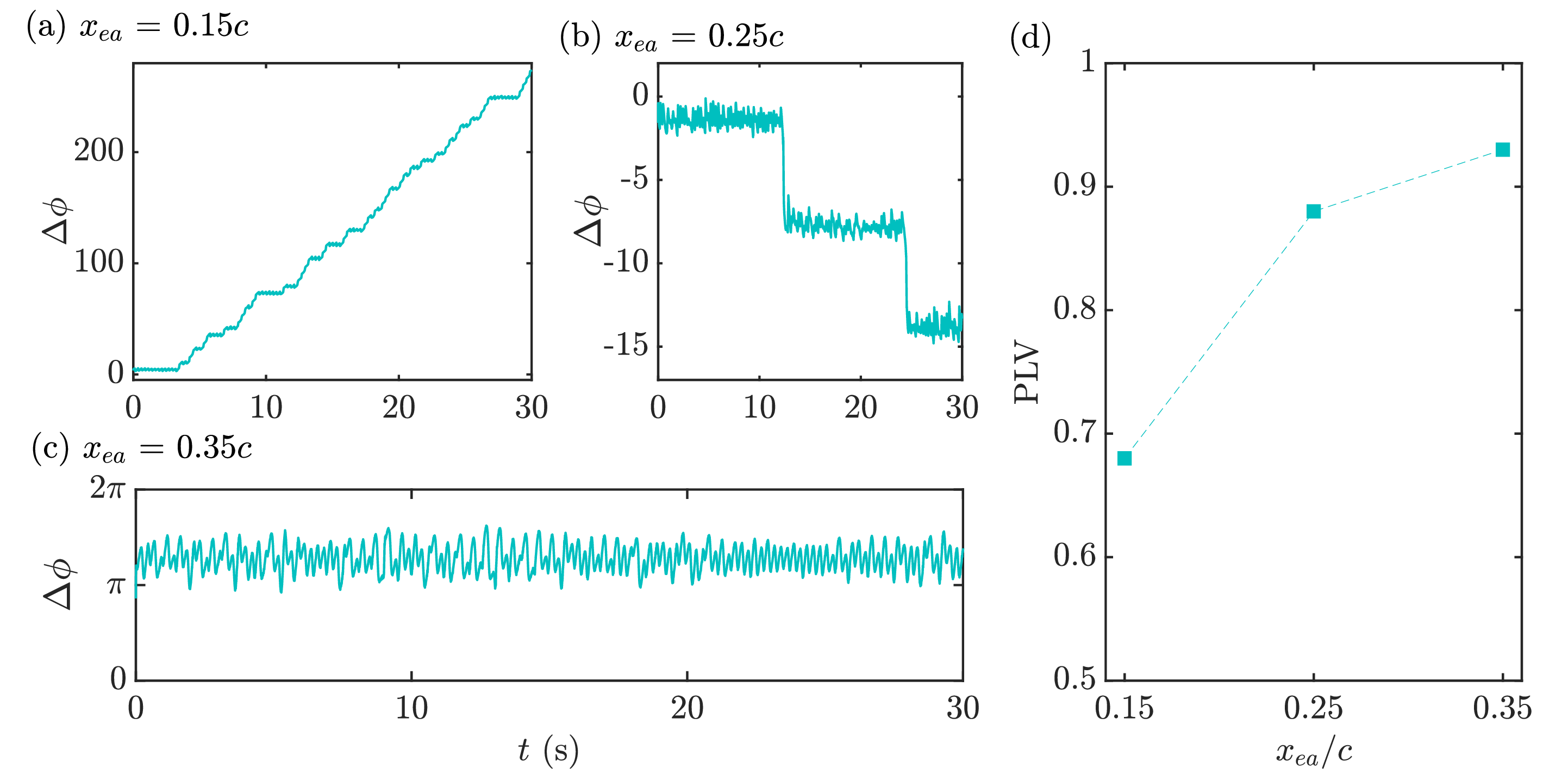}
\caption{The synchronization dynamics at $U_{fld}$ for various $x_{ea}$ positions for $\bar{\omega}$ = 0.57.}
\label{xea_phfldc4}
\end{figure}

Next, the pitch and plunge responses are observed at  $U_{fld}$ for three elastic axis positions (Fig.~\ref{xea_tfldc4}). Again, the pitch-dominant flutter frequencies are observed in all the cases. The pitch amplitudes for $x_{ea}$ = 0.15$c$ is approximately 7$^\circ$ (Fig.~\ref{xea_tfldc4}(a)) and close to 30$^\circ$ for the rest two cases (Fig.~\ref{xea_tfldc4}(b)-(c)). The synchronization dynamics for the three cases are shown in (Fig.~\ref{xea_phfldc4}). For $x_{ea}$ = 0.15$c$ the phase dynamics depicts an intermittent phase synchronization (Fig.~\ref{xea_phfldc4}(a)). Here the phase slips are frequent amidst short-lived phase synchronized plateaus. This synchronization is weaker as compared to that for $x_{ea}$ = 0.25$c$, where the dynamics is still characterized as an intermittent phase synchronization but the phase slips are far lesser with prominent phase synchronized regimes (Fig.~\ref{xea_phfldc4}(b)). For $x_{ea}$ = 0.35$c$, the phase difference is bounded, representing an even higher synchronization via a phase trapping (Fig.~\ref{xea_phfldc4}(c)). Hence the synchronization between the structural modes gradually increases as the elastic axis moves closer to the mass center. This trend can be quantitatively seen in the PLVs of the three cases (Fig.~\ref{xea_phfldc4}(c)). The PLV observed for $x_{ea}$ = 0.15$c$ is 0.68 - the least among three, which increases to 0.88 for $x_{ea}$ = 0.25$c$, and further to 0.93 for $x_{ea}$ = 0.35$c$.

To summarize, aeroelastic responses and their underlying synchronization are observed with three different $x_{ea}$ positions for $\bar{\omega}$ = 0.57. Shifting the elastic axis closer to the leading edge leads to a delayed flutter onset and weaker synchronization. On the contrary, shifting $x_{ea}$ to 0.35$c$ leads to an earlier flutter onset and a higher synchronization. The increase in pitch amplitudes as the elastic axis moves closer to the mass center can be explained by observing the static imbalance for these cases (see Table~\ref{eimbc4}). The static imbalance for $x_{ea}$ = 0.35$c$ is the least, denoting the higher degree of aerodynamic coupling which justifies the increase in the instability in this case \cite{goyaniuk2020pitch}. On the contrary, the static imbalance is highest for $x_{ea}$ = 0.15$c$, indicating much lower coupling through the aerodynamics, resulting in a rather stable dynamical response. This section focuses on the parametric analysis involving variations in $x_{ea}$ solely for $\bar{\omega}$ = 0.57. However, it's crucial to explore how the elastic axis location affects the aeroelastic behaviour for other $\bar{\omega}$ cases considered in Section~\ref{eow}. To that end, the analysis with different $x_{ea}$ is extended to cover these additional $\bar{\omega}$ instances, and the findings are succinctly outlined next.

\subsection{Change in elastic axis position at other frequency ratios}

\begin{figure}
\centering
\includegraphics[width=5in, height=2.2in]{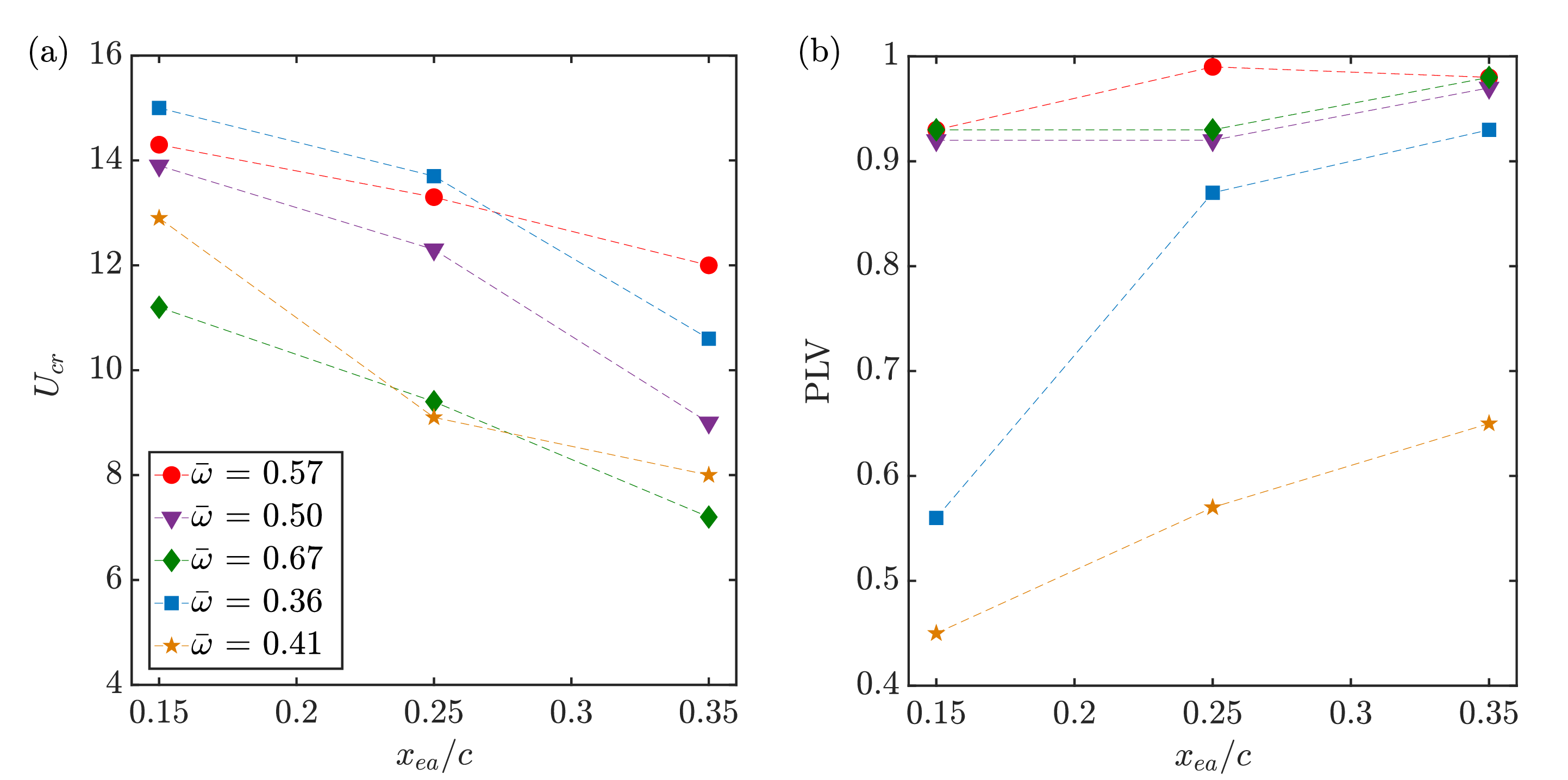}
\caption{Variation of (a) $U_{cr}$, and (b) PLV, with $x_{ea}$ for different $\bar{\omega}$.}
\label{ucr_xea_allc4}
\end{figure}

\begin{figure}
\centering
\includegraphics[width=5.2in, height=2.2in]{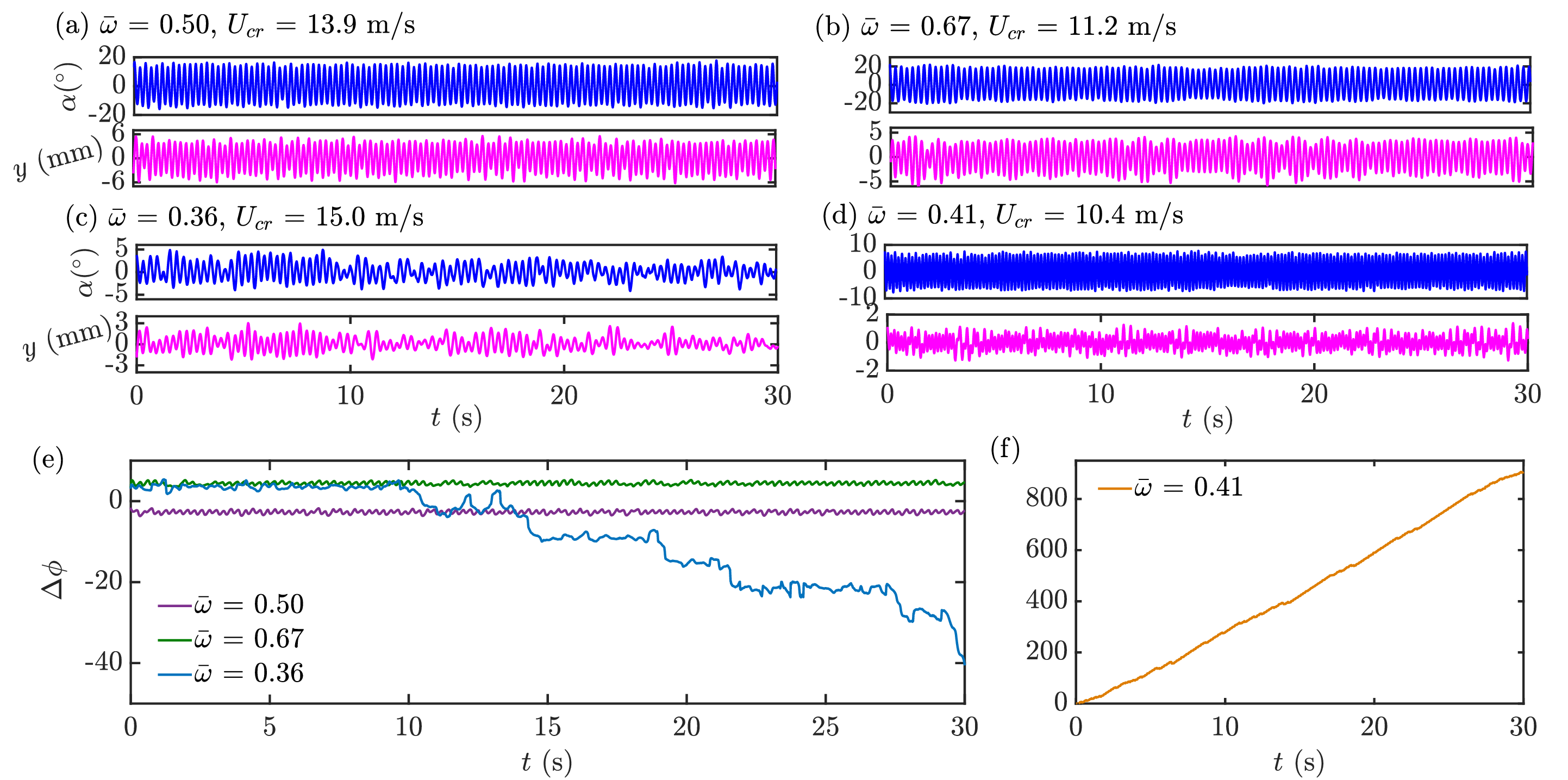}
\caption{Aeroelastic and RPV time responses for different $\bar{\omega}$ at $U_{cr}$ for $x_{ea}$ = 0.15$c$.}
\label{ucr_15c_allc4}
\end{figure}

\begin{figure}
\centering
\includegraphics[width=5.2in, height=3in]{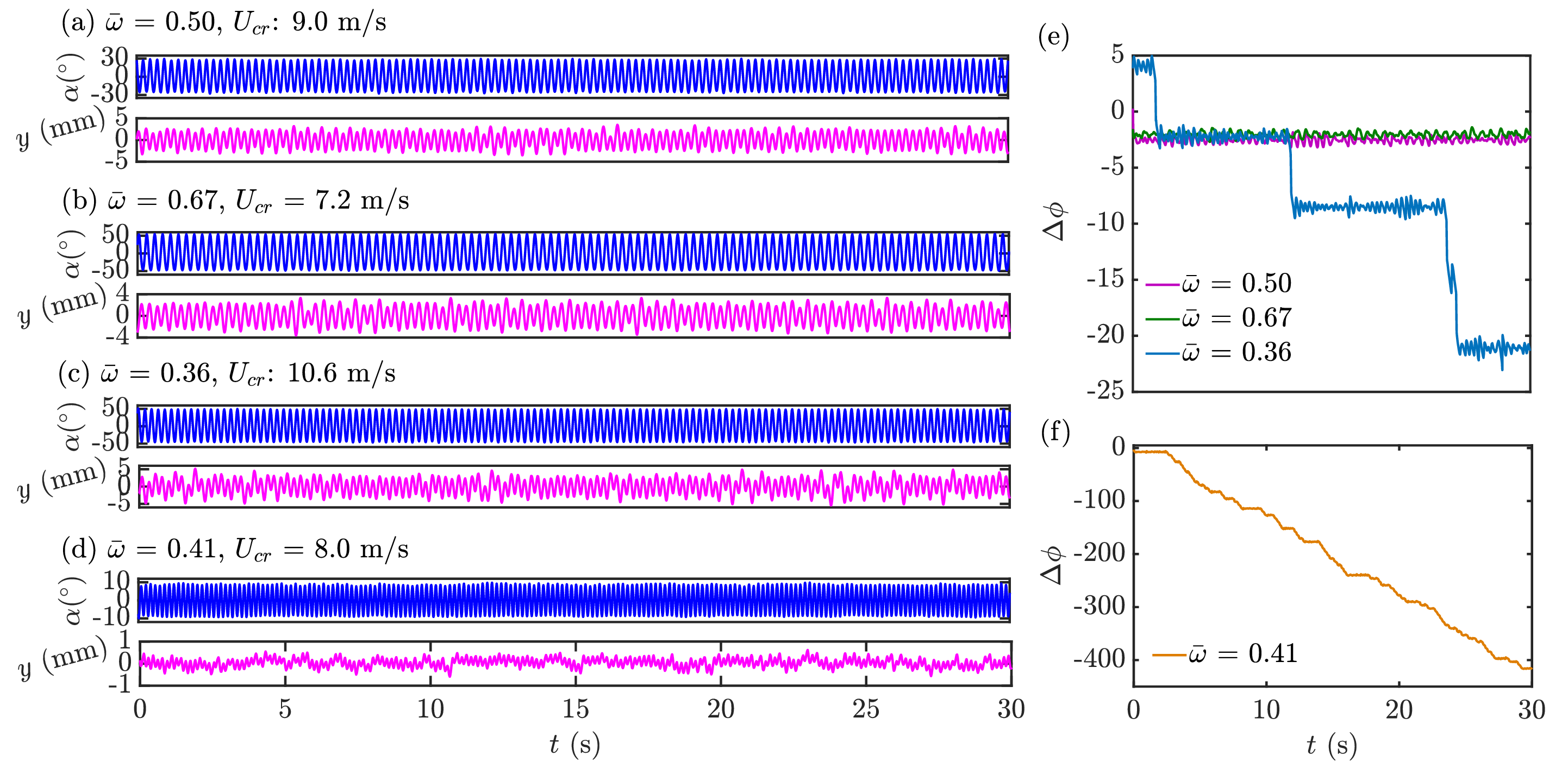}
\caption{Aeroelastic and RPV time responses for different $\bar{\omega}$ at $U_{cr}$ for $x_{ea}$ = 0.35$c$.}
\label{ucr_35c_allc4}
\end{figure}

The variation of $U_{cr}$ with the elastic axis location is shown for different $\bar{\omega}$ in Fig.~\ref{ucr_xea_allc4}(a). A similar trend is seen as observed for $\bar{\omega}$ = 0.57, which means that the Hopf point lowers as the elastic axis moves closer to the mass center. Corresponding synchronization strength in terms of PLV also depicts a consistent trend (Fig.~\ref{ucr_xea_allc4}(b)), where the synchronization is increased when the elastic axis location moves closer to the mass center. The aeroelastic responses and corresponding synchronization at $U_{cr}$ are shown for $x_{ea}$ = 0.15$c$ in Fig.~\ref{ucr_15c_allc4}, and for $x_{ea}$ = 0.35$c$ in Fig.~\ref{ucr_35c_allc4}. Akin to $\bar{\omega}$ = 0.57, for $\bar{\omega}$ = 0.67, and 0.50 as well, the PLVs are above 0.90 for all $x_{ea}$ values at the Hopf point - indicating strong synchronization even for elastic axis locations closer to the leading edge  (Fig.~\ref{ucr_xea_allc4}(b)). 

At $x_{ea}$ = 0.15$c$, a bounded RPV is observed for $\bar{\omega}$ = 0.50 and 0.67 depicting a phase trapping, with pitch and plunge amplitudes are approximately 20$^\circ$ and 5 mm (Fig.~\ref{ucr_15c_allc4}(a),(b) and (e)), respectively. At $x_{ea}$ = 0.35$c$, we again observe the pitch and plunge modes to be strongly synchronized for both $\bar{\omega}$ = 0.50 and 0.67 cases (Fig.~\ref{ucr_35c_allc4}(e)). Interestingly, the three $\bar{\omega}$ cases (0.57, 0.50, and, 0.67) are grouped in Category I (see Section~\ref{cat1c4}), as they show similar dynamics during the parametric variation analysis with $\bar{\omega}$ about a fixed elastic axis location (0.25$c$). Here, changing the elastic axis position for these three cases also yields similar synchronization behaviour, which is in line with the expectations. 

However, for the other two cases of $\bar{\omega}$ (0.36 and 0.41), which are presented in Category II (see Section~\ref{cat2c4}), we observe a lower synchronization as compared to the other three cases (Fig.~\ref{ucr_xea_allc4}(b)). For $\bar{\omega}$ = 0.36 at $x_{ea}$ = 0.15$c$, the aeroelastic responses are aperiodic with very small amplitudes and corresponding RPV time response shows an intermittent phase synchronization with a low PLV of 0.56 (Fig.~\ref{ucr_15c_allc4}(c) and (e)). For $\bar{\omega}$ = 0.41, an even lower synchronization strength is seen with PLV of 0.45, with pitch amplitude below 10$^\circ$ and plunge amplitude close to 1.5 mm (Fig.~\ref{ucr_15c_allc4}(d) and (f)). The $\Delta{\phi}$ monotonously increases with time for this case depicting an asynchronous behaviour. At $x_{ea}$ = 0.35$c$, the pitch amplitudes are significantly increased for $\bar{\omega}$ = 0.36 - which has also been seen for other $\bar{\omega}$ cases (see Fig.~\ref{ucr_35c_allc4}), except for the case with $\bar{\omega}$ = 0.41. High pitch stiffness in case of $\bar{\omega}$ = 0.41 restricts the pitch amplitudes to grow even at an $x_{ea}$ closer to the mass center. At this elastic axis location, synchronization is strong with PLV = 0.93 (Fig.~\ref{ucr_xea_allc4}(b)) for $\bar{\omega}$ = 0.36, with extremely rare occurrences of phase slips amidst the phase synchronized regime (Fig.~\ref{ucr_35c_allc4}(e)). For $\bar{\omega}$ = 0.41 case, the synchronization is relatively much weaker with a PLV of 0.65 (Fig.~\ref{ucr_xea_allc4}(b)) and characterized as an intermittent phase synchronization (Fig.~\ref{ucr_35c_allc4}(f)). Here we can see that the phase synchronization regimes are very frequently interrupted by the phase slips (Fig.~\ref{ucr_35c_allc4}(f)). 

\begin{figure}
\centering
\includegraphics[width=5in, height=2.2in]{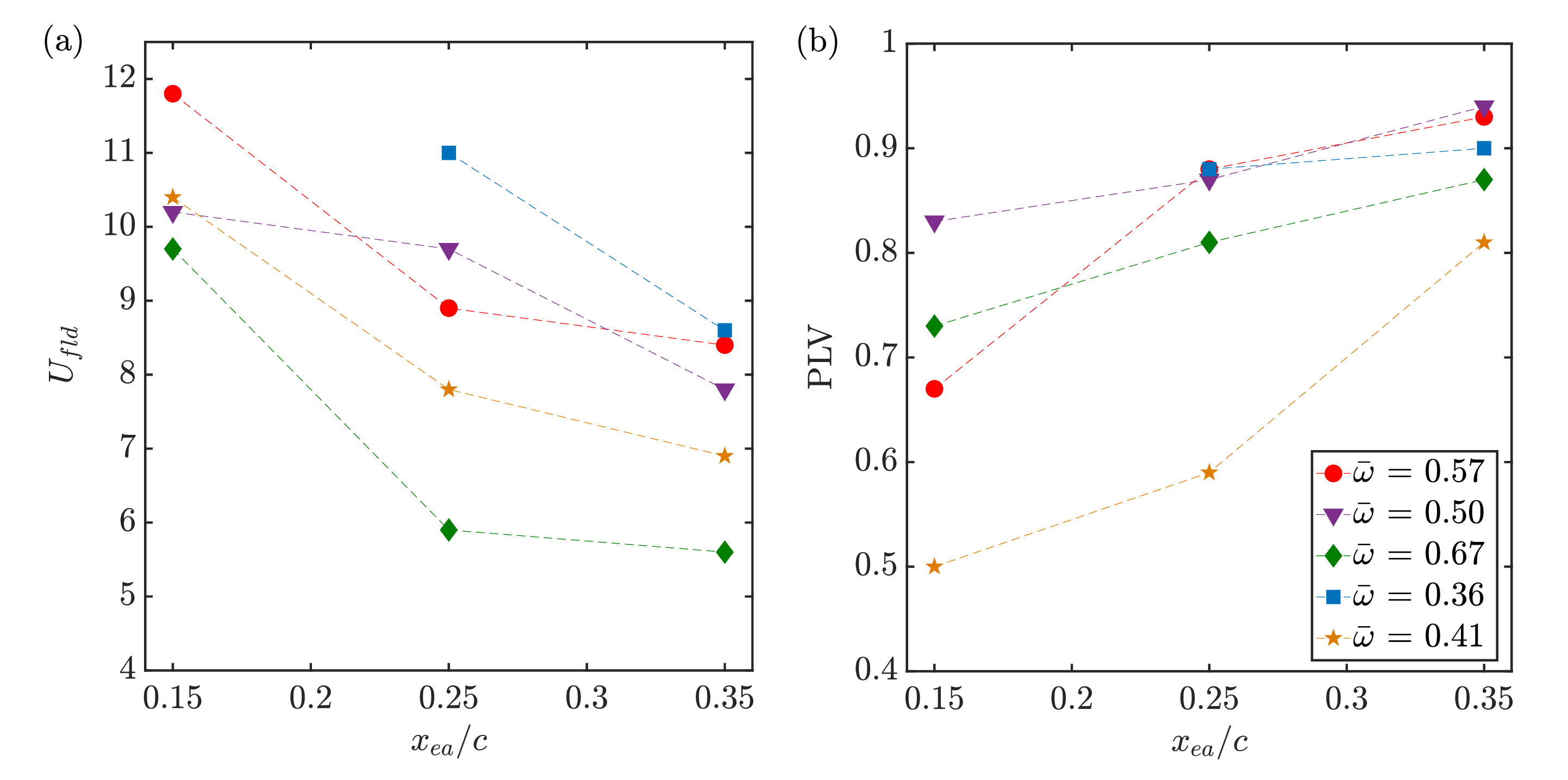}
\caption{Variation of (a) $U_{fld}$, and (b) PLV, with $x_{ea}$ for different $\bar{\omega}$.}
\label{ufld_xea_allc4}
\end{figure}

Next, we take a look at the variation of $U_{fld}$ with elastic axis positions along with the corresponding PLV variation (see Fig.~\ref{ufld_xea_allc4}). Here again, the fold point is observed to reduce as the elastic axis moves closer to the mass center - a similar trend that is observed at $U_{cr}$. Interestingly, for $\bar{\omega}$ = 0.36, no fold point is observed at $x_{ea}$ = 0.15$c$ (Fig.~\ref{ufld_xea_allc4}(a)), which means that this is the only case where subcriticality is not present. Note that at $x_{ea}$ = 0.25$c$ for $\bar{\omega}$ = 0.36, the onset of flutter is marked by a classical flutter (Case II in Section~\ref{cat2c4}). Here, the stall flutter is triggered by the increased amplitude at higher speeds due to a 2:1 internal resonance. Changing the $x_{ea}$ from 0.25$c$ to 0.15$c$ results in a reduced aerodynamic coupling, leading to a decrease in flutter amplitude, as demonstrated across all cases. Consequently, the system does not undergo stall flutter even at higher speeds for parameters $x_{ea}$ = 0.15$c$ and $\bar{\omega}$ = 0.36. This is possibly the reason why the bifurcation, in this case, is supercritical Hopf instead of a Subcritical Hopf, which is commonly observed bifurcation in classical flutter studies \cite{lee1999nonlinear,poirel2003random}.

\begin{figure}
\centering
\includegraphics[width=5.2in, height=2.4in]{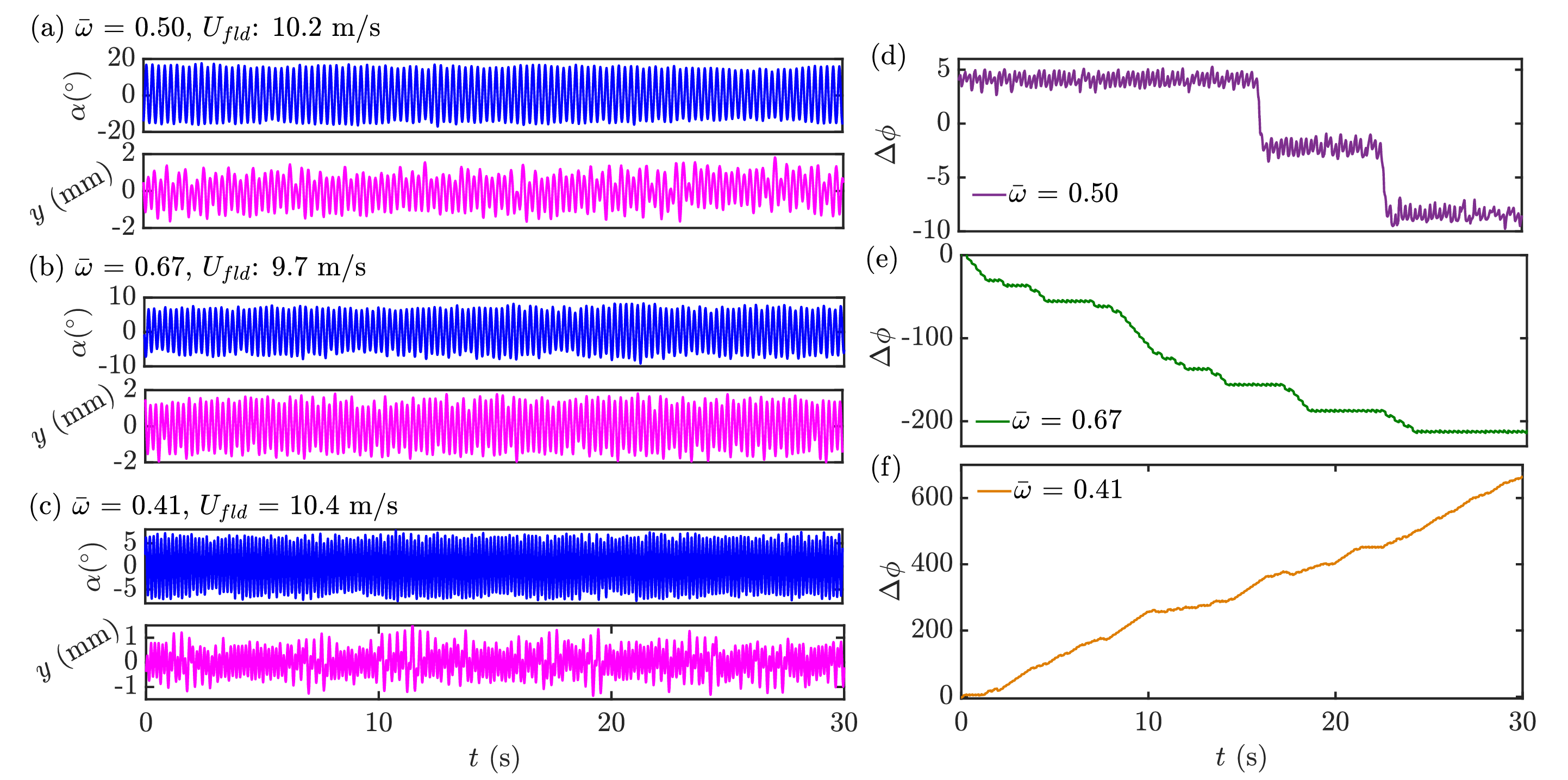}
\caption{Aeroelastic and RPV time responses for different $\bar{\omega}$ at $U_{fld}$ for $x_{ea}$ = 0.15$c$.}
\label{ufld_15c_allc4}
\end{figure}

The PLVs are seen to have a gradual increase as $x_{ea}$ increases (Fig.~\ref{ufld_xea_allc4}(b)). The least amount of synchronization is again found to be for the case $\bar{\omega}$ = 0.41. Figure~\ref{ufld_15c_allc4} shows the pitch-plunge responses at $U_{fld}$ and $x_{ea}$ = 0.15$c$ for different $\bar{\omega}$ along with the corresponding synchronization behaviour. The highest synchronization is seen to be for $\bar{\omega}$ = 0.50 with a PLV of 0.88 (Fig.~\ref{ufld_xea_allc4}(b)) where RPV variation shows an intermittent phase synchronization with prolonged regimes of phase synchronization amidst a few phase slips (Fig.~\ref{ufld_15c_allc4}(d)). The pitch amplitude is close to 20$^\circ$ in this case. For $\bar{\omega}$ = 0.67, PLV is reduced to 0.73 (Fig.~\ref{ufld_xea_allc4}(b)) with an intermittent phase synchronization, having multiple phase slips (Fig.~\ref{ufld_15c_allc4}(e)). The pitch amplitudes are below 10$^\circ$ in this case. For $\bar{\omega}$ = 0.41, the $\Delta{\phi}$ variation shows asynchronous behaviour (Fig.~\ref{ufld_15c_allc4}(f)) with a PLV of 0.51 (Fig.~\ref{ufld_xea_allc4}(b)). 

\begin{figure}
\centering
\includegraphics[width=5.2in, height=2.4in]{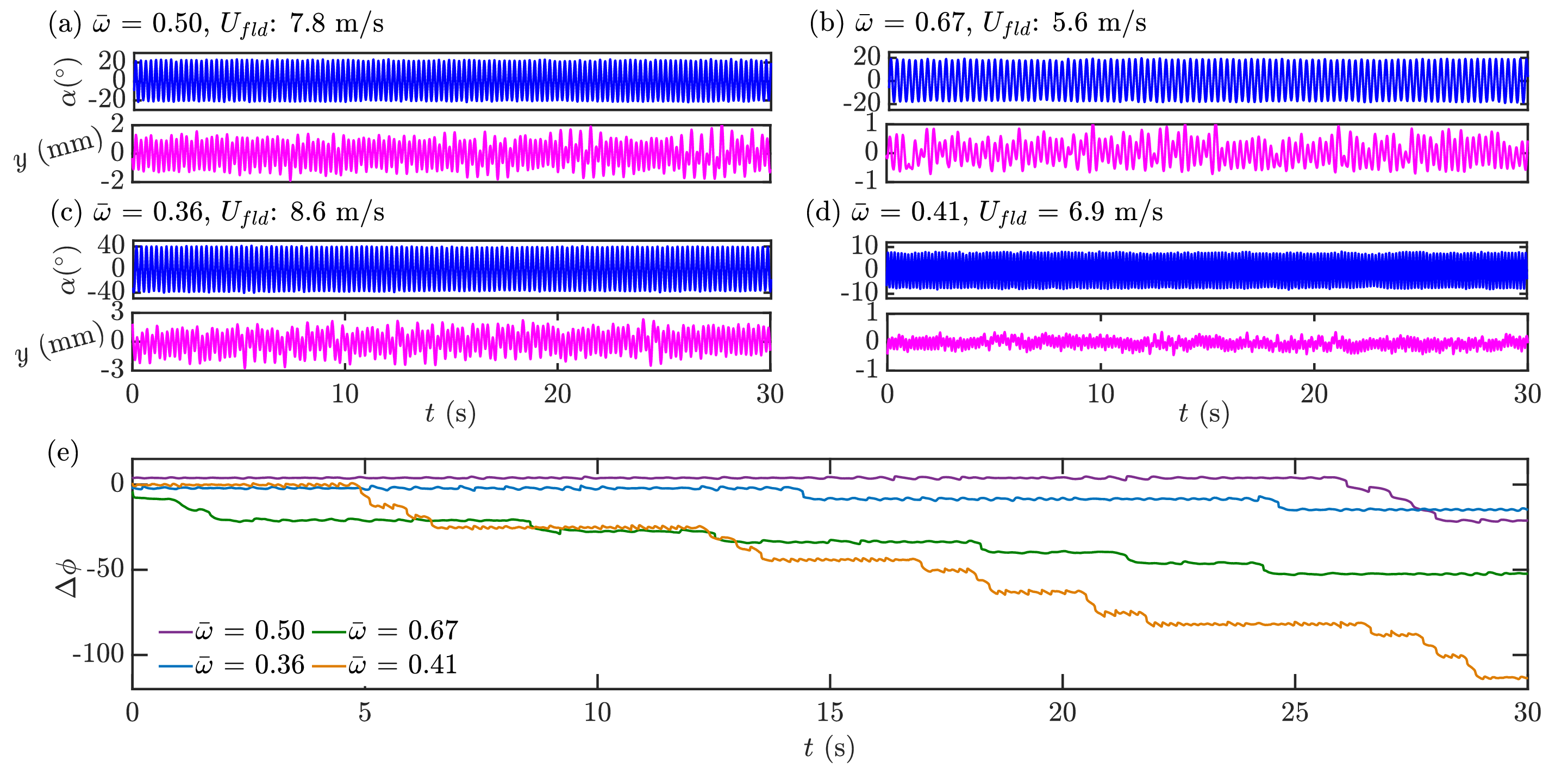}
\caption{Aeroelastic and RPV time responses for different $\bar{\omega}$ at $U_{fld}$ for $x_{ea}$ = 0.35$c$.}
\label{ufld_35c_allc4}
\end{figure}

Increasing the $x_{ea}$ to 0.35$c$ at the fold point, we observe a significant increase in pitch amplitudes due to increased aerodynamic coupling (Fig.~\ref{ufld_35c_allc4}(a)-(d)). The synchronization is much higher for all $\bar{\omega}$ values as compared to that at $x_{ea}$ to 0.15$c$ (Fig.~\ref{ufld_35c_allc4}(e)). Here, an intermittent phase synchronization is observed for all $\bar{\omega}$ cases with $\bar{\omega}$ = 0.50 having the highest PLV of 0.94 and $\bar{\omega}$ = 0.41 having the least PLV of 0.81. Nevertheless, the pitch and plunge modes are strongly synchronized which has been seen for all cases at $x_{ea}$ = 0.35$c$. A summary of the underlying synchronization characteristics at each elastic axis location under consideration is tabulated for various cases in Table~\ref{ea_sum}.

\begin{table}
\caption{\label{ea_sum}Summary of the synchronization behaviour observed for different $\bar{\omega}$ at various elastic axis locations (pt - phase trapping, as - asynchrony, ips - intermittent phase synchronization).}
\centering
\begin{tabular}{c| c c c| c c c}
\hline
\hline
 & \multicolumn{3}{c|}{synchronization at $U_{cr}$} & \multicolumn{3}{c}{synchronization at $U_{fld}$} \\
$\bar{\omega}$ $\downarrow$ / $x_{ea}$ $\rightarrow$& 0.15$c$ & 0.25$c$ & 0.35$c$ & 0.15$c$ & 0.25$c$ & 0.35$c$ \\
\hline
0.57 & pt & pt & pt & ips & ips & ips \\
0.50 & pt & pt & pt & ips & ips & ips\\
0.67 & pt & ips & pt & ips & ips & ips\\
0.36 & ips & pt & ips & - & ips & ips\\
0.41 & as & ips & ips & as & ips & ips\\
\hline
\hline
\end{tabular}
\end{table}

\section{\label{pv_soc4} Effect of parametric variation under stochastic conditions}
The findings presented thus far in this paper are obtained under sterile flow. However, in field scenarios, the flow is not sterile but rather influenced by noise. For aeroelastic systems operating at low speeds, it is essential to understand the behaviour of the aeroelastic system under fluctuating inflow hand in hand with that under sterile flow. It is well understood that noise has the potential to influence the response characteristics of dynamic systems \cite{landa2000changes,venkatramani2016precursors,venkatramani2017physical}. The presence of noise-induced random oscillations has been shown previously \cite{tripathi2022experimental}, where the pitch and plunge modes are weakly synchronized by the input noise. Noise strongly influences the synchronization behaviour even at the onset of stall flutter which is marked by multiple phase slips. To this end, we briefly extend the parametric analysis undertaken in this paper under noisy inflow conditions. Note that the input noise is purely attributed to the fan noise generated in the blowing conditions. Consequently, it's important to emphasize that this study should not be misconstrued as an exploration of parametric uncertainties. Rather, experiments are repeated for some salient cases from Sections~\ref{eow} and \ref{eoea} under fluctuating inflow conditions which are presented in this section.

\subsection{\label{ea_soc4} Elastic axis variation}
The aeroelastic response analysis under stochastic inflow conditions is first conducted for different $x_{ea}$ at $\bar{\omega}$ = 0.57 (Case I in Section~\ref{cat1c4}). Tripathi \textit{et al.}\cite{tripathi2022experimental} demonstrates the transitions in aeroelastic response dynamics under stochastic inflow for $x_{ea}$ = 0.25$c$. Here we present the aeroelastic response dynamics for two different elastic axis locations, $x_{ea}$ = 0.15$c$ and 0.35$c$ under blowing conditions. 

\begin{figure}
\begin{center} \includegraphics[width=5in,height=2.2in]{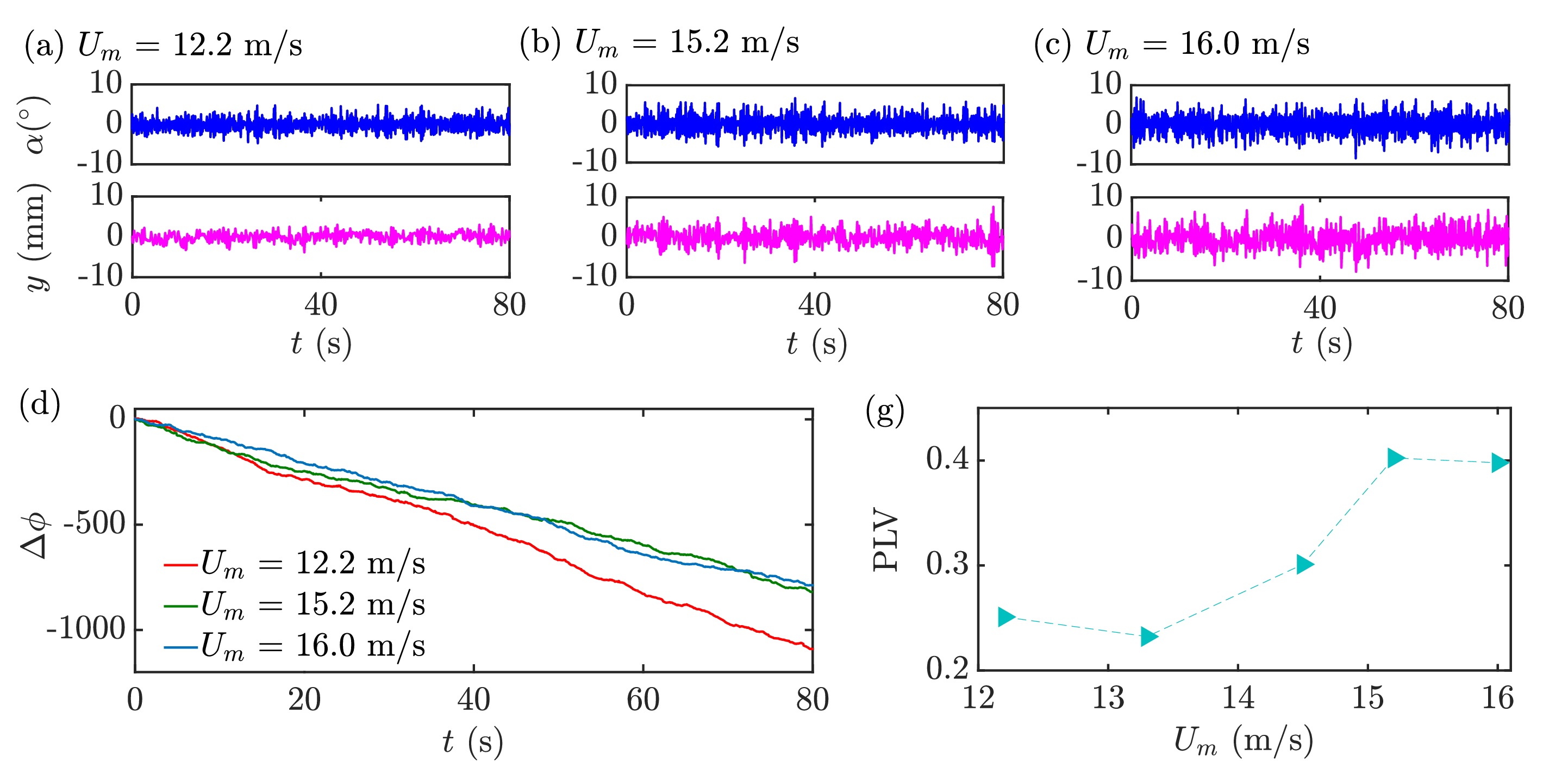}
 \caption{\label{hist_sto15c} Aeroelastic response and relative phase time histories in blowing conditions for $x_{ea}$ = 0.15$c$.}
\end{center}
\end{figure}

For $x_{ea}$ = 0.15$c$, the small amplitude noise induced random oscillations (NIRO) are observed at $U_m$ = 12.2 m/s (see Fig.~\ref{hist_sto15c}(a)), below the corresponding $U_{cr}$ = 14.3 m/s observed under deterministic flows for same parameters (see Fig.~\ref{ucr_xea_allc4}). The NIROs are further observed up to 16.0 m/s (see Fig.~\ref{hist_sto15c}(b)-(c)). However, the dynamics does not change into well-developed LCOs. This dynamics is expected as we observe very low amplitude oscillation under deterministic flows for the same parameters (see Fig.~\ref{xea_thpfc4}(a)). The relative phase between pitch and plunge modes has a monotonous decay in the range $U_m$ = 12.2 m/s - 16.0 m/s (see Fig.~\ref{hist_sto15c}(d)), depicting an asynchrony between the two modes. It can be seen that the PLV is very low within the considered range of $U_m$. At $U_m$ = 12.2 m/s, the PLV is 0.25 which increases up to 0.40 at 16.0 m/s (see Fig.~\ref{hist_sto15c}(d)). Overall the synchronization under stochastic inflow is very weak for $x_{ea}$ = 0.15$c$ due to the combined effect of two key factors, the first being the input noise, and the second is high static imbalance (0.048 kg-m). Noise, even a weak one, might cause significant dynamical transitions \cite{landa2000changes}. However, complete visualization of the stochastic flow field can only provide deeper insight into how the aeroelastic responses and synchronization are affected in the current scenario, which is outside the purview of the current experimental study.

\begin{figure}
\begin{center} \includegraphics[width=5.2in,height=2.2in]{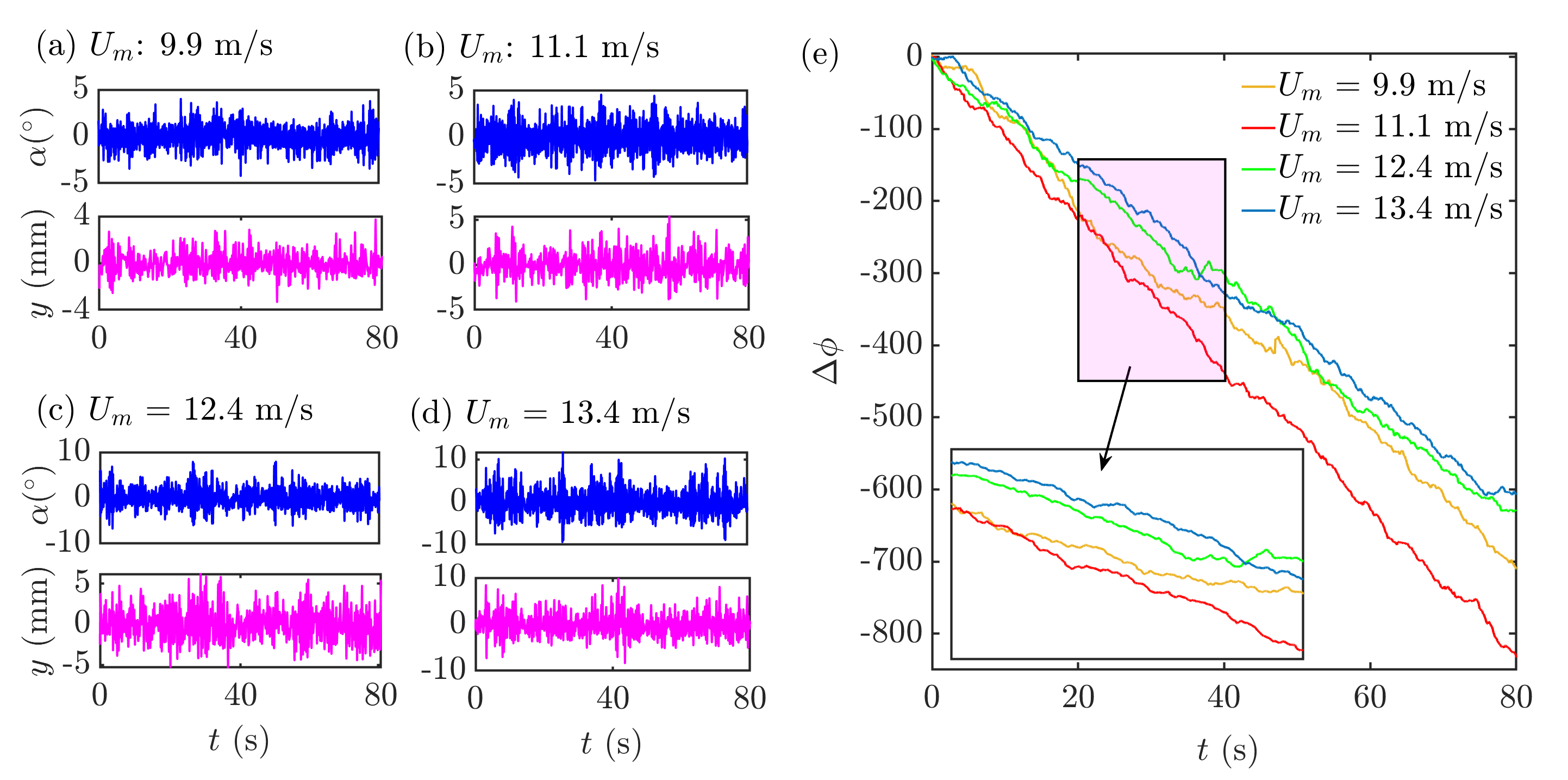}
\includegraphics[width=5.2in,height=2.2in]{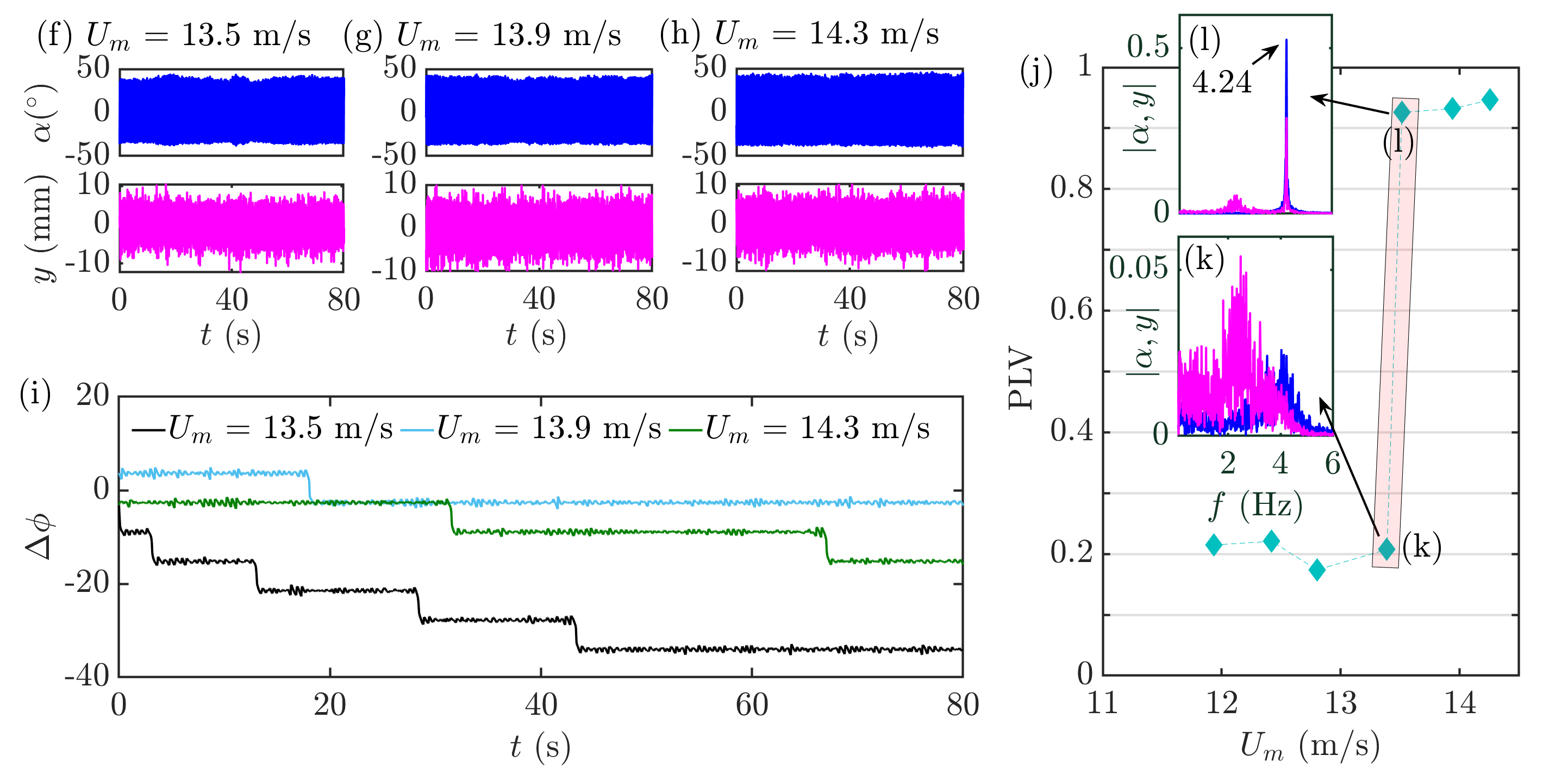}
 \caption{\label{histpre_blow35c} Time history and phase in blowing conditions for $x_{ea}$ = 0.35$c$.}
\end{center}
\end{figure}

Next, the stochastically influenced aeroelastic behavior is analyzed for $x_{ea}$ = 0.35$c$. In this case, the first instance of instability is observed at $U_m$ = 9.9 m/s (Fig.~\ref{histpre_blow35c}(a)). Note that for this case under the deterministic inflow, the  Hopf point is observed at 12.0 m/s. This instability is again characterized as small amplitude NIRO and persists up to $U_m$ = 13.4 m/s (Fig.~\ref{histpre_blow35c}(a)-(d)). The pitch amplitudes reach from $\approx$5$^\circ$ at $U_m$ = 9.9 m/s to $\approx$10$^\circ$ at $U_m$ = 13.4 m/s while the plunge amplitudes reach from $\approx$4 mm to $\approx$10 mm in this range. The corresponding synchronization dynamics are shown in Fig.~\ref{histpre_blow35c}(e), which depict an asynchrony between pitch and plunge. This synchronization behaviour is quite similar to the one observed for $x_{ea}$ = 0.15$c$ and is typical of a small amplitude NIRO as explained in Tripathi \textit{et al.}\cite{tripathi2022experimental}.

At $U_m$ = 13.5 m/s, the small amplitude NIROs transition to well-developed LCOs (see Fig.~\ref{histpre_blow35c}(f)). The pitch amplitudes suddenly grow up to $\approx$40$^\circ$ while the plunge LCOs are approximately 10 mm in amplitude. These high amplitude LCOs are observed further up to $U_m$ = 14.3 m/s, beyond which the analysis could not be carried out due to such violent pitch amplitudes. The synchronization analysis shows an intermittent phase synchronization for the regime $U_m$ = 13.5 - 14.3 m/s (Fig.~\ref{histpre_blow35c}(i)). At $U_m$ = 13.5 m/s, the long plateau of phase synchronization regimes are observed with a limited number of \textit{noise-induced} phase slips. A description of these noise-induced phase slips is extensively provided in Tripathi \textit{et al.}\cite{tripathi2022experimental}.

Upon increasing the $U_m$ up to 14.3 m/s, the noise-induced phase slips are gradually reduced depicting a further increase in synchronization. The synchronization behaviour can be summarized through PLV observations throughout the considered flow speed range. For $U_m$ = 9.9 m/s - 13.4 m/s, the PLV is very small (ranging between 0.17 - 0.22) depicting an asynchronous behaviour. This behaviour is consistent with the pre-stall flutter behaviour explained in Tripathi \textit{et al.}\cite{tripathi2022experimental}. However, a sudden jump in PLV is observed as the small amplitude NIROs transition into fully developed stall flutter oscillations at $U_m$ = 13.5 m/s. The PLV at $U_m$ = 13.5 m/s is 0.93 indicating a strong synchronization between the two modes, which further increases up to 0.95 at $U_m$ = 14.3 m/s. The sudden jump in synchronization between $U_m$ = 13.4 m/s with PLV 0.21 and $U_m$ = 13.5 m/s with PLV 0.93 corresponds to the dynamic transition from small-amplitude NIRO to high-amplitude LCOs. This can be observed from the FFTs given for pitch and plunge modes at $U_m$ = 13.4 m/s (Fig.~\ref{histpre_blow35c}(k)) and $U_m$ = 13.5 m/s (Fig.~\ref{histpre_blow35c}(l)). At $U_m$ = 13.4 m/s, the frequency responses of the two modes show a broadband, which is typical of a random noisy aeroelastic response shown in Fig.~\ref{histpre_blow35c}(d). While At $U_m$ = 13.5 m/s, a single dominant peak appears at 4.24 Hz (Fig.~\ref{histpre_blow35c}(l)) which is close to $f_{\alpha}$. This confirms the presence of stall flutter, which is evident from violent pitch oscillations at this $U_m$ (see Fig.~\ref{histpre_blow35c}(f)). During the stall flutter, the plunge responses are visibly more noisy as compared to the pitch responses, as shown in Fig.~\ref{histpre_blow35c}(l) and this possibly causes the phase slips during the stall flutter (see Fig.~\ref{histpre_blow35c}(i)).

\begin{figure}
\begin{center} \includegraphics[width=5in,height=2.2in]{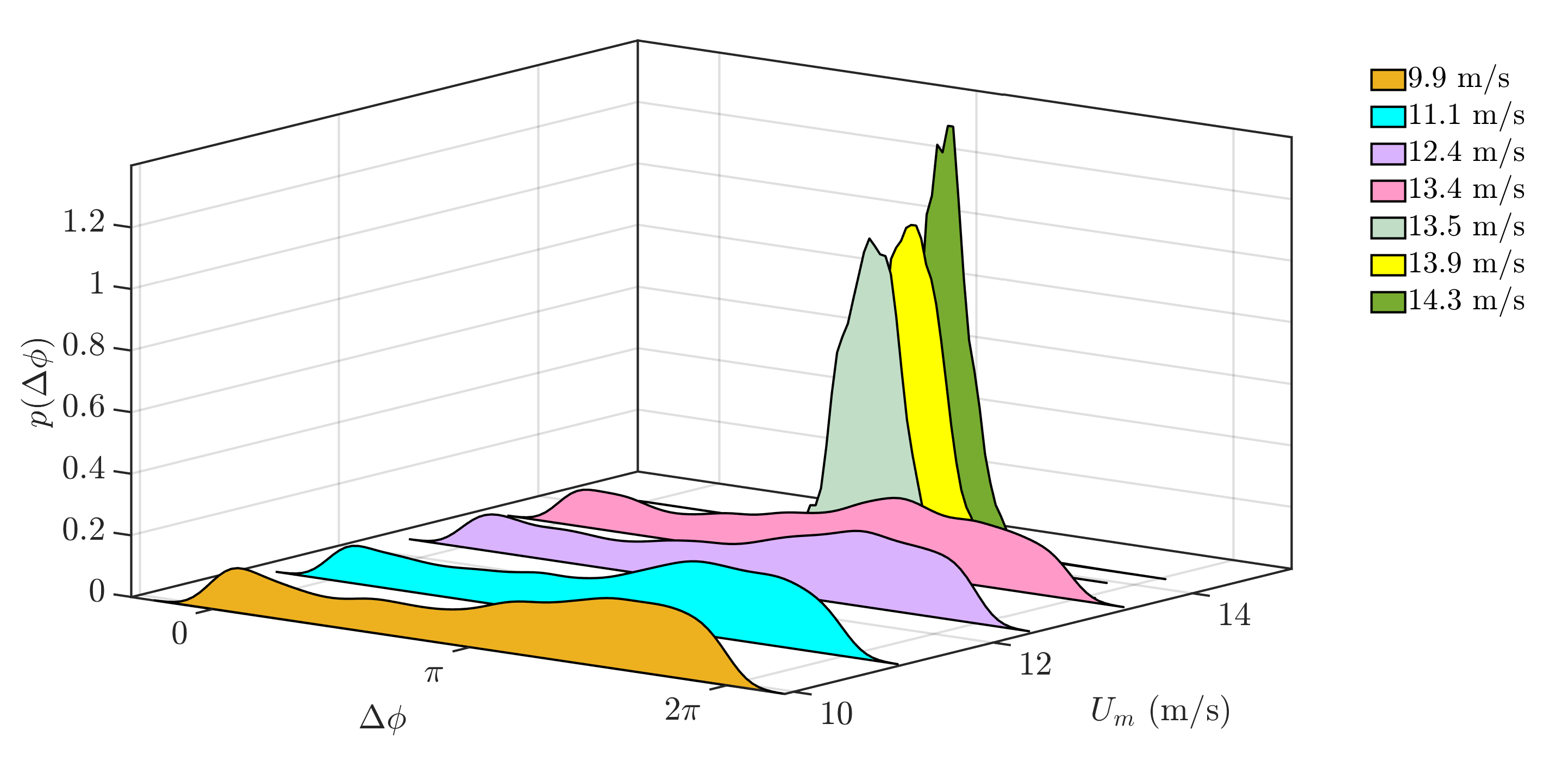}
 \caption{\label{waterfall35} Waterfall plot showing the cyclic distribution of $\Delta{\phi}$ via its probability density function for $x_{ea}$ = 0.35$c$.}
\end{center}
\end{figure}

A waterfall plot is also shown (Fig.~\ref{waterfall35}) to observe the changes in relative phase distribution at $x_{ea}$ = 0.35$c$. It can be seen that $\Delta{\phi}$ is uniformly distributed between 0 to 2$\pi$ resulting in a flat probability distribution ($p(\Delta{\phi})$) during small-amplitude NIRO responses ($U_m$ = 9.9 m/s - 13.4 m/s). A sudden change in the $p(\Delta{\phi})$ topology is observed at $U_m$ = 13.5 m/s when the distribution becomes highly concentrated about $\pi$. In Tripathi \textit{et al.}\cite{tripathi2022experimental}, a similar behaviour was observed during the transition from NIRO to random LCOs, except the $p(\Delta{\phi})$ was more concentrated around 0 and 2$\pi$ during NIRO in that case.

\subsection{\label{fr_soc4} Frequency ratio variation}
The effect of input noise from the flow is so far observed for different elastic axis positions keeping $\bar{\omega}$ the same (\textit{i.e.}, 0.57). 
The wind tunnel experiments under stochastic inflow are further conducted for two additional $\bar{\omega}$ cases: 0.36 and 0.41, keeping the elastic axis position constant at the quarter-chord ($x_{ea}$ = 0.25$c$). 

\begin{figure}
\begin{center} \includegraphics[width=5in,height=2.2in]{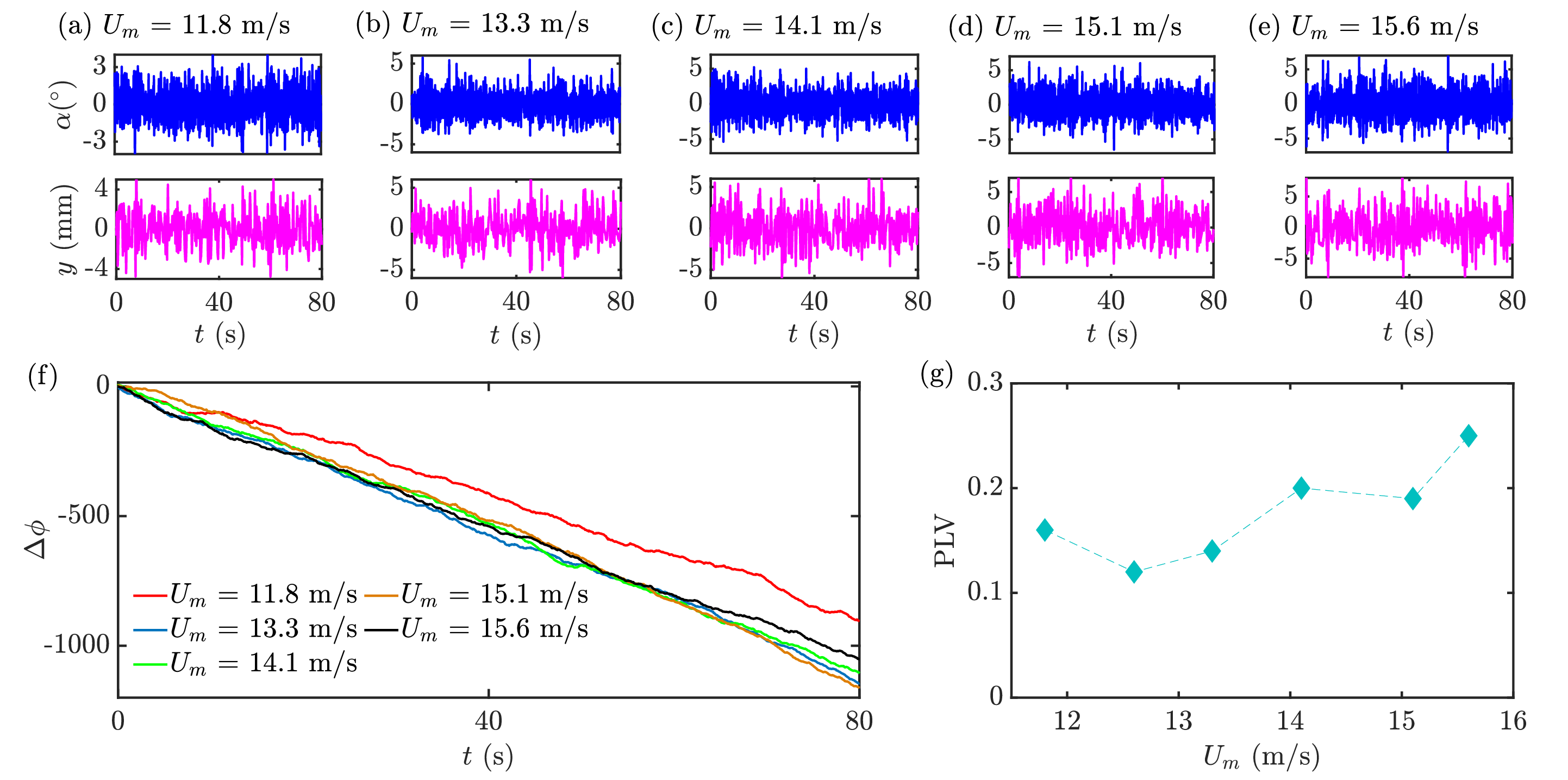}
 \caption{\label{y4_blow} Time histories of pitch-plunge responses and their relative phase values in blowing conditions for $\bar{\omega}$ = 0.36 at $x_{ea}$ = 0.25$c$.}
\end{center}
\end{figure}

For $\bar{\omega}$ = 0.36 ($U_{cr}$ = 12.3 m/s), the first instance of small amplitude NIROs is observed at $U_m$ = 11.8 m/s (see Fig.~\ref{y4_blow}(a)). However, increasing the $U_m$ further up to 15.6 m/s we only observe a marginal increase in amplitudes of pitch and plunge modes (Fig.~\ref{y4_blow}(b)-(e)). Corresponding synchronization dynamics can be seen in Fig.~\ref{y4_blow}(f), showing monotonous decay in $\Delta{\phi}$. This indicates asynchronous behaviour throughout the flow speed regimes considered for the analysis. The PLVs range between 0.12 - 0.25 for $U_m$ = 11.8 - 15.6 m/s, indicating an extremely weak synchronization under noise. 

\begin{figure}
\begin{center} \includegraphics[width=5in,height=2.2in]{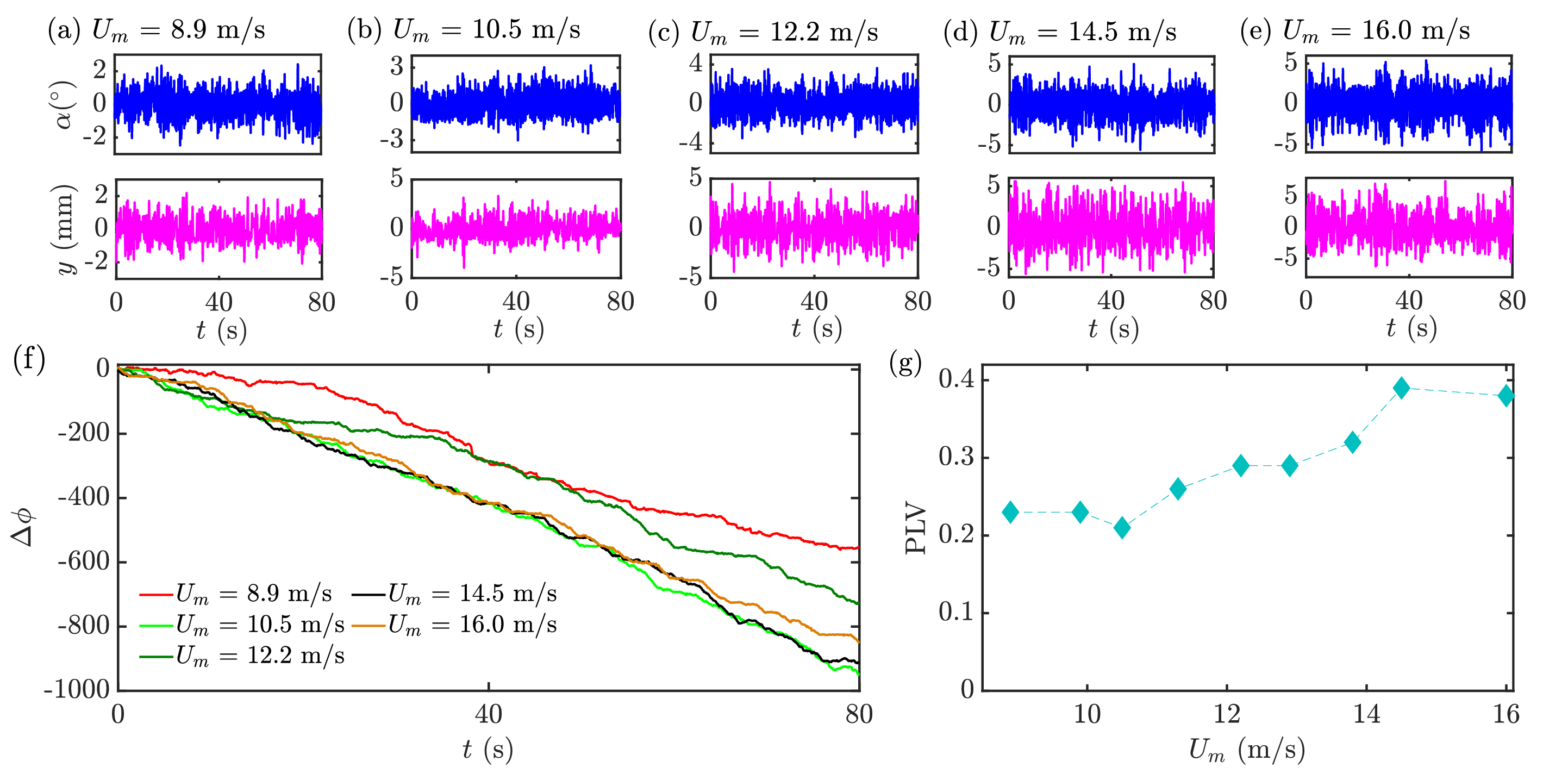}
 \caption{\label{2y_blow} Time histories of pitch-plunge responses and their relative phase values in blowing conditions for $\bar{\omega}$ = 0.41 at $x_{ea}$ = 0.25$c$.}
\end{center}
\end{figure}

Even for $\bar{\omega}$ = 0.41, similar behaviour is observed (see Fig.~\ref{2y_blow}). The small amplitude NIROs start to appear at $U_m$ = 8.9 m/s (Fig.~\ref{2y_blow}(a)), which is slightly below the corresponding $U_{cr}$ (9.1 m/s). Increasing the $U_m$ from 8.9 m/s to 16.0 m/s, there is a very small change in pitch and plunge amplitudes (Fig.~\ref{2y_blow}(b)-(e)). The maximum pitch and plunge amplitudes in this case are close to 5$^\circ$ and 5 mm, respectively. Phase drifts are observed in synchronization dynamics, which are indicative of asynchrony (Fig.~\ref{2y_blow}(f)). The PLVs in this case are slightly higher than those observed for $\bar{\omega}$ = 0.36 case (Fig.~\ref{y4_blow}), ranging between 0.21 - 0.38 for $U_m$ between 8.9 - 16.0 m/s.

For both $\bar{\omega}$ = 0.36 and 0.41 cases, small amplitude NIROs are observed within the considered flow-speed range. The transition from small amplitude NIROs to well-developed LCOs is not observed here which is seen for $\bar{\omega}$ = 0.57 (for $x_{ea}$ = 0.25$c$ and 0.35$c$). However, this is not very surprising as in both cases ($\bar{\omega}$ = 0.36 and 0.41) under deterministic flows, the LCO amplitude at flutter onset is not too high (approximately 10$^\circ$ for $\bar{\omega}$ = 0.36 and 6$^\circ$ for $\bar{\omega}$ = 0.41 in pitch) unlike the case $\bar{\omega}$ = 0.57 where pitch LCO amplitudes at flutter onset are much higher (close to 20$^\circ$ for $x_{ea}$ = 0.25$c$ and 40$^\circ$ for $x_{ea}$ = 0.35$c$). As a result, for both cases ($\bar{\omega}$ = 0.36 and 0.41) we observe asynchronous behaviour under stochastic inflow. In both cases, phase synchronization is observed when experiments are performed under deterministic conditions. Note that in $\bar{\omega}$ = 0.57 cases also, the noise jeopardizes the synchronization between the pitch and plunge, however, the loss in synchronization is considerably more severe in the present scenarios, (\textit{i.e.}, $\bar{\omega}$ = 0.36 and 0.41). The effect of noise on synchronization in a dynamical system is multifaceted, capable of both enhancing and disrupting \cite{boccaletti2002synchronization}. The synchronization in dynamical systems depends on the interplay between multiple factors \cite{pikovsky1997coherence,hu2000phase,zhou2002noisein,boccaletti2002synchronization,antonio2015nonlinearity} such as noise characteristics, nonlinearity, type of coupling, etc. Hence, it will be premature to comment on how the synchronization is affected by the stochastic inflow in the present scenario under different structural parameters without a proper characterization of input-noise and corresponding stochastic flow field. The salient cases presented in this section are summarized in Table~\ref{sto_sum}.
 
\begin{table}
\caption{\label{sto_sum}Summary of the aeroelastic behaviour observed under stochastic inflow (as - asynchrony, ips - intermittent phase synchronization, NA - not applicable).}
\centering
\begin{tabular}{c| c c c | c c}
\hline
\hline
 & \multicolumn{3}{c|}{$\bar{\omega}$ = 0.57} & $\bar{\omega}$ = 0.36 & $\bar{\omega}$ = 0.41 \\
$x_{ea}$ $\rightarrow$ & 0.15$c$ & 0.25$c$ & 0.35$c$ & 0.25$c$ & 0.25$c$ \\
\hline
NIRO onset (m/s) & 12.2 & 12.8 & 9.9& 11.8 & 8.9 \\
Synchronization & as & as & as & as & as \\
\hline
Stall flutter onset (m/s) & NA & 15.0 & 13.5 & NA & NA \\
Synchronization & - & as & ips & - & - \\
\hline
\hline
\end{tabular}
\end{table}

\section{Conclusion}

The onset and characteristics of flutter phenomena are significantly influenced by structural parameters. Hence, understanding the interactions between these parameters and aerodynamic forces is pivotal in predicting and mitigating flutter instabilities. In this paper a parametric study via wind tunnel experimentation is conducted, exploring the effect of variations in frequency ratio and elastic axis location on response dynamics of a 2-DoF aeroelastic system. 
The study reveals diverse dynamic behaviours resulting from the coupled interaction of aerodynamic and structural nonlinearities at critical parameters. Further comprehension of aeroelastic response dynamics is attained through a synchronization framework. It is worth noting that parameters like non-dimensional mass ratio and static AoA are not considered for the present parametric study due to experimental limitations and we glean into our paramteric insights only from variations in frequency ratio and elastic axis location.

To that end, the key findings from this study are summarized as follows:

\begin{itemize}
    \item For $\bar{\omega}$ $\geq$ 0.5, the aeroelastic responses are pitch dominant - indicating stall flutter. Plunge acts as a driven or passive mode.
    \item For $\bar{\omega}$ $<$ 0.5, the plunge mode actively participates in the flutter mechanism. This is manifested via a classical flutter or a 2:1 internal resonance.
    \item Strong synchronization is observed for $\bar{\omega}$ $\geq$ 0.5 cases either via a phase trapping or via an intermittent phase synchronization.
    \item Synchronization is lost during the internal resonance. However, for the case where the system moves away from internal resonance ($\bar{\omega}$ = 0.36), synchronization is again established.
    \item Moving the elastic axis towards the mass center results in the advancement of flutter limits and an increase in LCO amplitudes - particularly in pitch due to increased aerodynamic coupling.
    \item The synchronization is reduced as the elastic axis moves closer to the leading edge and away from the mass center.
    \item Noise-induced random oscillations with small amplitudes are observed under fluctuating inflow conditions. These oscillations start to appear at mean flow speeds below $U_{cr}$ (under deterministic flow with the same structural parameters) and persist even at higher mean speeds than $U_{cr}$.
    \item The phases of pitch and plunge modes are weakly synchronized during the noise-induced oscillations and the overall synchronization dynamics can be characterized as asynchrony.
    \item Only for elastic axis location 0.35$c$ (for $\bar{\omega}$ = 0.57 case), the small amplitude random oscillations culminate into well-developed stall flutter under stochastic inflow. During this, a sudden jump in synchronization dynamics - from asynchronous to intermittent phase synchronization, is observed between the two oscillating modes as the small amplitude random oscillations evolve into high amplitude stall flutter.
\end{itemize}

\appendix
\section{\label{A0} Uncertainty quantification of response data}
\begin{figure}
\begin{center}
\includegraphics[width=5.6in,height=3in]{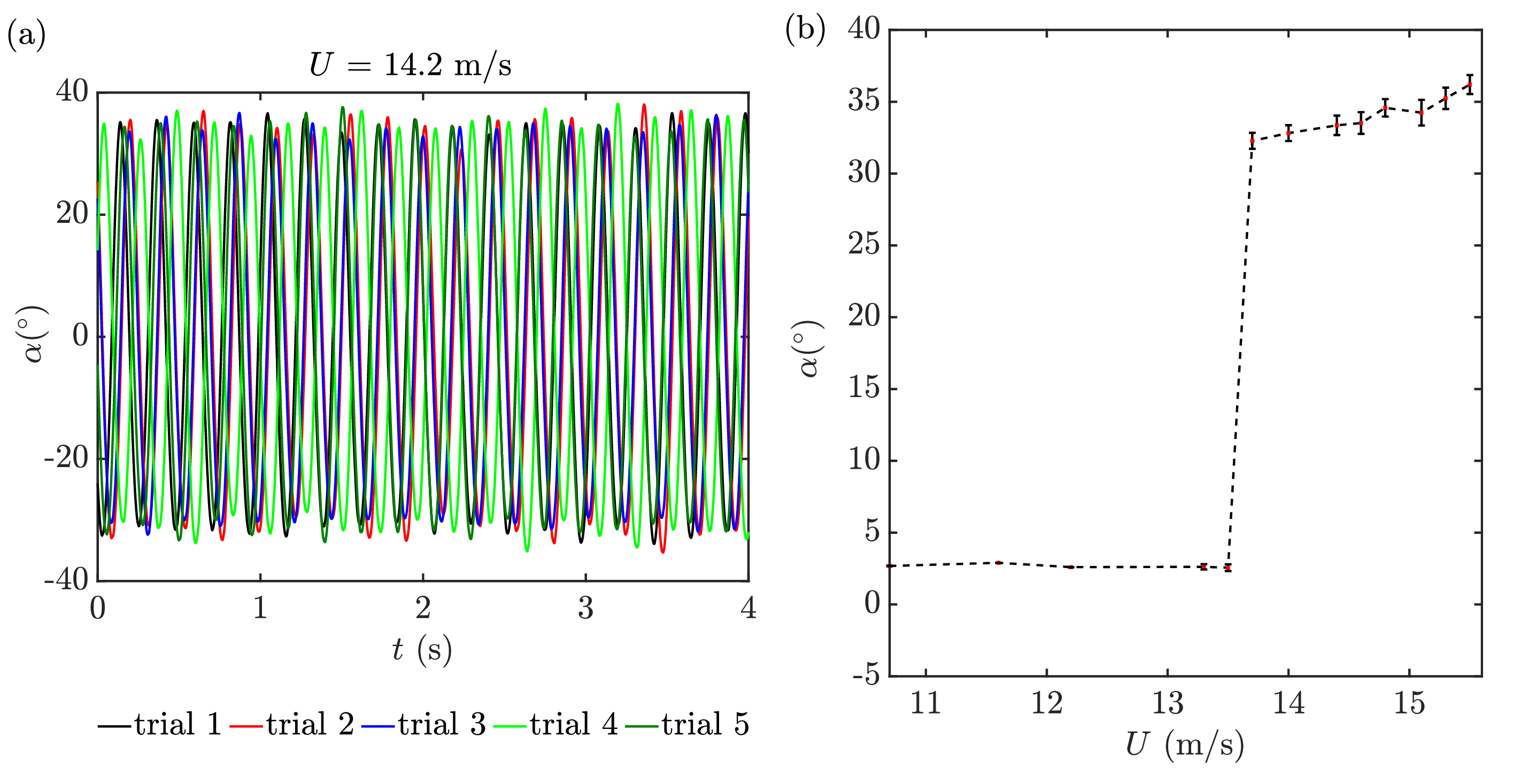}
 \caption{\label{unc_pi} Uncertainty quantification: (a) Sample pitch responses from five trials conducted under specific system parameters and flow speed, (b) Sample pitch bifurcation plot with error bars.}
\end{center}
\end{figure}

The uncertainty in flow speed measurement is found to be within $\pm$ 1$\%$ in suction mode and within $\pm$ 5$\%$ in blowing mode of wind tunnel experimentation. The maximum standard deviation in the measurement of pitch and plunge responses is 1. To illustrate this, sample pitch responses from five different trials conducted under specific system parameters and flow speed are shown in Fig.~\ref{unc_pi}(a). It is evident that the amplitudes closely match, indicating good repeatability from the displacement sensors. Additionally, a sample pitch bifurcation diagram in suction mode (see Fig.~\ref{unc_pi}(b)), with flow speed as the control parameter is depicted with error bars. A maximum standard deviation of 0.89 is observed for the mean amplitude of 34$^{\circ}$ at 15.1 m/s. Note that this mean pitch amplitude falls within the range observed for the majority of experiments in the study.

\section{\label{A1} Frequency-specific synchronization analysis for frequency ratio 0.41}

\begin{figure}
\centering
\includegraphics[width=5in, height=2in]{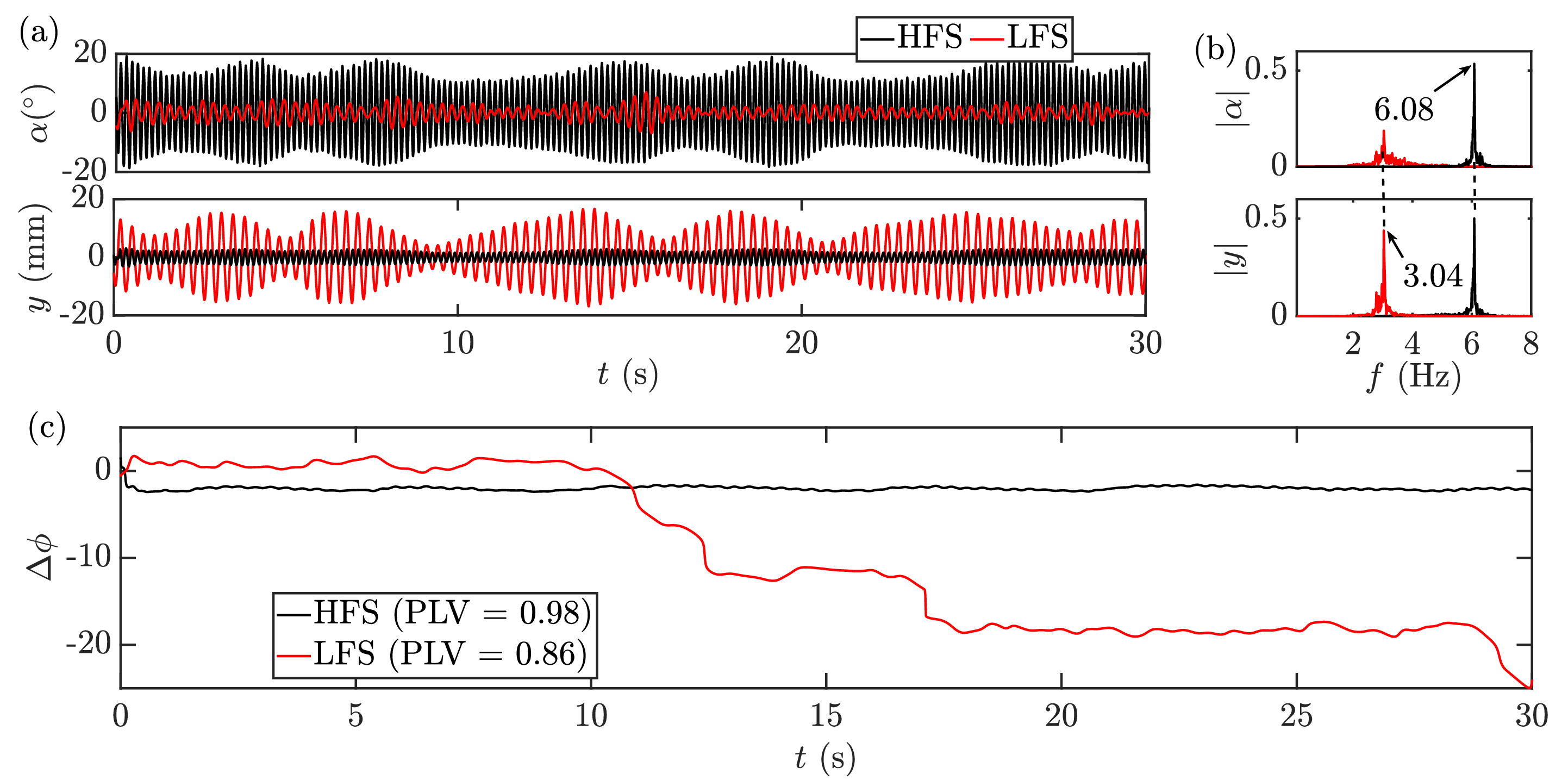}
\caption{Frequency-specific synchronization analysis at $U$ = 14.9 m/s. (a) HFS and LFS signals for pitch and plunge, (b) corresponding frequencies, and (c) Synchronization dynamics.}
\label{FSA_2yc4}
\end{figure}

A brief frequency-specific synchronization analysis is carried out for the case presented in Section~\ref{cat2c4} (Case V: $\bar{\omega}$ = 0.41 and $x_{ea}$ = 0.25c) at $U$ = 14.9 m/s (see Fig.~\ref{thist2yc4}(d)). The original signal (OS) is split into a higher frequency signal (HFS) (5.00 - 10.00 Hz) and a lower frequency signal (LFS) (0.10 -
5.00 Hz) using a band-pass filter for both pitch and plunge modes (see Fig.~\ref{FSA_2yc4}(a)). Figure~\ref{FSA_2yc4}(b) shows the frequency response for each of these signals with each signal carrying only a single dominant frequency. In Fig.~\ref{FSA_2yc4}(c) the synchronization dynamics between HFS-HFS and LFS-LFS of pitch and plunge modes are shown. It can be seen that the HFS of pitch and plunge are phase-locked with a constant RPV over time. The corresponding PLV is 0.98 indicating a perfect synchronization. Even for the LFS, the PLV is observed to be 0.86 with an intermittent phase synchronization, indicating a strong synchronization again. Recall that the synchronization between the original pitch and plunge signals is very weak for this case with a PLV of 0.10 (see Fig.~\ref{phist_2yc4}). This exercise reveals that despite the apparent asynchrony between the pitch and plunge modes, a strong underlying synchronization between inherent signals exists, due to which the motion is sustained with two frequencies.

\section*{Data availability}
The data that support the findings of this study are available from the corresponding author upon reasonable request.
\section*{\label{sec:level6}Acknowledgment}
The last author is immensely grateful for the financial support from the SERB core research grant (CRG/2022/001609) for this research work.
 \bibliographystyle{elsarticle-num} 
 \bibliography{cas-refs}





\end{document}